\documentclass[twocolumn]{aastex702}
\usepackage{amstext}
\usepackage{graphicx}
\usepackage{color}
\usepackage{soul}
\usepackage{amsmath}
\usepackage{amssymb}
\usepackage{booktabs}

\begin{document}

\title{The MASSIVE SURVEY XXI: Local Variations in the Stellar Initial Mass Function of MASSIVE Early-Type Galaxies}

\journalinfo{Submitted to The Astrophysical Journal}

\shorttitle{Local IMF Variations in MASSIVE ETGs}
\shortauthors{Gu et al.}

\author[0000-0002-4267-9344]{Meng Gu}
\email{menggu@tsinghua.edu.cn}
\affiliation{Department of Astronomy, Tsinghua University, Beijing 100084, People's Republic of China}
\affiliation{Hong Kong Institute for Astronomy \& Astrophysics, Pokfulam Road, The University of Hong Kong}

\author[0000-0002-5612-3427]{Jenny E. Greene}
\email{jgreene@astro.princeton.edu}
\affiliation{Department of Astrophysical Sciences, Princeton University, Princeton, NJ, USA}

\author[0000-0001-7769-8660]{Andrew B. Newman}
\email{anewman@carnegiescience.edu}
\affiliation{The Observatories of the Carnegie Institution for Science, 
Pasadena, CA, USA}

\author[0000-0002-4430-102X]{Chung-Pei Ma}
\email{cpma@berkeley.edu}
\affiliation{Department of Astronomy and Department of Physics, University of California at Berkeley, Berkeley, CA, USA}

\author[0000-0002-5213-3548]{John P. Blakeslee}
\email{john.blakeslee@noirlab.edu}
\affiliation{NSF NOIRLab, 950 N. Cherry Avenue, Tucson, AZ 85719, USA}
\affiliation{Steward Observatory, University of Arizona, 933 N. Cherry Avenue, Tucson, AZ 85721}


\begin{abstract}

Extensive evidence suggests that the stellar initial mass function (IMF) varies 
among early-type galaxies (ETGs), but spatially resolved studies 
within individual galaxies are limited in sample size.  We investigate radial 
variations in the low-mass ($\leq1M_{\odot}$) IMF and its connection to stellar 
populations in 37 nearby massive ETGs from the MASSIVE survey. Using high-quality 
Magellan/LDSS$-3$ long-slit spectroscopy spanning $0.4\mu$m$-1.01\mu$m, we 
extract spectra in radial bins reaching outermost radii of $0.2$--$1.1R_{\rm e}$ 
across the sample.  We find that the IMF becomes less bottom-heavy with increasing 
radius in most galaxies.  The sample-averaged IMF mismatch parameter,  
$\alpha_{\rm IMF}=(M/L)/(M/L)_{\rm Kroupa}$, decreases from $2.16$ within 
$R_{\rm e}/8$ to $1.74$ in the $R_{\rm e}/4$--$R_{\rm e}/2$ bin, with 
galaxy-to-galaxy scatters of $0.50$ and $0.42$, respectively.  Thus, the average 
IMF remains more bottom-heavy than Kroupa and approximately Salpeter-like or
more bottom-heavy over these radii. The radial gradients of $\log(\alpha_{\rm IMF})$ 
anti-correlate with the central value of $\alpha_{\rm IMF}$, indicating that 
galaxies with more bottom-heavy central IMFs decline more steeply toward less 
bottom-heavy, approximately Salpeter-like values at larger radii. We find mild 
positive local correlations between $\alpha_{\rm IMF}$ and stellar metallicity, 
but no significant local correlation with [Mg/Fe] or [Na/Fe].  Together with the 
approximately flat profiles of several [$\alpha$/Fe], this  
suggests that IMF variation in massive ETGs is more closely linked to metallicity 
than to the star-formation timescale traced by [$\alpha$/Fe].  Finally, the radial 
variation in stellar $M/L_r$ is dominated by the IMF gradient rather than by the 
stellar-population gradient. A fixed Kroupa IMF underestimates stellar masses by 
factors of $1.7$ and $1.5$ within $R_{\rm e}/2$ and $R_{\rm e}$ in massive ETGs.
\end{abstract} 

\keywords{galaxies: abundances --- galaxies: stellar content --- 
galaxies: formation --- galaxies: evolution --- galaxies: structure --- 
stars: mass function}
\section{Introduction}

The stellar initial mass function (IMF) is a fundamental quantity in astrophysics 
that describes the distribution of stellar masses at birth.  It plays a critical 
role in studies of star formation and the interstellar medium, and it affects 
various physical processes, including stellar evolution, chemical enrichment, 
and stellar feedback.  In the Milky Way, the IMF is commonly described in several 
standard forms \citep{Scalo1986, Bastian2010}, including those of 
\citet{Salpeter1955}, \citet{Kroupa2001}, and \citet{Chabrier2003}.  However, 
studies over the past decade have shown that the IMF is not necessarily universal.  
Star-counting studies of resolved Milky Way dwarf satellite 
galaxies provide evidence for variation in the low-mass IMF. Several ultra 
faint dwarf galaxies appear to have bottom-light IMFs relative to the 
Milky-Way disk, with possible metallicity dependence \citep{Geha2013, 
Gennaro2018}, and the degree of deviation from a Milky-Way-like IMF appears to vary from system to 
system \citep{Filion2024}.  Even within the Milky Way, recent studies suggest 
possible IMF variation, potentially linked to stellar metallicity \citep{Li2023}. 

The IMF directly affects stellar mass measurements in galaxies.  
Although high-mass stars are brighter and dominate the light, the low-mass 
part of the IMF dominates the stellar mass budget, making the IMF essential for 
accurately translating light into physical properties like stellar mass, star 
formation rate (SFR), or star formation history.  Accurate measurements of 
stellar mass also have profound implications for measurements of, e.g., dark matter fraction \citep{Sonnenfeld2015,Tortora2014,Mendel2020},
central supermassive black holes \citep{Simon2024, Thater2023}, and low-frequency gravitational waves due to binary supermassive black holes \citep{Liepold2024}. 
Early-type galaxies (ETGs) are valuable targets for investigating the form of 
the IMF at its low mass end ($\leq1M_{\odot}$). 

The main approaches used for studying the IMF in ETGs include dynamical modeling 
\citep[e.g.][]{Schwarzschild1979, Thomas2011, Dutton2012, Cappellari2013, Li2017, Liepold2020, McConnell2012,Shetty2014}, gravitational lensing 
\citep[e.g.][]{Spiniello2011, Treu2010, Newman2017,Leier2016,Sonnenfeld2019}, 
and stellar population synthesis (SPS) modeling 
\citep[e.g.][]{Cenarro2003, vanDokkum2010, vanDokkum2012, Conroy2012b, 
Villaume2017, Conroy2017,Tang2017}. In particular, the SPS method relies on 
the strengths of absorption features in the optical to near-infrared (NIR) 
that are sensitive to surface gravity \citep{Wing1969, Cohen1978, Cohen1979, Faber1980}.  
Unlike gravitational lensing or stellar kinematics, which constrain the total 
mass, SPS directly probes the stellar component and is crucial in decomposing 
different mass components. 

A general trend of a more bottom-heavy IMF (i.e., more abundant in 
low-mass stars) with increasing velocity dispersion or stellar mass in ETGs 
has been found with both stellar population studies 
\citep[e.g.][]{Ferreras2013,LaBarbera2013,Spiniello2014,Rosani2018,Lagattuta2017} 
and dynamical approaches 
\citep[e.g.][]{Cappellari2012,Dutton2012,Wegner2012,Lasker2013,Posacki2015,Li2017} 
(see \citet{Smith2020} for a review), challenging the idea of a universal IMF. 
Additionally, stellar populations are found to correlate with the IMF when their 
variations are assessed globally.  Metallicity has been found to correlate with 
IMF slope \citep{Conroy2012b, Geha2013}, however, many studies also conclude 
that metallicity cannot be the sole driver of IMF variation 
\citep{Martin-Navarro2019, Villaume2017, LaBarbera2019, Barbosa2021, Cheng2023}. 
$\alpha$ enhancement relative to Fe has been found to vary with IMF 
\citep{Conroy2012b,Smith2012}, suggesting that the approximate star-formation 
timescale may influence the IMF slope.  

In addition, different methods do not always yield consistent IMF constraints for individual objects. 
Although some studies find broad agreement among methods \citep{Conroy2013, 
Tortora2013,Wegner2012}, object-by-object discrepancies remain \citep{Smith2014,Newman2017}.
Several works \citep{Conroy2017,vanDokkum2017} emphasized that the apertures used 
in measurements are often inconsistent, which could be a key issue in the 
inconsistency among object-by-object studies. Therefore, studying the local 
variation or radial distribution of the IMF is crucial for connecting 
measurements with different methodologies, as well as for shedding light on 
the physical processes shaping the IMF.

Previous works have spatially resolved the low-mass IMF slopes 
\citep[e.g.][]{Conroy2017, vanDokkum2017, Parikh2018, Martin-Navarro2015c,
Lonoce2021, Alton2017, Alton2018} in nearby ETGs using the SPS method. 
However, due to the challenges in obtaining high-quality spectra extracted 
in spatial bins, particularly those in the outskirts of galaxies, studies 
of stellar IMF gradients have been limited to a small number of galaxies 
\citep{LaBarbera2019,Martin-Navarro2015c,vanDokkum2017}, or relied on stacked 
spectra \citep{Parikh2018}. Among these studies, whether the IMF generally 
becomes less bottom-heavy with increasing radius (defined as negative 
gradients) is still under debate. Some works 
\citep{Parikh2018, vanDokkum2017, Martin-Navarro2015c, Sarzi2018,Vaughan2018a} 
found negative radial trends, while others find no clear evidence for radial IMF 
variation \citep{Alton2017,Alton2018}.  The local drivers of IMF variation also 
remain uncertain.  \citet{Parikh2018} found a local IMF-$\sigma$ relation 
within galaxies that is steeper than the global relation, whereas 
\citet{vanDokkum2017} found that the local velocity dispersion is not a 
strong predictor of the local IMF.  Both works found that local metallicity 
is closely related to the IMF variations. \citet{vanDokkum2017} found no 
significant local IMF-[Mg/Fe] correlation, in contrast to the global 
IMF-[Mg/Fe] relation suggested by earlier studies.

Dynamical modeling has also been used to test radial IMF variations.
For M87, \citet{Oldham2018} inferred a radial IMF variation.  
\citet{Davis2017} used the kinematics of relaxed molecular gas discs as 
a dynamical tracer and derived spatially resolved $M/L$ for seven ETGs 
in the ATLAS$^{\rm 3D}$ sample, but did not find a significant connection 
between the IMF normalization and stellar population properties. 

The MASSIVE Survey \citep{Ma2014} is a volume-limited, multiwavelength 
survey of the most massive nearby ETGs that has provided a broad observational 
foundation for galaxy structure \citep{Goullaud2018,Quenneville2024}, 
stellar kinematics and dynamical structure \citep{Ene2019, Ene2020, Veale2017a, Veale2017b, Veale2018}, 
stellar populations \citep{Greene2015, Greene2019}, 
black hole masses \citep{Liepold2020, Dominiak2024, Pilawa2022, Pilawa2025}. 
As part of this survey, \citet[][Paper~I]{Gu2022} explored 
41 nearby massive galaxies using the deep Magellan/LDSS$-3$ optical-NIR 
long-slit spectra in the volume limited MASSIVE survey (D$\leq108$~Mpc, 
$\log(M_{\star}/M_{\odot})\geq 11.5$). Paper~I focused on the central regions 
and measured the IMF as well as stellar chemical abundance through full spectral 
modeling. The spectra contain many IMF sensitive features such as Na 
$\lambda0.82\mu m$, the Ca II triplet $\lambda0.86\mu$m, and the FeH band of 
Wing-Ford $\lambda0.99\mu$m. Paper~I revealed that the IMF mismatch parameter, 
$\alpha_{\rm IMF}=(M/L)/(M/L)_{\rm Kroupa}$, is positively correlated with 
both [Mg/Fe] and the estimated total metallicity [Z/H], and suggestively 
correlated with the effective stellar surface density $\Sigma_{\rm Kroupa}$.
Larger values of $\alpha_{\rm IMF}$ indicate a more bottom-heavy IMF, 
implying that ETGs with higher central metallicity and $\alpha$-abundance, 
as well as more compact systems, have more bottom-heavy IMFs in their 
central regions \citep{Martin-Navarro2015a}.

Our sample represents the most massive galaxies in the nearby universe.  In this 
work, we push the limits of the long-slit spectra and explore the local 
variations of the IMF in a sample of 37 massive ETGs. Radial coverage for the 
sample varies from 0.2 to 1.1 $R_{\rm e}$.  Based on the scaling relations 
between the IMF and stellar mass, we expect to find the most extreme IMF in 
ETGs in our sample.  We carefully choose the region to extract spectra and 
adopt two radial binning schemes (\S~2).  We adopt strict data selection criteria 
and measure the gradients of IMF and stellar population properties for each 
galaxy.  We explore correlations between the IMF and stellar population 
properties and galaxy compactness (\S~3).  In \S~4, we estimate the impact on 
stellar mass measurement, compare our results with previous studies, and discuss 
the physical implications and caveats of our results.  We conclude our paper in 
\S~5.
 
\begin{figure*}[th]
\centering
\includegraphics[width=\textwidth]{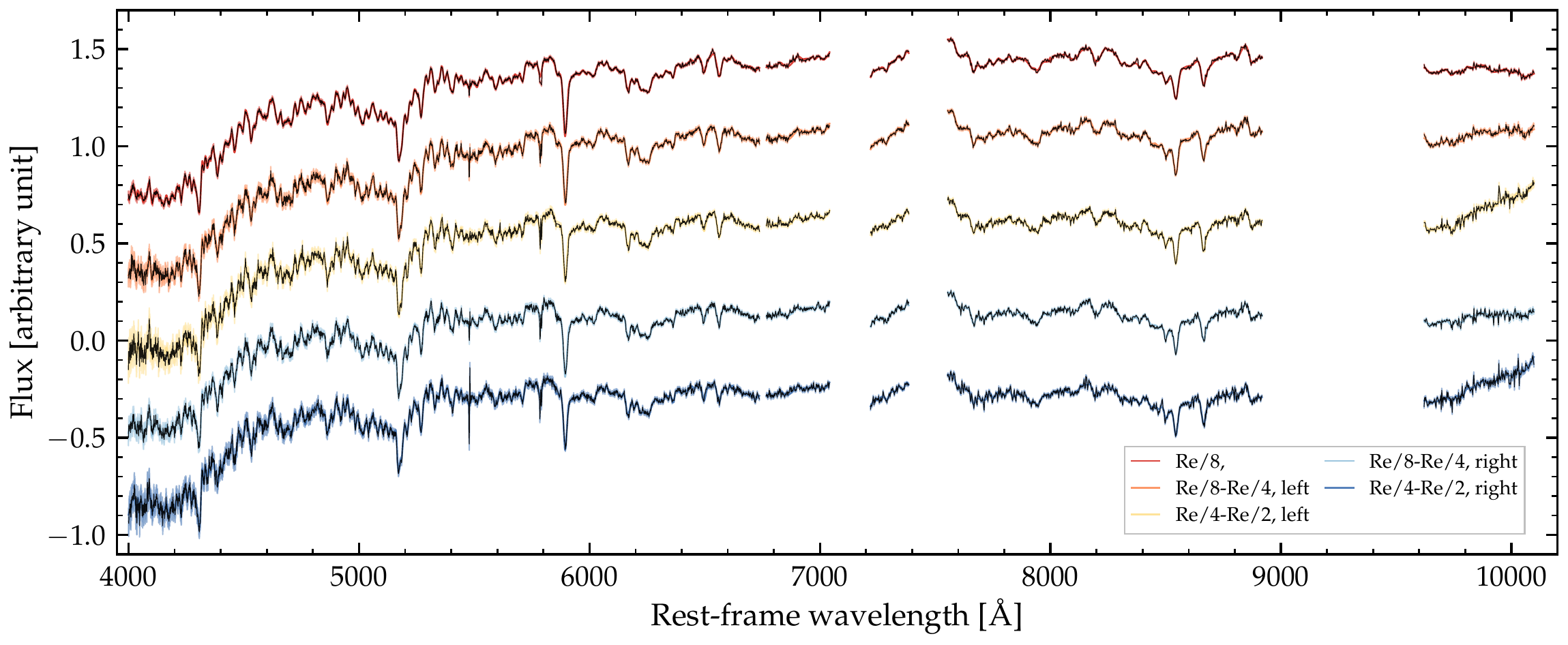}
\caption{
Observed spectra (black) and best-fit spectral models for NGC~0057 in five radial bins
defined by fractional effective radii. Gaps in the spectra
correspond to masked wavelength regions, primarily affected by telluric
absorption.
}
\label{fig_spectra}
\end{figure*}
\begin{figure*}[ht]
\centering 
\includegraphics[width=\textwidth]{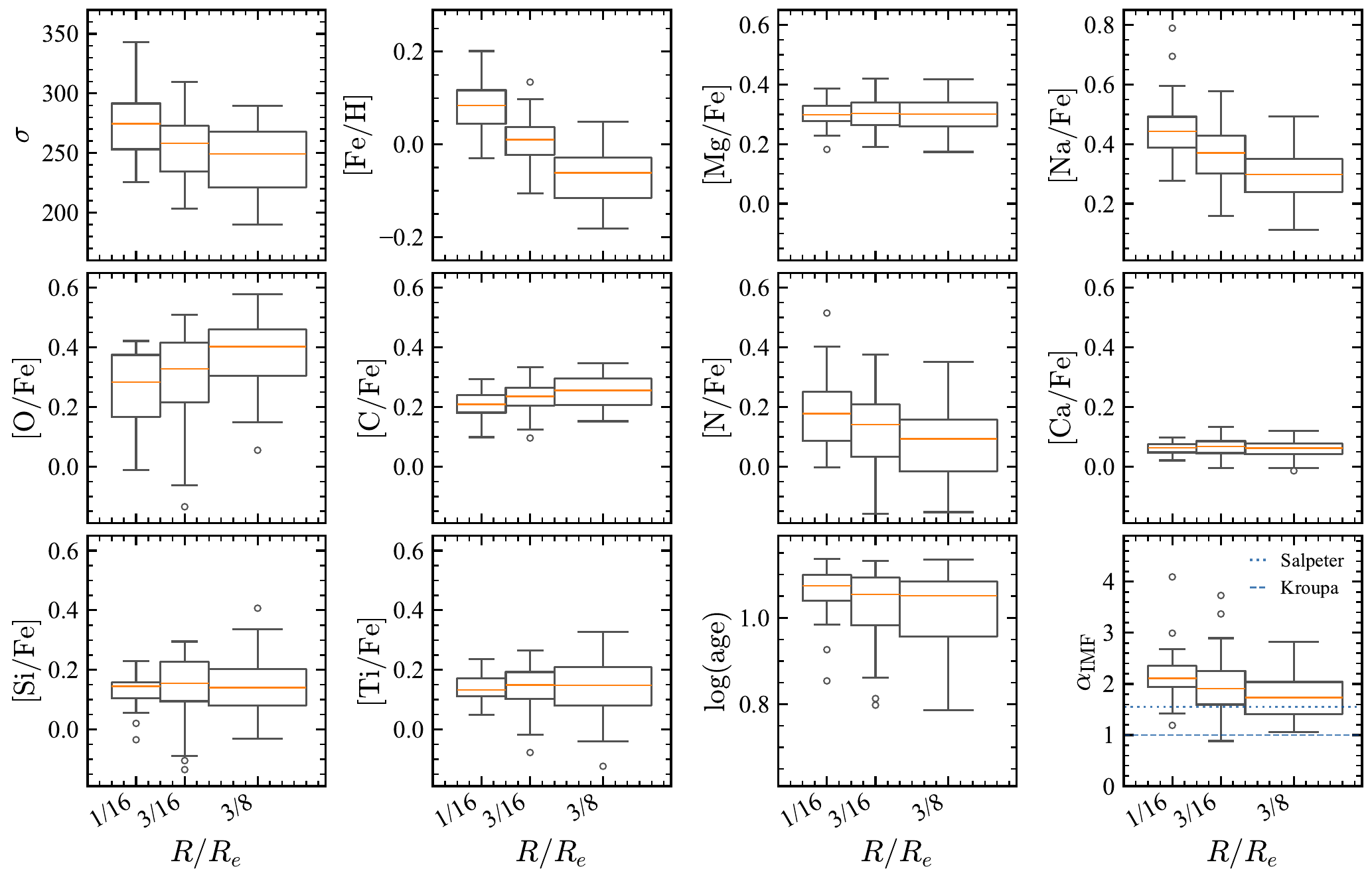}
\caption{
Distributions of the fitted stellar-population parameters in three radial bins 
defined by fractional effective radius. The central bin corresponds to 
$R<R_{\rm e}/8$; the outer bins correspond to $R_{\rm e}/8$--$R_{\rm e}/4$ and 
$R_{\rm e}/4$--$R_{\rm e}/2$.  Each box spans the interquartile range, with the 
orange line marking the median. The whiskers extend to the most extreme data 
points within $1.5$ times the interquartile range from the first and third 
quartiles; points outside this range are shown as outliers.
}
\label{fig_box}
\end{figure*}


\section{Data}

\subsection{Observations and Reduction}

In Paper~I, we presented the central properties of 41 ETGs in the MASSIVE survey, 
including their stellar population and IMF measurements. The observations were 
carried out using the Magellan/LDSS$-3$ optical-NIR spectrograph. The targets were 
selected from the volume-limited MASSIVE survey \citep{Ma2014} within 108~Mpc, 
and the spectra were reduced and extracted within the central $R_{\rm e}/8$ of 
the galaxies. A detailed description of the data reduction procedures can be found 
in Paper~I. 

The goal of this paper is to study the radial variation of galaxy properties and 
the stellar IMF. To achieve this, we extended our spectral extraction beyond 
$R_{\rm e}/8$ for each galaxy, ensuring that the slit orientations among different 
exposures were consistent. 
Of the 41 galaxies analyzed in Paper~I, 37 have usable radial measurements after 
applying the extraction and quality-selection criteria described below. 
Four galaxies are excluded in this radial profile work due to asymmetric or 
unstable radial measurements, and possible stellar-population or emission-line 
complications in the outer bins.  We reserve them for separate study.
In Table~2, we summarize the properties of the 37 galaxies, including slit 
orientation and radial coverage.

We adopt two binning schemes to extract spectra in radial bins. First, we extract 
spectra with radial bins defined by fractional effective radii, with three radial bins 
of $0-1/8R_{\rm e}$, $1/8-1/4R_{\rm e}$, and $1/4-1/2R_{\rm e}$. Additionally, we 
define the aperture in units of pixels and stack the spectra in bins with boundaries 
of 0, 3, 6, 12, 24, and 48 pixels.  Observations were taken with a 
$2\times$ spatial binning scheme, resulting in a final pixel scale of 0.378\arcsec~pixel$^{-1}$.
The ``pixel binning'' approach covered a radial range up to 18\arcsec~and provided the 
finest radial bins, which we use to calculate the radial gradients. We note that not 
all slits were aligned with the major axis of the galaxies.  
Following the extraction procedure in Paper~I \citep[Section~2.5]{Gu2022}, we
assign weights to spatial pixels as a function of projected distance from the
galaxy center to approximate circular-aperture measurements from the long-slit
spectra. For bins outside the central aperture, we perform the extraction
separately on the two sides of the galaxy center, resulting in two spectra per
radial bin.

\subsection{Spectral Modeling}

We use the absorption line fitter ({\tt alf}) \citep{Conroy2012a, Conroy2014, Conroy2018} 
to model the spectra, adopting the same set of parameters as in Paper~I. The 
code is equipped with the MIST stellar isochrones \citep{Choi2016} and the 
MILES$+$E-IRTF spectral library \citep{Villaume2017} with a wide $0.35$--$2.4\mu m$ 
wavelength coverage.  
The models also include theoretical response functions \citep{Kurucz1970, Kurucz1993}. 
These functions describe the fractional change in the model spectrum produced by 
varying the abundance of one element at a time.  The code is capable of fitting the 
stellar age, metallicity, a young component, 19 elemental abundances of Fe, O, C, 
N, Na, Mg, Si, K, Ca, Ti, V, Cr, Mn, Co, Ni, Cu, Sr, Ba, Eu, and a parametric 
form of the IMF.  The difference, compared to Paper~I, is that we adopt a double 
power-law IMF and fix the cutoff mass to 0.08 $M_{\odot}$, while in Paper~I the 
cutoff mass is set to be a free parameter. In Paper~I, we compared the results 
with two IMF parameterizations and found that the fixed low cutoff mass results in 
slightly higher $M/L$ but does not alter our conclusions. According to the mock 
test performed in \citet{Conroy2017}, fitting three IMF variables requires higher 
S/N than fitting two IMF variables.  We chose this IMF format based on the limited 
S/N in radial bins outside the central region of ETGs. Specifically, the two free 
parameters for the stellar IMF are the two logarithmic slopes in 
$dn/dm\propto M^{-\gamma}$ in the mass ranges 
$0.08 M_{\odot} < M < 0.5 M_{\odot}$ ({\tt imf1}) and $0.5 M_{\odot}<M<1 M_{\odot}$ 
({\tt imf2}). As examined in Paper~I, our conclusion is robust 
even if this IMF parameterization introduces a small systematic offset, and we 
expect that at least the relative gradients should be robust in this work.  
In Figure~\ref{fig_spectra} we present the observed spectra and best-fit model 
of NGC~0057 in each radial bin as an example.

\subsection{Selection Criteria}

To ensure the accuracy of our analysis, we select targets based on the 
signal-to-noise ratio (S/N) of the spectra. 
We first require S/N$\geq50$~\AA$^{-1}$ in the $8000$--$9000$~\AA\ region.
Because high S/N alone does not guarantee an acceptable fit over all wavelength
regions, we also apply a cut based on the fractional rms residuals, which 
provides a quantitative measure of the agreement between the observed 
spectrum and the model spectrum. 
We calculate the fractional root-mean-square (rms) deviation using 
\[\text{fractional rms} = \frac{\sqrt{\frac{1}{N}\sum_{i=1}^{N}{(y_{\text{obs}, i} - y_{\text{model}, i})^2}}}{\bar{y}_{\text{obs}}} \] 
Full spectral modeling is performed across six wavelength intervals: 
$0.40-0.47\mu$m, $0.47-0.57\mu$m, $0.57-0.67\mu$m, $0.67-0.80\mu$m, $0.80-0.892\mu$m, 
$0.962-1.01\mu$m. 
We compute the fractional rms separately in each interval and retain only spectra
with fractional rms below $3\%$ in every fitted interval.

This rms requirement primarily removes low-surface-brightness spectra from the
outer parts of galaxies that pass the S/N cut but show poor residuals in the
reddest fitting interval. For example, in the fourth and fifth pixel bins,
covering $4.5$--$9\arcsec$ and $9$--$18\arcsec$, respectively, $29\%$ and
$67\%$ of spectra that meet the S/N criterion are rejected because of high
fractional rms around the Wing-Ford feature in the $0.962$--$1.01\mu$m interval.
As discussed in Paper~I, this wavelength range is particularly sensitive to
residual scattered-light contamination \citep{Gu2022}.

\begin{figure*}
\centering
\includegraphics[width=\textwidth]{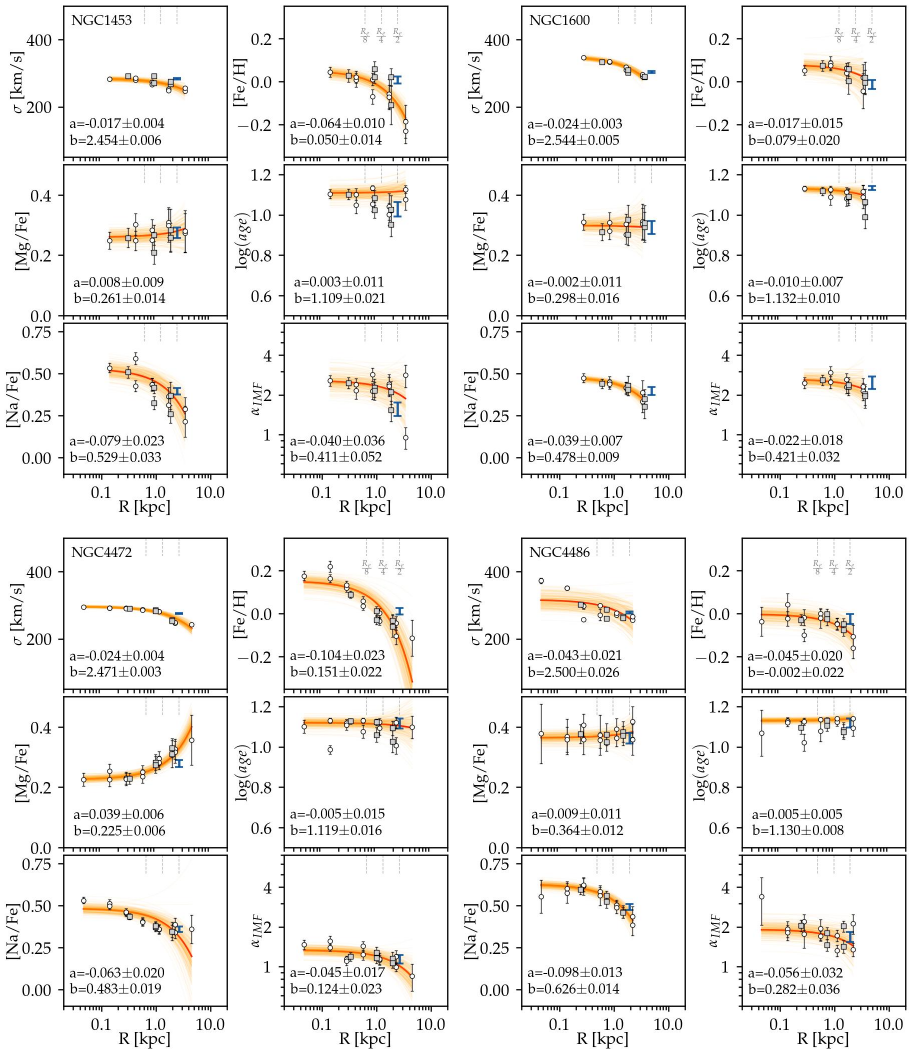}
\caption{
Radial profiles of stellar velocity dispersion, [Fe/H], [Mg/Fe], stellar age, 
[Na/Fe], and IMF mismatch parameter $\alpha_{\rm IMF}$ 
for four representative galaxies: NGC~1453, NGC~1600, NGC~4472, and NGC~4486.
Open black circles show measurements from the pixel-bin spectra, while light-gray 
filled squares show measurements from the fractional-$R_{\rm e}$ bins (see \S~2).
Blue vertical error bars mark the aperture-integrated measurements within 
$R_{\rm e}/2$, plotted at $R_{\rm e}/2$.  Red curves show the best-fit 
radial-gradient model, ${\rm Y}=a(r/{\rm 1~kpc})+b$, evaluated only over the 
radial range used in the fit for each galaxy; thin orange curves show random 
draws from the posterior distribution.  The printed values of $a$ and $b$ 
give the fitted slope and intercept, respectively. For $\sigma$ and 
$\alpha_{\rm IMF}$, the gradients are calculated based on the logarithmic 
value, although the profiles are plotted in linear units. 
The gradient measurements for all galaxies are provided in
Tables~\ref{tab:gradient} and \ref{tab:gradient_cont}.
Results based on different binning schemes are generally consistent with each 
other, and with previous measurements in an aperture of $\mathrm{R}_e/2$.
}
\label{fig_radial}
\end{figure*}
\noindent 

\section{Results}

In this section, we quantify how the stellar populations and IMF vary within 
the MASSIVE galaxies.  We begin with sample-averaged radial trends measured 
in fractional effective radius bins, which show which parameters change
systematically from galaxy centers to larger radii (\S~3.1).  We then divide 
the sample by central IMF mismatch parameter to test whether galaxies with 
more bottom-heavy central IMFs show different radial behavior (\S~3.2).  
Next, we present the radial profiles and gradient measurements for individual 
galaxies (\S~3.3).  We then examine how the measured gradients depend on 
central $\alpha_{\rm IMF}$ and galaxy compactness (\S~3.4). We test local 
relations between $\alpha_{\rm IMF}$ and stellar-population parameters across 
the radial bins (\S~3.5). Finally, we compare several individual galaxies with 
previous IMF-gradient studies (\S~3.6).

\begin{figure}[]
  \centering 
  \includegraphics[width=\columnwidth]{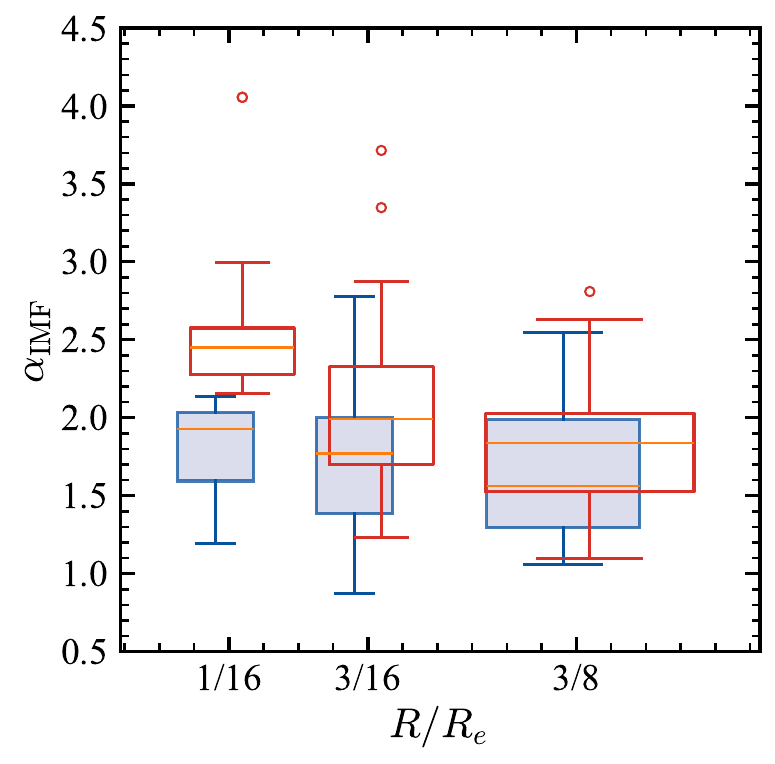}
  \caption{
Distribution of $\alpha_{\rm IMF}$ in three fractional-$R_{\rm e}$ 
radial bins after dividing the sample by the median central value 
measured within $R_{\rm e}/8$, $\alpha_{\rm IMF}=2.15$.  
With the division, we have a group (blue) with 
$\alpha_{\rm IMF}=1.81\pm0.27$ and the other (red) with 
$\alpha_{\rm IMF}=2.55\pm0.45$.  
For the two outer radial bins, measurements from the two sides of 
each galaxy are included separately when they pass the selection 
criteria.  Each box spans the interquartile range, with the horizontal 
line marking the median.  The whiskers extend to the extreme data 
points within $1.5$ times the interquartile range, and open circles 
show points outside this range.  The difference between the two 
central-IMF groups decreases with radius, suggesting that galaxies 
with different central IMF mismatch values become more similar 
toward larger radii.  
  }
\label{fig_divide}
\end{figure}

\begin{figure*}[t]
\centering
\includegraphics[width=15cm]{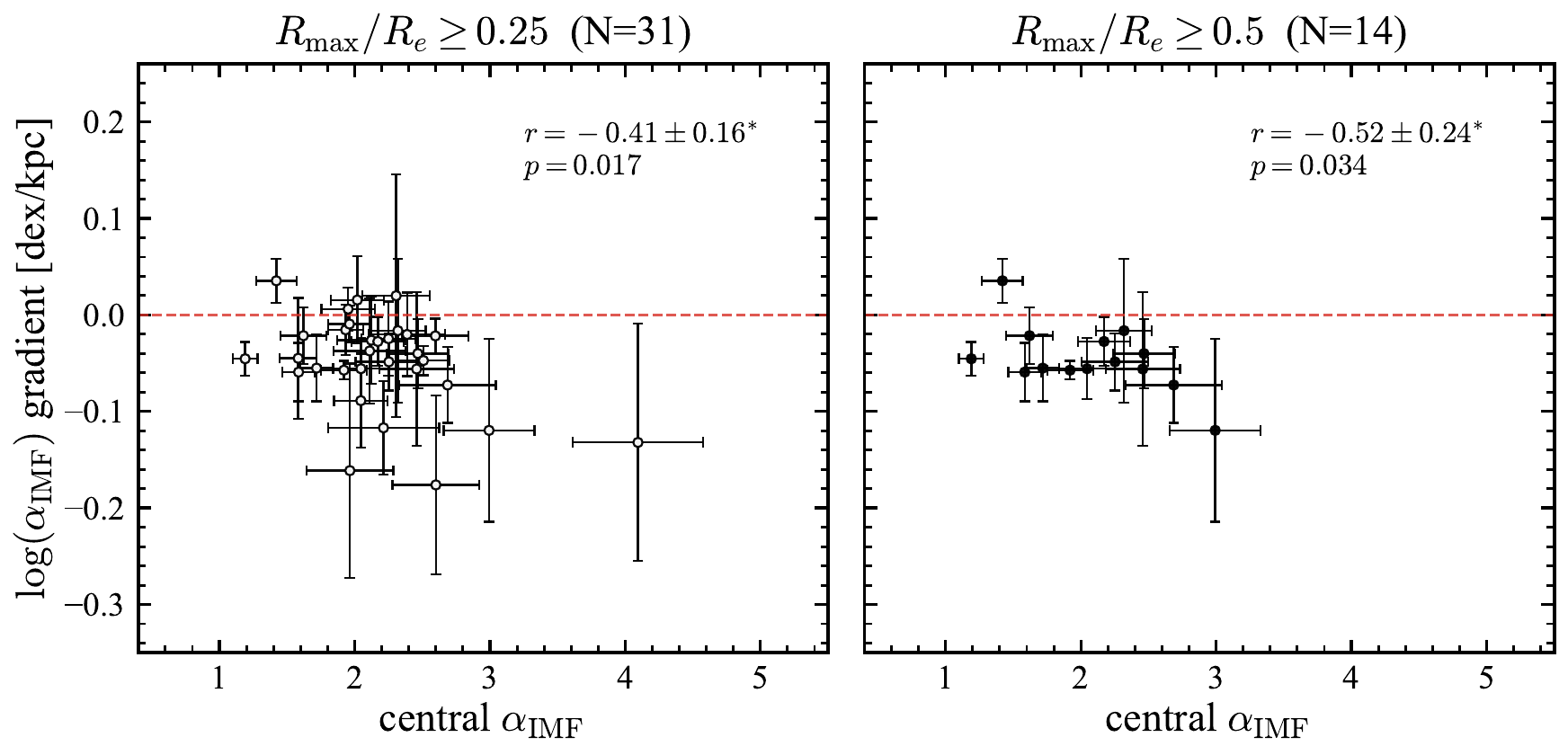}
\caption{
Radial gradient of $\log(\alpha_{\rm IMF})$ as a function of central 
$\alpha_{\rm IMF}$.  The left panel includes galaxies with radial coverage 
extending beyond $0.25R_{\rm e}$, while the right panel includes only galaxies 
with coverage beyond $0.5R_{\rm e}$.  We measure the central values using 
spectra stacked within $R_{\rm e}/8$, and the gradients are in units of 
dex~kpc$^{-1}$. The anti-correlation indicates that galaxies with more 
bottom-heavy central IMFs tend to have more negative IMF gradients, so 
their IMF mismatch parameters converge toward less bottom-heavy values at 
larger radii.
 }
\label{fig_gradvscent}
\end{figure*}

\subsection{Radial Trends of Stellar Populations and IMF}

We begin by presenting sample averaged radial trends of stellar populations and 
IMF using measurements in the fractional effective-radius bins. For each 
galaxy, spectra are extracted within $R_{\rm e}/8$, from $R_{\rm e}/8$ to
$R_{\rm e}/4$, and from $R_{\rm e}/4$ to $R_{\rm e}/2$.  We select galaxies 
with old central stellar ages ($>7$~Gyr). The two relatively young galaxies, 
NGC~1700 and NGC~3462 with central ages around 5~Gyr 
are excluded from this comparison and discussed in Appendix~\ref{app:young}.
After applying the selection 
criteria described in \S~2.3, the sample contains 35 measurements in the 
central bin, 68 measurements in the $R_{\rm e}/8$--$R_{\rm e}/4$ bin, 
and 60 measurements in the $R_{\rm e}/4$--$R_{\rm e}/2$ bin.
We refer to these 35 old galaxies as the fiducial sample and use them for
sample-averaged statistical analyses unless otherwise noted.

\begin{table}[!b]
\centering
\begin{tabular}{l|r|r|r}
\toprule
Frac. Re & $0-R_e/$8 & $R_e/8$-$R_e/4$ & $R_e/4$-$R_e/$2 \\
\midrule
$\sigma$ & 275 $\pm$ 30 & 257 $\pm$ 28 & 246 $\pm$ 27 \\
$\mathrm{[Fe/H]}$ & 0.08 $\pm$ 0.05 & 0.01 $\pm$ 0.05 & -0.07 $\pm$ 0.06 \\
$\mathrm{[Mg/Fe]}$ & 0.30 $\pm$ 0.04 & 0.30 $\pm$ 0.05 & 0.30 $\pm$ 0.05 \\
$\mathrm{[Na/Fe]}$ & 0.45 $\pm$ 0.10 & 0.37 $\pm$ 0.10 & 0.30 $\pm$ 0.09 \\
$\mathrm{[O/Fe]}$ & 0.27 $\pm$ 0.12 & 0.30 $\pm$ 0.14 & 0.38 $\pm$ 0.11 \\
$\mathrm{[C/Fe]}$ & 0.21 $\pm$ 0.04 & 0.23 $\pm$ 0.05 & 0.25 $\pm$ 0.05 \\
$\mathrm{[N/Fe]}$ & 0.18 $\pm$ 0.11 & 0.12 $\pm$ 0.13 & 0.08 $\pm$ 0.13 \\
$\mathrm{[Ca/Fe]}$ & 0.06 $\pm$ 0.02 & 0.06 $\pm$ 0.03 & 0.06 $\pm$ 0.03 \\
$\mathrm{[Si/Fe]}$ & 0.13 $\pm$ 0.05 & 0.15 $\pm$ 0.10 & 0.15 $\pm$ 0.09 \\
$\mathrm{[Ti/Fe]}$ & 0.14 $\pm$ 0.05 & 0.14 $\pm$ 0.06 & 0.15 $\pm$ 0.09 \\
$\log({\rm age})$ & 1.06 $\pm$ 0.06 & 1.03 $\pm$ 0.08 & 1.02 $\pm$ 0.09 \\
$\alpha_{\rm IMF}$ & 2.16 $\pm$ 0.50 & 1.96 $\pm$ 0.55 & 1.74 $\pm$ 0.42 \\
\bottomrule
\end{tabular}

\caption{Mean values and scatters of stellar-population and IMF parameters in
the three fractional effective-radius bins used in Figure~\ref{fig_box}.}
\label{tab1}
\end{table}

We summarize the distribution of galaxy properties in fractional $R_{\rm e}$ bins 
using box and whisker plots in Figure~\ref{fig_box}. These plots provide an overview 
of the distribution of 12 parameters, including $\sigma$, [Fe/H], [Mg/Fe], [Na/Fe], 
[O/Fe], [C/Fe], [N/Fe], [Ca/Fe], [Si/Fe], [Ti/Fe], $\log({\rm age})$, and the IMF 
mismatch parameter $\alpha_{\rm IMF}$.  The mean and standard deviation are 
summarized in Table~\ref{tab1}.  Figure~\ref{fig_box} presents a comprehensive 
summary of the general radial trends of several properties of the most massive 
ETGs in the nearby universe within their half $R_{\rm e}$. 

To assess whether each parameter shows a statistically significant radial trend 
across the sample, we apply an inverse-variance weighted one-way ANOVA test to 
the best-fit measurements shown in Figure~\ref{fig_box}. For each parameter, 
the measurements are grouped into three fractional radial bins: 
$0$--$R_{\rm e}/8$, $R_{\rm e}/8$--$R_{\rm e}/4$, and $R_{\rm e}/4$--$R_{\rm e}/2$. 
The central bin contributes one spectrum per galaxy, while the two outer bins 
can include spectra extracted separately from the two sides of the galaxy 
center, provided they pass the S/N and rms selection criteria described in \S~2.3. 

\begin{figure*}[t!]
\centering
\includegraphics[width=17cm]{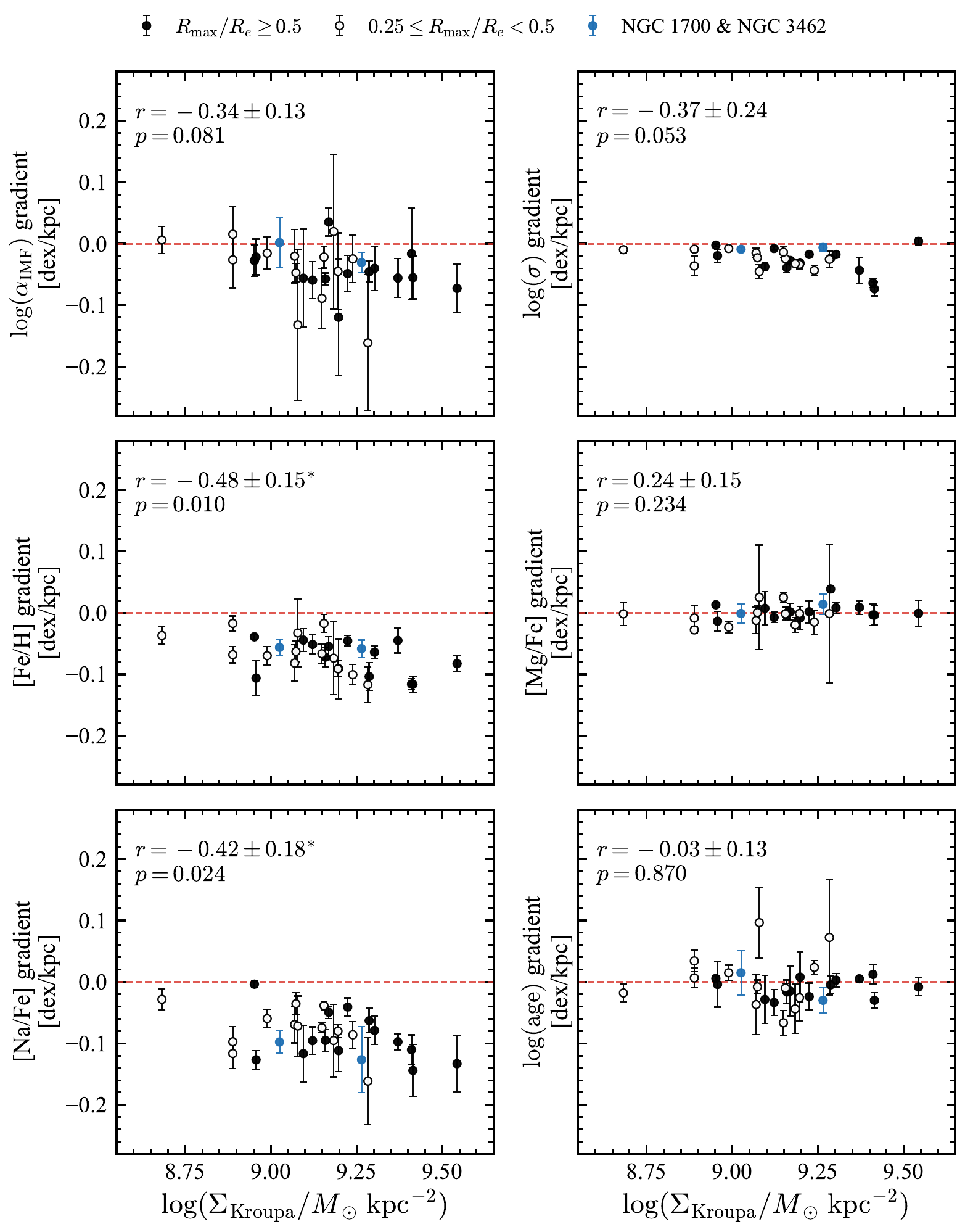}
\caption{
Radial gradients as a function of effective stellar surface density 
$\Sigma_{\rm Kroupa}$.  For the 27 old galaxies with radial coverage extending 
beyond $0.25R_{\rm e}$ and with surface-density measurements, we find statistically significant anti-correlations 
between compactness and the gradients of [Fe/H] and [Na/Fe].  The gradients 
of $\log(\alpha_{\rm IMF})$ and $\log(\sigma)$ also show a negative, though 
only marginally significant, correlation with compactness ($p=0.081$ and $p=0.053$, respectively).  We 
do not find statistically significant trends between 
$\log(\Sigma_{\rm Kroupa})$ and the gradients of [Mg/Fe] or 
$\log({\rm age})$.
}
\vspace{10mm}
\label{fig_gradient_compactness}
\end{figure*}

The ANOVA test evaluates whether the mean values of a given parameter differ 
significantly among these three radial bins. Using a significance threshold 
of $p<0.05$, we find significant evidence for sample-averaged radial variation 
in $\sigma$, [Fe/H], [Na/Fe], [O/Fe], [C/Fe], and $\alpha_{\rm IMF}$; all of 
these parameters have ANOVA $p$-values below $0.01$.
Based on the median values shown in Figure~\ref{fig_box}, $\sigma$, [Fe/H], 
[Na/Fe], and $\alpha_{\rm IMF}$ decrease with radius, while [O/Fe] and [C/Fe] 
increase with radius.

For parameters with ANOVA $p$-value $<5\%$, we then apply Tukey's honestly
significant difference (HSD) test to identify which pairs of radial bins differ.
For $\alpha_{\rm IMF}$, the Tukey HSD test shows that the two inner bins are not
statistically distinct from each other.  The outermost bin,
$R_{\rm e}/4$--$R_{\rm e}/2$, is significantly different from both inner bins.
This indicates that the sample-averaged IMF mismatch parameter becomes
significantly lower in the outer bin, rather than showing a statistically
resolved change between the central bin and the $R_{\rm e}/8$--$R_{\rm e}/4$ bin.

For [Mg/Fe], [Ca/Fe], [Si/Fe], [Ti/Fe], and $\log({\rm age})$, the ANOVA
$p$-values are greater than 0.05.  We therefore find no evidence for
sample-averaged radial variation in these quantities over the radial range
probed here.

The abundance ratios of $\alpha$ elements, including Mg, O, Ca, Si, and Ti, 
relative to Fe are traditionally regarded as an indicator of star-formation 
timescale, as they are sensitive to  the time delay between core-collapse and 
Type Ia supernovae \citep[e.g.][]{Tinsley1979}.  
In our sample, the radial behavior of the IMF mismatch parameter is closer 
to that of [Fe/H] and [Na/Fe] than to the $\alpha$-enhancement.
[Mg/Fe], [Ca/Fe], [Ti/Fe], and [Si/Fe] are approximately flat with radius, while
[Fe/H], [Na/Fe], and $\alpha_{\rm IMF}$ decline. Although O is formally 
an $\alpha$ element, the fitted [O/Fe] profile rises with radius and differs 
from the behavior of Mg, Ca, Ti, and Si.  We treat this result with caution 
because the optical spectral features that constrain O can also depend on C 
and N abundances through molecular absorption, making the inferred [O/Fe] 
potentially covariant with C and N abundances \citep{Ting2018}.  Our results 
indicate that these two sets of parameters are likely shaped by different 
physical processes, suggesting that radial IMF variation is more closely 
associated with metallicity-sensitive stellar-population changes than with 
the nearly flat [Mg/Fe] profile, often used as a proxy for star-formation 
timescale. 

\subsection{Dividing the Sample based on Central IMF}

In this section, we explore whether galaxies with different central IMF 
mismatch parameters, $\alpha_{\rm IMF}$, show different radial IMF behavior. 
In Figure~\ref{fig_divide}, we categorize these galaxies into two distinct 
groups. For this analysis, we use the measurements that reach the outermost 
fractional radial bin, $R_{\rm e}/4$--$R_{\rm e}/2$.  This bin contains 60 
side measurements from 32 galaxies; 28 galaxies contribute measurements on 
both sides of the galaxy center. 

We divide the side measurements into two groups according to the central IMF
mismatch parameter of their host galaxy, measured within $R_{\rm e}/8$.  The
division is made at the median central value, $\alpha_{\rm IMF}=2.15$.  The
low-central-IMF group has a mean central value of
$\alpha_{\rm IMF}=1.81\pm0.27$, while the high-central-IMF group has
$\alpha_{\rm IMF}=2.55\pm0.45$. These central distributions are separated by
construction and are shown in different colors in Figure~\ref{fig_divide}.

As we examine the radial profiles toward larger radii, the two distributions
move closer together.  In the outermost radial bin, $R_{\rm e}/4$--$R_{\rm e}/2$, 
we find $\alpha_{\rm IMF}=1.80\pm0.40$ and $1.63\pm0.42$ for the high- and
low-central-IMF groups, respectively.  This suggests that the IMF mismatch
parameters converge toward Salpeter-like values by $\sim R_{\rm e}/2$, despite
their different central normalizations.  The result has implications for mass
measurements, which we discuss further in \S~4.

\subsection{Gradients in Individual Galaxies}

We first fit the spectra extracted in pixel bins for each galaxy after 
applying the selection criteria described in \S~2.3. These spectral fits 
provide the radial-bin measurements of stellar velocity dispersion, 
stellar-population parameters, and $\alpha_{\rm IMF}$. We then fit these 
radial-bin measurements as a function of deprojected radius to determine 
the gradients listed in Tables~\ref{tab:gradient} and \ref{tab:gradient_cont}. In 
Figure~\ref{fig_radial}, we show four representative galaxies and their radial 
profiles of six parameters: stellar velocity dispersion, age, metallicity, 
[Mg/Fe], [Na/Fe], and the IMF mismatch parameter $\alpha_{\rm IMF}$.  The 
figure includes measurements from both fractional-$R_{\rm e}$ bins and the pixel 
bins described in \S~2.1. The two binning schemes generally give consistent 
results. We also compare the radial-bin measurements with the aperture-integrated
measurement within $R_{\rm e}/2$, shown as a blue error bar at $R_{\rm e}/2$.  
The locations of $R_{\rm e}/8$, $R_{\rm e}/4$ and $R_{\rm e}/2$ are marked 
with short gray ticks.  The radial range available for gradient calculation 
is shown in Table~\ref{tab:gradient}.

To calculate the radial gradient of each parameter, we apply a robust linear
regression using a Student's $t$-distribution likelihood, which is less 
sensitive to outliers than a standard Gaussian likelihood. 
The fitted relation is 

\[Y = a\times\frac{r}{\rm 1kpc}+b \] 

Here, $Y$ denotes one of the fitted properties:
$\log(\sigma~{\rm [km~s^{-1}]})$,  $\log({\rm age/Gyr})$, [Fe/H], [X/Fe] 
and $\log(\alpha_{\rm IMF})$. 
Because the slit orientation varies among galaxies, we convert the projected 
slit radius to the corresponding semi-major-axis radius of the galaxy 
isophote and use this deprojected radius, $r$, in kpc.  The coefficient $a$ 
is therefore the radial gradient, in units of dex kpc$^{-1}$, and $b$ is 
the intercept.

The Student's $t$ likelihood includes the degrees of freedom, $\nu$, as an
additional free parameter. It controls the heaviness of the distribution's 
tails and affects the robustness of the model against outliers. In 
Figure~\ref{fig_radial} the thick red curves show the radial profile model 
constructed from the posterior mean slopes and intercept.  The printed 
values show the mean and $1\sigma$ uncertainties of the slope and intercept 
for each galaxy.  Thin orange curves show 200 random draws from the posterior 
distributions, showing the range of allowed radial profiles.

Many galaxies show declining profiles in $\sigma$, [Fe/H], and [Na/Fe].  The
IMF gradients are also predominantly flat or negative: in the 
fiducial sample, 30 galaxies have declining $\log(\alpha_{\rm IMF})$ profiles
and 5 have positive gradients. 
It is important to note that the radial coverage differs from galaxy
to galaxy due to differences in the signal-to-noise (S/N) ratio of the data, so 
these fitted gradients should be interpreted only over the measured
radial range listed in Table~\ref{tab:gradient}. Obtaining higher quality data 
with larger apertures in the future will enhance our understanding of these gradients.

\subsection{Distribution of Gradients}

The gradients measured for individual galaxies are listed in
Tables~\ref{tab:gradient} and \ref{tab:gradient_cont}.  We focus on how the IMF gradients
depend on central IMF mismatch and galaxy compactness, which show the clearest
trends.  For completeness, Appendix~\ref{app:gradient_sigma} shows the radial
gradients of $\log(\sigma)$, $\log({\rm age})$, [Fe/H], [Mg/Fe], [Na/Fe], and
$\log(\alpha_{\rm IMF})$ as a function of central velocity dispersion.  All
gradients are measured with respect to deprojected radius and are reported in
units of dex~kpc$^{-1}$.

In Figure~\ref{fig_gradvscent}, we show the gradient of $\log(\alpha_{\rm IMF})$ 
as a function of central $\alpha_{\rm IMF}$ for two radial-coverage cuts.  For 
the fiducial sample, the left panel includes galaxies with 
$R_{\max}/R_{\rm e}\geq0.25$, while the right panel includes only galaxies with 
$R_{\max}/R_{\rm e}\geq0.5$.  In both cases, the IMF gradient anti-correlates 
with central $\alpha_{\rm IMF}$, indicating that galaxies with more bottom-heavy 
central IMFs tend to have steeper declining IMF profiles. Our findings suggest 
that the gradient becomes more negative with higher central values, implying 
that the IMF converges toward the outskirts.  This behavior is consistent with 
the sample-averaged radial trend shown in Figure~\ref{fig_box}, supporting the 
idea that the range of IMF mismatch values narrows at larger radii.  Measurements 
extending to larger radii will be needed to test whether this convergence 
continues into the outskirts.

\begin{deluxetable*}{lcccccccccc}
\tabletypesize{\scriptsize}
\tablewidth{0pt}
\tablecaption{Radial gradient measurements.\label{tab:gradient}}

\tablehead{
\colhead{Name} & \colhead{Range} & \colhead{Range} &
\colhead{PA$_{\rm gal}$} & \colhead{PA$_{\rm slit}$} &
\multicolumn{2}{c}{$\log(\sigma)$} &
\multicolumn{2}{c}{[Fe/H]} &
\multicolumn{2}{c}{[Mg/Fe]} \\
\colhead{} & \colhead{[$R_{\rm e}$]} & \colhead{[kpc]} &
\colhead{[deg]} & \colhead{[deg]} &
\colhead{$a$} & \colhead{$b$} &
\colhead{$a$} & \colhead{$b$} &
\colhead{$a$} & \colhead{$b$}
}

\startdata
 NGC~0057 & 0.45 & 2.48 & 40.9 & 162.9 & $-0.043\pm0.008$ & $2.490\pm0.010$ & $-0.101\pm0.017$ & $0.140\pm0.020$ & $-0.015\pm0.020$ & $0.355\pm0.022$ \\
 NGC~0080 & 0.33 & 2.74 & 8.9 & 169.9 & $-0.036\pm0.016$ & $2.399\pm0.016$ & $-0.068\pm0.013$ & $0.092\pm0.015$ & $-0.009\pm0.021$ & $0.294\pm0.024$ \\
 NGC~0533 & 0.28 & 2.90 & 50.1 & 179.0 & $-0.009\pm0.005$ & $2.405\pm0.009$ & $-0.017\pm0.012$ & $0.034\pm0.019$ & $-0.028\pm0.005$ & $0.367\pm0.008$ \\
 NGC~0741 & 0.30 & 2.53 & 83.8 & 138.4 & $-0.008\pm0.003$ & $2.426\pm0.003$ & $-0.070\pm0.015$ & $0.114\pm0.019$ & $-0.023\pm0.010$ & $0.295\pm0.012$ \\
 NGC~1016 & 0.35 & 2.89 & 41.7 & 2.3 & $-0.015\pm0.007$ & $2.465\pm0.009$ & $-0.082\pm0.030$ & $0.192\pm0.042$ & $-0.012\pm0.021$ & $0.330\pm0.028$ \\
 NGC~1453 & 0.69 & 3.39 & 34.6 & 151.5 & $-0.017\pm0.004$ & $2.454\pm0.005$ & $-0.064\pm0.010$ & $0.050\pm0.014$ & $0.008\pm0.009$ & $0.261\pm0.014$ \\
 NGC~1600 & 0.35 & 3.35 & 8.9 & 122.5 & $-0.024\pm0.003$ & $2.544\pm0.005$ & $-0.017\pm0.015$ & $0.079\pm0.020$ & $-0.002\pm0.011$ & $0.298\pm0.016$ \\
 NGC~1700 & 1.06 & 4.38 & 105.2 & 49.4 & $-0.006\pm0.006$ & $2.358\pm0.007$ & $-0.058\pm0.014$ & $0.177\pm0.018$ & $0.014\pm0.017$ & $0.206\pm0.021$ \\
 NGC~2418 & 0.26 & 1.35 & 45.1 & 7.8 & $-0.025\pm0.014$ & $2.376\pm0.012$ & $-0.117\pm0.029$ & $0.124\pm0.023$ & $-0.001\pm0.113$ & $0.259\pm0.081$ \\
 NGC~2513 & 0.41 & 2.54 & 171.0 & 23.2 & $-0.034\pm0.007$ & $2.456\pm0.008$ & $-0.091\pm0.013$ & $0.157\pm0.013$ & $-0.001\pm0.002$ & $0.283\pm0.003$ \\
 NGC~2672 & 0.35 & 2.36 & 123.9 & 164.8 & $-0.014\pm0.009$ & $2.421\pm0.009$ & $-0.067\pm0.016$ & $0.136\pm0.017$ & $0.025\pm0.008$ & $0.274\pm0.009$ \\
 NGC~3462 & 0.75 & 4.35 & 49.0 & 149.3 & $-0.009\pm0.003$ & $2.355\pm0.005$ & $-0.056\pm0.013$ & $0.235\pm0.029$ & $-0.001\pm0.015$ & $0.177\pm0.033$ \\
 NGC~3615 & 0.63 & 4.03 & 42.2 & 174.1 & $-0.027\pm0.005$ & $2.426\pm0.010$ & $-0.055\pm0.016$ & $0.150\pm0.026$ & $0.002\pm0.014$ & $0.202\pm0.027$ \\
 NGC~3842 & 0.40 & 3.04 & 171.9 & 32.4 & $-0.023\pm0.002$ & $2.461\pm0.004$ & $-0.063\pm0.017$ & $0.074\pm0.026$ & $0.001\pm0.012$ & $0.339\pm0.021$ \\
 NGC~3862 & 0.25 & 1.64 & 154.8 & 156.6 & $-0.045\pm0.011$ & $2.425\pm0.013$ & $-0.033\pm0.056$ & $0.106\pm0.061$ & $0.025\pm0.085$ & $0.288\pm0.091$ \\
 NGC~3937 & 0.52 & 3.60 & 23.4 & 165.9 & $-0.037\pm0.006$ & $2.468\pm0.006$ & $-0.044\pm0.019$ & $0.105\pm0.019$ & $0.007\pm0.027$ & $0.260\pm0.037$ \\
 NGC~4055 & 0.34 & 1.77 & -- & 155.9 & $-0.069\pm0.045$ & $2.431\pm0.048$ & $-0.061\pm0.026$ & $0.081\pm0.030$ & $-0.005\pm0.073$ & $0.296\pm0.077$ \\
 NGC~4073 & 0.54 & 6.77 & 101.1 & 47.2 & $-0.002\pm0.002$ & $2.466\pm0.005$ & $-0.039\pm0.005$ & $0.081\pm0.012$ & $0.013\pm0.004$ & $0.309\pm0.009$ \\
 NGC~4472 & 0.85 & 4.52 & 142.7 & 17.9 & $-0.024\pm0.004$ & $2.471\pm0.003$ & $-0.104\pm0.023$ & $0.151\pm0.022$ & $0.039\pm0.006$ & $0.225\pm0.006$ \\
 NGC~4486 & 0.57 & 2.21 & 152.0 & 35.3 & $-0.043\pm0.021$ & $2.500\pm0.026$ & $-0.045\pm0.020$ & $-0.003\pm0.022$ & $0.009\pm0.011$ & $0.364\pm0.012$ \\
 NGC~4555 & 0.70 & 4.50 & 118.8 & 6.6 & $-0.017\pm0.003$ & $2.508\pm0.007$ & $-0.045\pm0.009$ & $0.130\pm0.015$ & $0.002\pm0.018$ & $0.321\pm0.036$ \\
 NGC~4649 & 0.70 & 2.75 & 116.0 & 38.3 & $-0.073\pm0.011$ & $2.543\pm0.009$ & $-0.116\pm0.013$ & $0.161\pm0.009$ & $-0.004\pm0.016$ & $0.336\pm0.012$ \\
 NGC~4839 & 0.41 & 5.39 & 62.9 & 165.4 & $-0.010\pm0.005$ & $2.424\pm0.010$ & $-0.037\pm0.015$ & $0.011\pm0.025$ & $-0.002\pm0.019$ & $0.342\pm0.033$ \\
 NGC~4874 & 0.21 & 3.38 & 62.8 & 18.5 & $-0.029\pm0.005$ & $2.429\pm0.010$ & $-0.027\pm0.018$ & $0.045\pm0.023$ & $0.008\pm0.015$ & $0.306\pm0.024$ \\
 NGC~5490 & 0.65 & 2.50 & 7.2 & 36.7 & $-0.064\pm0.007$ & $2.538\pm0.008$ & $-0.116\pm0.009$ & $0.103\pm0.010$ & $-0.004\pm0.017$ & $0.325\pm0.022$ \\
 NGC~6375 & 0.62 & 3.61 & 140.3 & 33.8 & $-0.032\pm0.006$ & $2.357\pm0.008$ & $-0.091\pm0.049$ & $0.128\pm0.068$ & $-0.008\pm0.019$ & $0.248\pm0.028$ \\
 NGC~6442 & 0.31 & 1.87 & 130.4 & 0.0 & $-0.032\pm0.005$ & $2.390\pm0.005$ & $-0.074\pm0.060$ & $0.091\pm0.063$ & $-0.020\pm0.012$ & $0.271\pm0.014$ \\
 NGC~6482 & 0.67 & 2.31 & 58.3 & 173.6 & $0.004\pm0.006$ & $2.436\pm0.008$ & $-0.083\pm0.013$ & $0.097\pm0.015$ & $-0.001\pm0.021$ & $0.315\pm0.028$ \\
 NGC~7052 & 0.43 & 2.89 & -- & 173.0 & $-0.007\pm0.008$ & $2.442\pm0.013$ & $-0.062\pm0.017$ & $0.058\pm0.020$ & $-0.013\pm0.022$ & $0.305\pm0.033$ \\
 NGC~7619 & 0.57 & 3.08 & 47.3 & 135.6 & $-0.039\pm0.008$ & $2.491\pm0.012$ & $-0.072\pm0.016$ & $0.139\pm0.022$ & $-0.001\pm0.009$ & $0.300\pm0.014$ \\
 NGC~7626 & 0.62 & 3.58 & 19.1 & 152.7 & $-0.008\pm0.003$ & $2.425\pm0.004$ & $-0.051\pm0.016$ & $0.096\pm0.023$ & $-0.007\pm0.008$ & $0.356\pm0.013$ \\
 UGC~10918 & 0.41 & 3.64 & -- & 35.1 & $-0.004\pm0.010$ & $2.399\pm0.020$ & $-0.001\pm0.016$ & $-0.080\pm0.030$ & $-0.021\pm0.016$ & $0.319\pm0.033$ \\
 NGC~7436 & 0.18 & 1.84 & 41.7 & 167.9 & $-0.050\pm0.012$ & $2.482\pm0.015$ & $-0.061\pm0.058$ & $0.116\pm0.077$ & $-0.020\pm0.072$ & $0.425\pm0.083$ \\
 NGC~7556 & 0.15 & 1.70 & 120.1 & 125.5 & $-0.029\pm0.007$ & $2.392\pm0.007$ & $-0.082\pm0.060$ & $0.094\pm0.058$ & $-0.025\pm0.028$ & $0.336\pm0.024$ \\
 NGC~7386 & 0.54 & 4.14 & 143.9 & 164.9 & $-0.019\pm0.011$ & $2.464\pm0.015$ & $-0.106\pm0.028$ & $0.154\pm0.036$ & $-0.014\pm0.016$ & $0.296\pm0.022$ \\
 NGC~0547 & 0.23 & 2.90 & 83.2 & 142.0 & $-0.021\pm0.009$ & $2.380\pm0.015$ & $-0.087\pm0.022$ & $0.117\pm0.034$ & $-0.005\pm0.008$ & $0.330\pm0.013$ \\
 NGC~0545 & 0.40 & 4.12 & 58.3 & 142.0 & $-0.009\pm0.003$ & $2.404\pm0.010$ & $-0.035\pm0.004$ & $0.162\pm0.017$ & $-0.002\pm0.004$ & $0.294\pm0.018$ \\

\enddata
\tablecomments{
Note. Column entries are given as best-fit value $\pm1\sigma$ uncertainty.
The table gives the radial coverage used for the gradient fit, the galaxy and slit
position angles, and the fitted slopes and intercepts for the first set of stellar
population parameters.  The fitted model is
${\rm Y}=a(r/{\rm 1~kpc})+b$, where $r$ is the deprojected radius in kpc.
For $\sigma$, the fitted quantity is $\log(\sigma/{\rm km~s^{-1}})$.
The galaxy position angle is from the Siena Galaxy Atlas 2020 \citep{Moustakas2023};
for targets without a PA measurement, no deprojection correction is applied.
Position angles are folded into the range $0$--$180^\circ$.
}
\end{deluxetable*}

\begin{deluxetable*}{lcccccccccc}
\tabletypesize{\scriptsize}
\tablewidth{0pt}
\tablecaption{Radial gradient measurements, continued.\label{tab:gradient_cont}}

\tablehead{
\colhead{Name} &
\multicolumn{2}{c}{$\log({\rm age})$} &
\multicolumn{2}{c}{[Na/Fe]} &
\multicolumn{2}{c}{$\log(\alpha_{\rm IMF})$} &
\colhead{$\alpha_{\rm IMF,r}$} &
\colhead{$(M/L)_{K_s}^{\rm Kr}$} &
\colhead{$\alpha_{\rm IMF,K_s}$} \\
\colhead{} &
\colhead{$a$} & \colhead{$b$} &
\colhead{$a$} & \colhead{$b$} &
\colhead{$a$} & \colhead{$b$} &
\colhead{$\leq R_{\rm e}/2$} &
\colhead{$\leq R_{\rm e}$} &
\colhead{$\leq R_{\rm e}$}
}

\startdata
 NGC~0057 & $0.024\pm0.011$ & $1.075\pm0.023$ & $-0.086\pm0.022$ & $0.495\pm0.024$ & $-0.025\pm0.038$ & $0.378\pm0.049$ & $2.14\pm0.21$ & $0.92\pm0.02$ & $1.76\pm0.20$ \\
 NGC~0080 & $0.034\pm0.017$ & $1.008\pm0.030$ & $-0.117\pm0.025$ & $0.457\pm0.022$ & $0.015\pm0.045$ & $0.309\pm0.056$ & $2.23\pm0.29$ & $0.78\pm0.08$ & $2.19\pm0.51$ \\
 NGC~0533 & $0.006\pm0.016$ & $1.079\pm0.027$ & $-0.097\pm0.025$ & $0.542\pm0.035$ & $-0.026\pm0.045$ & $0.366\pm0.078$ & $2.44\pm0.28$ & $0.86\pm0.03$ & $2.04\pm0.27$ \\
 NGC~0741 & $0.015\pm0.012$ & $1.075\pm0.026$ & $-0.060\pm0.015$ & $0.382\pm0.018$ & $-0.015\pm0.026$ & $0.316\pm0.038$ & $1.88\pm0.15$ & $0.84\pm0.03$ & $1.69\pm0.20$ \\
 NGC~1016 & $-0.037\pm0.049$ & $1.094\pm0.068$ & $-0.070\pm0.030$ & $0.529\pm0.038$ & $-0.020\pm0.043$ & $0.378\pm0.058$ & $2.06\pm0.26$ & $0.88\pm0.02$ & $1.44\pm0.15$ \\
 NGC~1453 & $0.003\pm0.011$ & $1.109\pm0.021$ & $-0.079\pm0.023$ & $0.529\pm0.033$ & $-0.040\pm0.036$ & $0.411\pm0.052$ & $1.57\pm0.18$ & $0.87\pm0.04$ & $1.82\pm0.23$ \\
 NGC~1600 & $-0.010\pm0.007$ & $1.132\pm0.010$ & $-0.038\pm0.007$ & $0.478\pm0.009$ & $-0.022\pm0.018$ & $0.421\pm0.032$ & $2.48\pm0.28$ & $0.89\pm0.09$ & $1.76\pm0.44$ \\
 NGC~1700 & $-0.030\pm0.020$ & $0.677\pm0.020$ & $-0.127\pm0.054$ & $0.332\pm0.066$ & $-0.031\pm0.017$ & $0.219\pm0.028$ & $1.55\pm0.17$ & $0.49\pm0.02$ & $1.13\pm0.16$ \\
 NGC~2418 & $0.072\pm0.094$ & $0.917\pm0.084$ & $-0.162\pm0.071$ & $0.539\pm0.046$ & $-0.161\pm0.111$ & $0.390\pm0.083$ & $1.62\pm0.29$ & $0.86\pm0.06$ & $1.48\pm0.24$ \\
 NGC~2513 & $-0.026\pm0.038$ & $1.056\pm0.041$ & $-0.080\pm0.011$ & $0.492\pm0.012$ & $-0.045\pm0.062$ & $0.168\pm0.078$ & $1.30\pm0.12$ & $0.89\pm0.02$ & $1.15\pm0.13$ \\
 NGC~2672 & $-0.067\pm0.020$ & $1.121\pm0.020$ & $-0.075\pm0.008$ & $0.500\pm0.007$ & $-0.089\pm0.049$ & $0.328\pm0.055$ & $1.65\pm0.22$ & $0.90\pm0.03$ & $1.23\pm0.17$ \\
 NGC~3462 & $0.015\pm0.036$ & $0.658\pm0.058$ & $-0.098\pm0.018$ & $0.366\pm0.034$ & $0.002\pm0.040$ & $0.264\pm0.084$ & $2.31\pm0.38$ & $0.54\pm0.03$ & $1.81\pm0.34$ \\
 NGC~3615 & $-0.015\pm0.040$ & $0.973\pm0.077$ & $-0.050\pm0.010$ & $0.298\pm0.018$ & $0.035\pm0.023$ & $0.176\pm0.043$ & $1.85\pm0.22$ & $0.67\pm0.04$ & $1.82\pm0.33$ \\
 NGC~3842 & $-0.008\pm0.011$ & $1.117\pm0.019$ & $-0.035\pm0.018$ & $0.541\pm0.029$ & $-0.047\pm0.015$ & $0.438\pm0.021$ & $2.24\pm0.22$ & $0.92\pm0.03$ & $1.55\pm0.19$ \\
 NGC~3862 & $0.097\pm0.058$ & $0.838\pm0.057$ & $-0.072\pm0.049$ & $0.530\pm0.054$ & $-0.132\pm0.123$ & $0.628\pm0.157$ & $2.97\pm0.38$ & $0.94\pm0.02$ & $1.67\pm0.23$ \\
 NGC~3937 & $-0.029\pm0.039$ & $1.034\pm0.050$ & $-0.117\pm0.046$ & $0.395\pm0.052$ & $-0.056\pm0.080$ & $0.389\pm0.113$ & $1.94\pm0.27$ & $0.83\pm0.05$ & $2.20\pm0.29$ \\
 NGC~4055 & $0.019\pm0.088$ & $1.058\pm0.117$ & $-0.025\pm0.115$ & $0.291\pm0.126$ & $-0.176\pm0.092$ & $0.447\pm0.102$ & $1.53\pm0.24$ & $0.82\pm0.02$ & $1.52\pm0.16$ \\
 NGC~4073 & $0.006\pm0.005$ & $1.087\pm0.022$ & $-0.004\pm0.006$ & $0.441\pm0.015$ & $-0.028\pm0.025$ & $0.380\pm0.074$ & $1.74\pm0.22$ & $0.92\pm0.03$ & $1.39\pm0.19$ \\
 NGC~4472 & $-0.005\pm0.015$ & $1.119\pm0.016$ & $-0.063\pm0.020$ & $0.482\pm0.019$ & $-0.045\pm0.017$ & $0.124\pm0.023$ & $1.14\pm0.09$ & $0.78\pm0.01$ & $1.40\pm0.11$ \\
 NGC~4486 & $0.005\pm0.005$ & $1.130\pm0.008$ & $-0.098\pm0.013$ & $0.626\pm0.013$ & $-0.056\pm0.032$ & $0.281\pm0.036$ & $1.68\pm0.15$ & $0.82\pm0.06$ & $1.69\pm0.26$ \\
 NGC~4555 & $-0.024\pm0.022$ & $0.941\pm0.038$ & $-0.041\pm0.015$ & $0.431\pm0.026$ & $-0.049\pm0.030$ & $0.310\pm0.076$ & $1.92\pm0.24$ & $0.68\pm0.05$ & $1.72\pm0.38$ \\
 NGC~4649 & $-0.030\pm0.012$ & $1.143\pm0.005$ & $-0.144\pm0.042$ & $0.696\pm0.038$ & $-0.055\pm0.035$ & $0.255\pm0.038$ & $1.68\pm0.12$ & $0.79\pm0.04$ & $1.38\pm0.15$ \\
 NGC~4839 & $-0.018\pm0.014$ & $1.110\pm0.026$ & $-0.028\pm0.017$ & $0.388\pm0.034$ & $0.006\pm0.022$ & $0.275\pm0.044$ & $2.08\pm0.27$ & $0.79\pm0.05$ & $1.57\pm0.38$ \\
 NGC~4874 & $-0.028\pm0.036$ & $1.073\pm0.057$ & $-0.017\pm0.027$ & $0.377\pm0.040$ & $-0.065\pm0.038$ & $0.222\pm0.076$ & $1.15\pm0.24$ & $0.78\pm0.07$ & $0.97\pm0.25$ \\
 NGC~5490 & $0.012\pm0.015$ & $1.087\pm0.024$ & $-0.110\pm0.024$ & $0.551\pm0.026$ & $-0.016\pm0.075$ & $0.299\pm0.083$ & $1.83\pm0.14$ & $0.73\pm0.03$ & $1.61\pm0.24$ \\
 NGC~6375 & $0.008\pm0.040$ & $1.066\pm0.062$ & $-0.112\pm0.034$ & $0.423\pm0.055$ & $-0.120\pm0.095$ & $0.499\pm0.126$ & $2.10\pm0.25$ & $0.90\pm0.04$ & $1.61\pm0.23$ \\
 NGC~6442 & $-0.044\pm0.040$ & $1.121\pm0.043$ & $-0.095\pm0.059$ & $0.357\pm0.059$ & $0.020\pm0.126$ & $0.366\pm0.146$ & $2.42\pm0.31$ & $0.86\pm0.03$ & $1.65\pm0.17$ \\
 NGC~6482 & $-0.008\pm0.015$ & $1.054\pm0.017$ & $-0.133\pm0.045$ & $0.668\pm0.061$ & $-0.073\pm0.039$ & $0.381\pm0.051$ & $2.01\pm0.22$ & $0.87\pm0.05$ & $1.49\pm0.16$ \\
 NGC~7052 & $0.013\pm0.029$ & $1.087\pm0.048$ & $-0.084\pm0.028$ & $0.486\pm0.043$ & $-0.117\pm0.048$ & $0.400\pm0.061$ & $1.66\pm0.22$ & $1.01\pm0.02$ & $0.98\pm0.10$ \\
 NGC~7619 & $-0.016\pm0.019$ & $0.994\pm0.025$ & $-0.095\pm0.017$ & $0.563\pm0.023$ & $-0.057\pm0.009$ & $0.290\pm0.015$ & $1.56\pm0.16$ & $0.74\pm0.02$ & $1.38\pm0.15$ \\
 NGC~7626 & $-0.034\pm0.021$ & $1.004\pm0.031$ & $-0.096\pm0.022$ & $0.555\pm0.031$ & $-0.059\pm0.030$ & $0.252\pm0.049$ & $1.42\pm0.11$ & $0.72\pm0.03$ & $1.33\pm0.17$ \\
 UGC~10918 & $-0.001\pm0.021$ & $1.118\pm0.036$ & $-0.043\pm0.021$ & $0.408\pm0.041$ & $-0.037\pm0.054$ & $0.333\pm0.103$ & $1.75\pm0.21$ & $0.96\pm0.02$ & $1.20\pm0.14$ \\
 NGC~7436 & $-0.021\pm0.185$ & $1.091\pm0.257$ & $-0.194\pm0.116$ & $0.771\pm0.134$ & $0.008\pm0.105$ & $0.325\pm0.136$ & $1.89\pm0.27$ & $0.90\pm0.02$ & $1.67\pm0.21$ \\
 NGC~7556 & $0.019\pm0.055$ & $1.073\pm0.049$ & $-0.131\pm0.121$ & $0.495\pm0.099$ & $-0.007\pm0.138$ & $0.238\pm0.128$ & $1.91\pm0.49$ & $0.79\pm0.07$ & $1.44\pm0.41$ \\
 NGC~7386 & $-0.004\pm0.037$ & $1.019\pm0.067$ & $-0.127\pm0.015$ & $0.516\pm0.018$ & $-0.021\pm0.029$ & $0.165\pm0.048$ & $1.48\pm0.20$ & $0.66\pm0.04$ & $1.44\pm0.29$ \\
 NGC~0547 & $-0.004\pm0.013$ & $1.084\pm0.022$ & $-0.055\pm0.008$ & $0.440\pm0.013$ & $-0.016\pm0.028$ & $0.327\pm0.043$ & $2.07\pm0.21$ & $0.76\pm0.06$ & $1.78\pm0.33$ \\
 NGC~0545 & $0.003\pm0.003$ & $1.062\pm0.016$ & $-0.027\pm0.005$ & $0.459\pm0.022$ & $-0.009\pm0.019$ & $0.314\pm0.087$ & $1.84\pm0.17$ & $0.61\pm0.11$ & $1.79\pm0.73$ \\

\enddata
\tablecomments{
Note. This table shows the remaining stellar-population gradients and the
aperture-integrated mass-to-light ratios.  The $r$-band column lists the
IMF-mismatch measurement within $R_{\rm e}/2$.
The CFHT/WIRCam $K_s$ columns show the corresponding predicted values within
$R_{\rm e}$.  For $\alpha_{\rm IMF}$, the fitted gradient quantity is
$\log(\alpha_{\rm IMF})$, where
$\alpha_{\rm IMF}=(M/L)/(M/L)_{\rm Kroupa}$. 
}
\end{deluxetable*}
\linenumbers
\twocolumngrid

\begin{figure*}[t]
\vskip 0.15cm
\centering
\includegraphics[width=15cm]{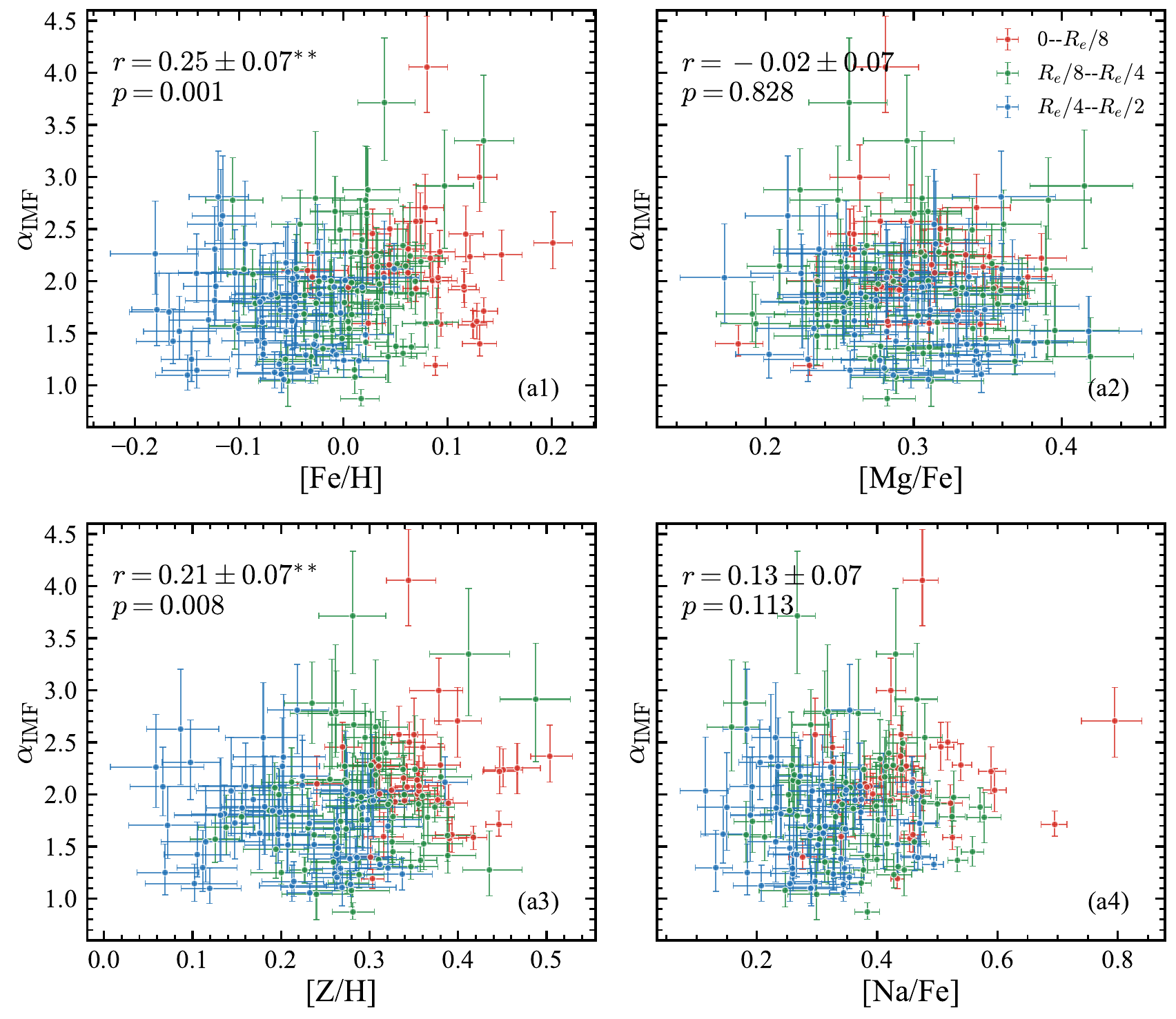}
\caption{
Local relations between $\alpha_{\rm IMF}$ and stellar-population parameters measured in radial bins. 
Each point represents one radial-bin measurement, and colors indicate the 
fractional-$R_{\rm e}$ bin. Pearson correlation coefficients and $p$-values 
are computed between
$\log(\alpha_{\rm IMF})$ and [Fe/H], [Mg/Fe], [Z/H], and [Na/Fe], using
measurements from all three radial bins. 
We find mild positive local correlations of $\alpha_{\rm IMF}$
with [Fe/H] and [Z/H], but no significant local correlation with [Mg/Fe] or
[Na/Fe].
 }
\label{fig_localcorr}
\end{figure*}
\noindent 
\begin{figure*}[ht!]
\centering
\includegraphics[width=\textwidth]{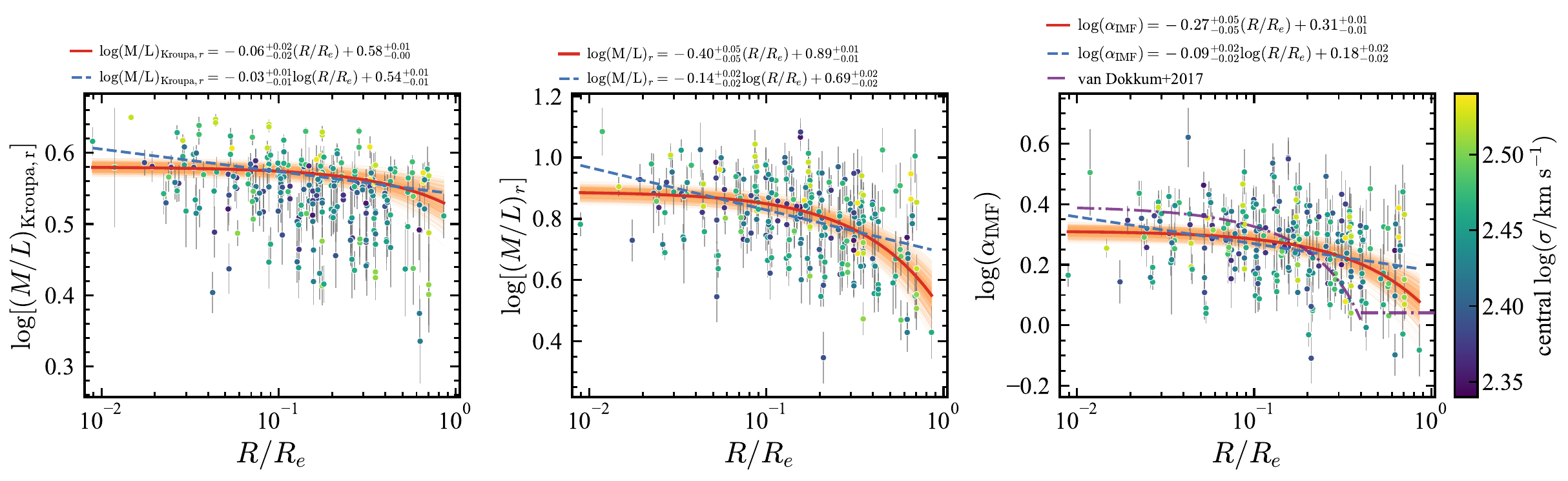}
\includegraphics[width=\textwidth]{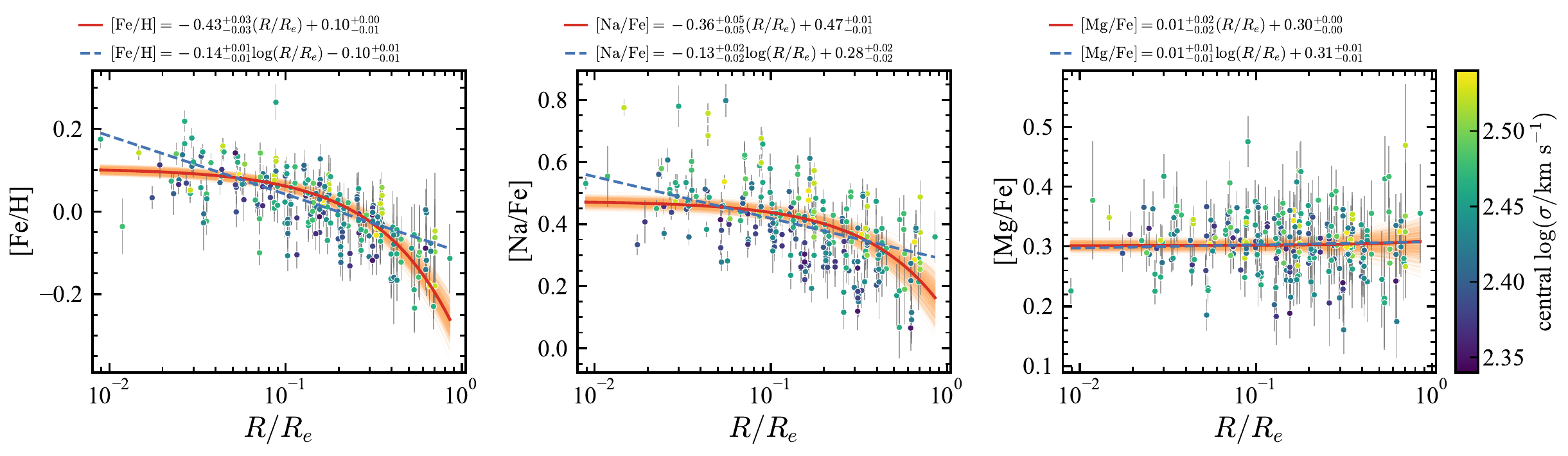}
\caption{
Radial trends from all individual radial ``pixel bins'' measurements in the main 
comparison sample.  Top panel shows the Kroupa-IMF $M/L_r$, flexible-IMF 
$M/L_r$, and $\alpha_{\rm IMF}$. The bottom panels show [Fe/H], [Na/Fe], and 
[Mg/Fe].  Points are colored by central velocity dispersion.  Red solid curves 
show Student-$t$ robust fits linear in $R/R_{\rm e}$, while blue dashed curves 
show Student-$t$ robust fits linear in $\log(R/R_{\rm e})$.  Orange curves show 
posterior draws.  The purple curve in the $\alpha_{\rm IMF}$ panel shows the 
relation from \citet{vanDokkum2017} for comparison.
}
\label{fig_gradsfromall}
\end{figure*}

Figure~\ref{fig_gradient_compactness} shows the same gradients as a function of
effective stellar surface density, $\Sigma_{\rm Kroupa}$, which traces galaxy
compactness.  For the 27 old galaxies with radial coverage reaching at least
$0.25R_{\rm e}$ and with $\Sigma_{\rm Kroupa}$ measurements, we find
statistically significant anti-correlations between compactness and the
gradients of [Fe/H] and [Na/Fe].  The gradients of $\log(\alpha_{\rm IMF})$ and
$\log(\sigma)$ also show negative but only marginal correlations with
compactness.  We do not find significant correlations between compactness and
the gradients of [Mg/Fe] or $\log({\rm age})$.
As an exploratory check, we also asked whether the weakest IMF gradients 
are preferentially found in low-compactness galaxies and found a suggestive trend. 
Splitting the sample at $\log(\Sigma_{\rm Kroupa}/M_{\odot}\,{\rm kpc}^{-2})=9.07$, 
only 1 of 7 galaxies below this value has a $\log(\alpha_{\rm IMF})$ gradient more than 
$1\sigma$ below zero, compared with 14 of 20 galaxies above this value.

Taken together, these comparisons show that the clearest trend is between the
IMF gradient and the central IMF normalization: galaxies with larger central
$\alpha_{\rm IMF}$ have more negative $\log(\alpha_{\rm IMF})$ gradients,
implying that the central diversity in IMF mismatch decreases with radius.
Compactness shows significant anti-correlations with the [Fe/H] and [Na/Fe]
gradients, while the corresponding trend with the IMF gradient is only marginal
in the current sample.  These results suggest that IMF gradients are closely
connected to the central IMF normalization, and may also be linked to the radial
behavior of metallicity-sensitive stellar-population parameters.  We therefore
next examine the local relations between $\alpha_{\rm IMF}$ and stellar-population 
properties in individual radial bins.


\subsection{Local Variation of Stellar Populations and IMF}

In our previous work (Paper~I), we found that in the central $R_{\rm e}/8$ aperture 
of massive ETGs, [Fe/H], [Mg/Fe], and [Na/Fe] all appeared to correlate with
$\alpha_{\rm IMF}$. Those central correlations suggested possible connections
between the IMF, metallicity, and star-formation history.  Here we test whether
similar relations are present locally, using the radial-bin measurements across
the galaxies.

In Figure~\ref{fig_localcorr}, we investigate local correlations between
$\alpha_{\rm IMF}$ and stellar-population parameters using the fractional
$R_{\rm e}$ radial bins. Each point represents one radial-bin measurement; for
the two outer bins, measurements from the two sides of a galaxy are included
separately when they pass the selection criteria.  We calculate Pearson
correlation coefficients between $\log(\alpha_{\rm IMF})$ and [Fe/H], [Mg/Fe],
[Z/H], and [Na/Fe] using all radial-bin measurements. 

When all radial bins are combined, $\alpha_{\rm IMF}$ shows mild but significant
positive correlations with [Fe/H] and [Z/H].  We do not find significant local
correlations with [Mg/Fe] or [Na/Fe].   
However, the [Fe/H] and [Z/H] correlations are not significant when 
the measurements are examined separately within individual radial bins.  
This suggests that the combined-bin correlations are driven largely by common
radial gradients within galaxies rather than by a strong galaxy-to-galaxy
relation at fixed radius.

The gradient behavior provides additional context.  [Fe/H] and [Na/Fe] both
decline with radius in the individual-galaxy gradient fits, similar to
$\alpha_{\rm IMF}$ for most galaxies, whereas [Mg/Fe], [Ca/Fe], [Ti/Fe], and
[Si/Fe] are approximately flat in the sample-averaged radial bins.  These results
suggest that IMF variation is more closely associated with metallicity-sensitive
stellar-population trends than with the nearly flat $\alpha$-enhancement
profiles traced by [Mg/Fe], which is often interpreted as a proxy for 
star-formation timescale.

\subsection{Comparison with Previous Studies}

The IMF in early-type galaxies has been explored extensively \citep{Smith2020} 
using both SPS and dynamical measurements.  In this section, we 
compare our results with recent studies that have investigated the same targets, 
focusing on the measurements of IMF radial profiles.

\citet{vanDokkum2017} studied the IMF gradient in six ETGs using Keck spectroscopy 
extending to 1$R_{\rm e}$ with {\tt alf}.  They found that the IMF in the central 
regions is typically more bottom-heavy.  They proposed an analytic formula for 
$\alpha_{\rm IMF}$ as a function of radius in units of $R_{\rm e}$. Their sample 
includes NGC~1600, and our results for that galaxy agree in general with their 
findings.  For data points within $0.1R_{\rm e}$, the $\alpha_{\rm IMF}$ in 
\citet{vanDokkum2017} ranges from around 2.6 to 3. Considering that the 
$R_{\rm e}$ in \citet{vanDokkum2017} is about 2.4 times the $R_{\rm e}$ in 
this work, we compare our results in the $R_{\rm e}/8-R_{\rm e}/4$ radial bin, 
where the two sides yield $\alpha_{\rm IMF}=2.31_{-0.32}^{+0.32}$ and 
$\alpha_{\rm IMF}=2.38_{-0.34}^{+0.32}$. In the $R_{\rm e}/4-R_{\rm e}/2$ radial 
bin, the two sides yield $\alpha_{\rm IMF}=2.03_{-0.39}^{+0.38}$ and 
$\alpha_{\rm IMF}=1.95_{-0.38}^{+0.36}$, 
falling within the $\alpha_{\rm IMF}$ range between their $0.1R_{\rm e}$ and 
$0.2R_{\rm e}$ data points as shown in their Figure~11. We conclude that 
our results for NGC1600 are consistent 
with each other, even though our measurements are derived from different observations 
using different instruments and slit angles. \citet{vanDokkum2017} provides an analytic 
formula for $\alpha_{\rm IMF}$ as a function of radius in units of $R_{\rm e}$:
\begin{equation}
 \log(\alpha_{\rm IMF})=-0.283^{+0.044}_{-0.046}(\frac{R}{R_{\rm e}}) + 0.328^{
 +0.010}_{-0.010}
\end{equation}

In Figure~\ref{fig_gradsfromall}, we derive sample-averaged radial
trends from the pixel-bin measurements of the 35 galaxies. Most
measurements lie within $1R_{\rm e}$.  Red curves show Student-$t$ robust fits
linear in $R/R_{\rm e}$, orange curves show posterior draws, and blue dashed
curves show fits linear in $\log(R/R_{\rm e})$. We find an offset from 
the galaxy centers to $1R_{\rm e}$ of 
$\Delta {\rm [Fe/H]}=-0.43^{+0.03}_{-0.03}$~dex while the [Mg/Fe] offset is
consistent with zero,  $\Delta {\rm [Mg/Fe]}=+0.01\pm0.02$~dex. The IMF mismatch
also decreases, with $\Delta\log(\alpha_{\rm IMF})=-0.27^{+0.05}_{-0.05}$ dex.  

As shown in the right panel of 
Fig~\ref{fig_gradsfromall}, at $\frac{R}{R_{\rm e}}=0.1$, our fitting results suggest an 
$\alpha_{\rm IMF}\sim1.93$ while \citet{vanDokkum2017} predicts $\alpha_{\rm IMF}\sim2.0$. 
Thus, our results are consistent in the central regions.  At $\frac{R}{R_{\rm e}}=0.3$, we measure 
an $\alpha_{\rm IMF}\sim1.70$ while \citet{vanDokkum2017} predicts $\alpha_{\rm IMF}\sim1.75$. 

The IMF of NGC~4486 (M87) has been studied extensively, making it a useful
benchmark for comparison.  Here we focus on previous work that directly
constrains radial IMF or stellar $M_\star/L$ variations in this galaxy
\citep{Sarzi2018, Oldham2018, Simon2024}.
The most direct comparison is \citet{Sarzi2018}, who used MUSE absorption-line 
measurements to infer a bottom-heavy IMF in the center of M87 that becomes close 
to Milky-Way-like by $\sim0.4R_{\rm e}$. Evaluating our best-fit radial profile at 
5\arcsec, 10\arcsec, and 40\arcsec, we obtain $\alpha_{\rm IMF}=1.82$, 1.72, and 1.26, 
respectively.  These values are broadly consistent with the negative IMF gradient 
inferred by \citet{Sarzi2018}, despite the different data sets, wavelength coverage, 
and fitting methods. A recent MUSE-based analysis by \citet{Parikh2024} also found 
good agreement between the stellar-population and dynamical $M_\star/L$ profiles for M87,
supporting the presence of a radially decreasing stellar $M_\star/L$.

Independent dynamical studies also support a radially decreasing stellar $M_\star/L$ in 
M87.  \citet{Oldham2018} found an IMF consistent with Salpeter-like in the central 
$\sim0.5$~kpc and Chabrier-like by $\sim3$~kpc.  Evaluating our best-fit radial profile 
at 0.5~kpc and 3~kpc, we infer $\alpha_{\rm IMF}=1.79$ and $1.30$, respectively, 
which are marginally consistent with their constraints.  More recently, 
\citet{Simon2024} showed that allowing for radial stellar $M_\star/L$ variations 
affects dynamical measurements of the black-hole mass in M87.  Taken together, 
these studies support substantial radial variation in the stellar $M_\star/L$ of M87.  
Our results indicate an IMF contribution to this variation.
 
Similar negative IMF gradients have also been reported in other massive ETGs.
For example, \citet{Lonoce2021} found a bottom-heavy central IMF in M89 that
becomes less bottom-heavy toward larger radii, with consistent results from
index-based fitting and full spectral modeling. Using {\tt alf}, \citet{Lonoce2023} also found 
negative IMF gradients in the central regions of the two brightest cluster galaxies in Hydra~I.   
Although these studies target different galaxies, their results support the picture that massive
ETGs commonly have centrally enhanced, radially declining IMF mismatch parameters, consistent
with the trends found for most galaxies in our sample.

\section{Discussion}

We have presented radial stellar population and IMF constraints for 37 galaxies 
from the MASSIVE survey \citep{Ma2014}.  In this section, we discuss what these 
measurements imply for galaxy mass estimates and for the physical origin of IMF 
variation.  We first quantify how IMF gradients affect aperture-integrated 
stellar mass-to-light ratios (\S~4.1).  
Next, we place the measured gradients in the broader context of 
stellar-population gradients and galaxy structure (\S~4.2).
Finally, we discuss the role of metallicity in IMF variation 
and outline the main caveats and future directions (\S~4.3--\S~4.4).

\subsection{Implications for Mass Measurements}

In this section, we investigate the potential influence of IMF variation on 
galaxy stellar mass measurements.  The presence of IMF gradients offers a 
possible solution to reconcile IMF measurements derived from stellar 
population synthesis modeling with methods such as dynamical modeling and 
strong lensing.  To examine the discrepancy, we derive radial gradients 
of mass-to-light ratio ($M/L$) for both the best-fit IMF and a Kroupa IMF 
using the fiducial sample.
We adopt a robust linear regression 
method, described in \S~3.3, to fit $\log({\rm M/L}) - R$ relations, where $R$ 
is in units of $R_{\rm e}$, using ``pixel bin'' data of 35 galaxies in our 
sample. The results of the fitting are presented in Figure~\ref{fig_gradsfromall}. 
In particular, Figure~\ref{fig_gradsfromall} shows the gradients of 
$\alpha_{\rm IMF}$, the IMF mismatch parameter, in the upper right panel.  
Its radial profile indicates that the IMF plays an important role in the 
radial variation of the $M/L$.  

A key result from Figure~\ref{fig_gradsfromall} is that the radial 
variation in $M/L_r$ is dominated by the IMF mismatch gradient rather than by the 
fixed-IMF stellar-population gradient.  For the fit in linear $R/R_{\rm e}$, we 
find $d\log(M/L_{r,{\rm Kroupa}})/d(R/R_{\rm e})\approx-0.06$, whereas
$d\log(\alpha_{\rm IMF})/d(R/R_{\rm e})\approx-0.27$, indicating that the 
contribution from age, metallicity, and abundance variations at fixed Kroupa IMF 
is subdominant to the IMF-mismatch gradient and not enough to capture the full 
radial variation in stellar $M/L$.  For the massive ETGs studied here, accounting 
only for fixed-IMF stellar-population gradients would substantially underpredict 
the total stellar $M/L$ gradient.  Similar concerns about IMF-driven $M/L$ 
gradients and their impact on dynamical stellar masses have been discussed by
\citet{McConnell2013},
\citet{Oldham2018}, \citet{Davis2017}, \citet{Bernardi2018}, and \citet{Mehrgan2024}.

\begin{figure}[h]
\centering
\includegraphics[width=\columnwidth]{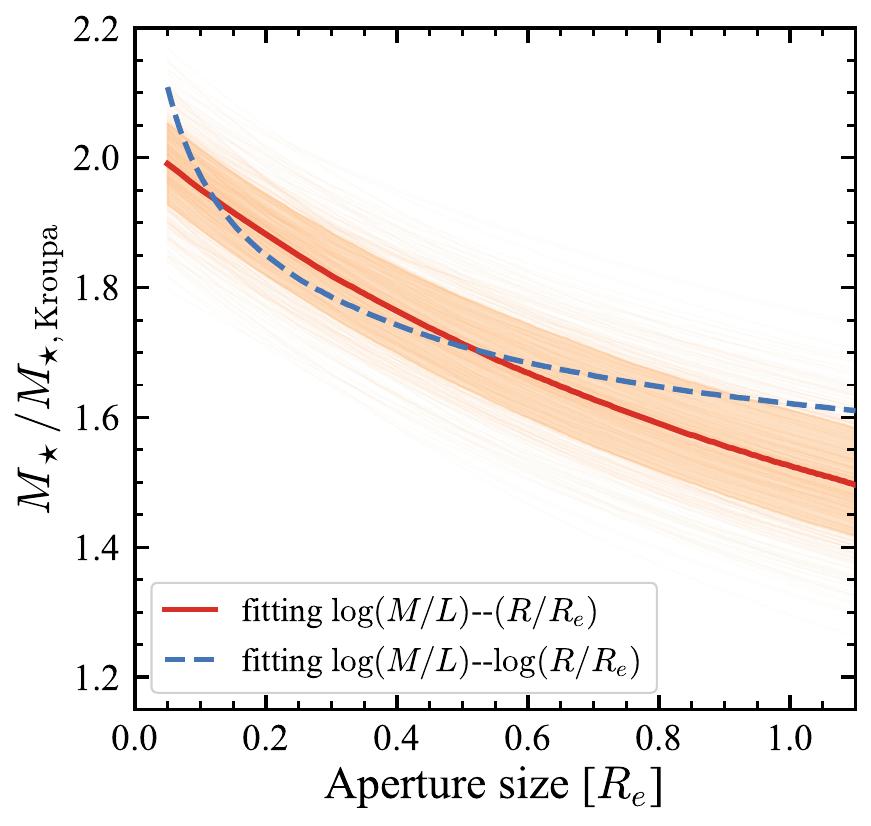}
\caption
{
Enclosed stellar mass ratio between the flexible-IMF and Kroupa-IMF models as a 
function of aperture size. The calculation uses the robust radial $M/L_r$ fits from 
Figure~\ref{fig_gradsfromall} and assumes an $n=4$ S$\text{\'{e}}$rsic light profile.  
The red curve shows the posterior median relation from a Student-$t$ regression of 
$Y=a(R/R_{\rm e})+b$.  Orange curves 
show posterior draws.  
The blue dashed curve shows 
robust fits of the form $Y=a\log_{10}(R/R_{\rm e})+b$.   The predicted mass ratios are
$1.71^{+0.07}_{-0.07}$ within $R_{\rm e}/2$ and
$1.53^{+0.09}_{-0.08}$ within $R_{\rm e}$ for the linear-radius fit, and
$1.71^{+0.11}_{-0.10}$ within $R_{\rm e}/2$ and
$1.62^{+0.09}_{-0.09}$ within $R_{\rm e}$ for the log-radius fit.
}
\label{fig_apermass}
\end{figure}

To assess the impact of the IMF on mass measurements, we estimate the total 
stellar mass by integrating the mass-to-light ratio profile and a light profile 
that follows an S$\text{\'{e}}$rsic \citep{Sersic1963} profile with 
$n=4$.  The intensity profile follows 
\begin{equation}
 I(R) = I_e \exp \left[-b_n \left(\left(\frac{R}{R_{\rm e}}\right)^{1/n} - 1\right)\right] 
\end{equation}
and $b_4=7.669$ \citep{Ciotti1991}.  The stellar mass in the given aperture is 
calculated as  
\begin{equation}
M(r)=\int_{0}^{R_{\rm aperture}} 2\pi I(r) (M/L)(r) r\,dr 
\end{equation}
The resulting enclosed mass-ratio profile is shown in Figure~\ref{fig_apermass},
indicating how much 
the stellar mass within a specific aperture will change with the IMF derived 
from the fiducial sample. The thin lines represent 500 
realizations based on the 
measurement uncertainties of the slope and intercept of $M/L$. 
Within half an effective 
radius, the factor is $1.71\pm0.07$, and if the aperture is one effective 
radius, the factor is around $1.53\pm0.09$.  
We note that the largest radial coverage is about 1.1 $R_{\rm e}$.  
The result emphasizes 
the importance of considering IMF gradients when making galaxy mass measurements.  
However, the estimate is  based on the most massive galaxies in the nearby 
universe, potentially the cases with the most extreme bottom-heavy IMF in their 
centers.  Mass measurements on lower mass ETGs should be more mildly affected by 
the assumption of a constant Kroupa IMF.

\citet{Liepold2024} compared stellar mass measurements of MASSIVE galaxies from this work with those obtained from detailed dynamical modeling.    
The average stellar masses from these two independent methods were found to agree within 7\%. The inferred galaxy stellar mass function (GSMF) above $M\ga 10^{11.3} M_\odot$ based on MASSIVE galaxies has a higher amplitude than earlier work that had assumed 
a Kroupa-like IMF. They found that a simple increase in stellar mass by a factor of $\alpha_{\rm IMF}\sim 1.7$ reported in this work would bring the earlier GSMF into better agreement with their GSMF.

\subsection{Stellar Populations and IMF Gradients}

Radial stellar population gradients in ETGs have long been used as fossil 
records of galaxy formation.  
Spatially resolved surveys across the Hubble sequence show that
age and metallicity profiles vary with galaxy mass, morphology, and star
formation history \citep{GonzalezDelgado2015,Goddard2017}.
Early long-slit studies found that elliptical 
galaxies have negative metallicity gradients \citep{Davies1993, Mehlert2003}.  
Subsequent long-slit studies and IFU work generally find that metallicity 
declines with radius, and age and alpha-enhancement gradients are weaker and 
more diverse \citep[e.g.][]{Kuntschner2010, Ferreras2019,Santucci2020, 
Rawle2010, Zibetti2020}.
In massive galaxies, stellar halos and BCGs are typically more metal poor at 
large radius, and alpha-enhancement often changes more weakly 
\citep{Greene2013,Greene2015, Loubser2012}. Individual brightest cluster 
galaxies can show more complex core to halo transitions \citep{Coccato2010}. 

Gradient strength does not simply increase with galaxy mass across the full ETG population.
For low and intermediate mass ETGs, metallicity gradients have been found to 
become more negative with increasing galaxy mass 
\citep[e.g.,][]{Spolaor2009, Spolaor2010}, or increasing central velocity 
dispersion \citep{Martin-Navarro2018}.  At the highest masses, however, the 
trend appears to weaken or turn over: massive ETGs show flatter gradients and 
larger scatter, which has been interpreted as evidence that mergers dilute 
pre-existing metallicity gradients \citep{Spolaor2010,Santucci2020,Kobayashi2004}. 
The connection between gradient strength and global galaxy properties remains 
unsettled. 

Simulations also provide useful context for interpreting stellar metallicity 
gradients.  The formation of nearby massive ETGs is commonly described by a 
two-phase formation scenario, bridging the observations of high-redshift 
($z\sim2$) compact massive galaxies \citep{vanDokkum2010, Oser2012} with 
spatially extended local massive galaxies \citep{vanderWel2014}. In this 
picture, compact, metal-rich centers form early through dissipative processes 
and the outer envelopes are assembled later through mergers and accretion. 
In such models, dissipative central star formation naturally produces negative 
metallicity gradients, while later major dry mergers tend to flatten pre-existing 
gradients \citep[e.g.][]{Larson1974, Kobayashi2004, DiMatteo2009, Cook2016, 
Rodriguez-Gomez2016}.  

These results suggest that metallicity gradients depend on both the early 
dissipative formation of the central regions and the later assembly history 
of the galaxy.  We have presented the gradients of stellar populations and 
the IMF of individual galaxies in Tables~\ref{tab:gradient} and \ref{tab:gradient_cont}
and throughout \S~3.
The large decline in [Fe/H], together with the nearly flat [Mg/Fe] profile,
suggests that the central and outer regions do not differ primarily in their
star-formation timescales traced by [Mg/Fe]. Instead, the
dominant radial contrast is in metal enrichment, consistent with central stars
forming in deeper potential wells where gas could be retained, recycled, and
enriched efficiently. The decline of $\log(\alpha_{\rm IMF})$ implies that 
the centrally concentrated, metal-rich component also formed with a larger 
fraction of low-mass stars. The relative offsets are 
$\Delta\log(\alpha_{\rm IMF})\simeq-0.27$ dex and 
$\Delta{\rm [Fe/H]}\simeq-0.43$ dex from the center to $1R_{\rm e}$.  
At the same time, the flat [Mg/Fe] profile implies that the outer regions are 
more metal poor and have less bottom-heavy IMFs, but they are not less 
alpha-enhanced.  Thus the IMF change should not 
be driven mainly by any changes in the duration of star formation.

A separate question is which global galaxy property is most closely associated
with gradient strength.  In our sample, central velocity 
dispersion (Appendix~\ref{app:gradient_sigma}) alone does not appear to be the primary driver.  
The clearest trends are with compactness, quantified by $\Sigma_{\rm Kroupa}$. 
More compact galaxies have significantly steeper [Fe/H] and [Na/Fe] gradients, while
$\log(\alpha_{\rm IMF})$ shows a similar but only marginal trend.  This suggests 
that compactness retains information about the degree of early dissipative buildup 
of the central regions more clearly than central velocity dispersion alone.

\subsection{The Role of Metallicity}

As presented in \S~3.5, when all radial-bin measurements are combined, 
we find mild positive correlations between local $\alpha_{\rm IMF}$ and the 
local metallicity indicators [Fe/H] and [Z/H].  However, these correlations are 
not significant when the measurements are examined separately within individual 
radial bins, suggesting that the combined-bin trends are partly driven by common 
radial gradients rather than by a tight one-to-one relation at fixed radius.   
We do not find significant local correlations between $\alpha_{\rm IMF}$ and 
[Mg/Fe] or [Na/Fe] in this analysis. Our local result suggests 
that the IMF is more closely related to metallicity than to [$\alpha$/Fe].
The significant scatter in the [Fe/H]-$\log(\alpha_{\rm IMF})$ relation 
also indicates that metallicity is 
unlikely to be the sole driver of IMF variation.

Several theoretical models predict that metallicity can influence the IMF mass 
scale. In the IGIMF framework, the galaxy-wide IMF depends on both metallicity 
and star-formation rate, and metal-rich massive galaxies can develop more 
bottom-heavy galaxy-wide IMFs \citep{Jerabkova2018}.  

Using models of collapsing dusty clouds, \citet{Sharda2022} show that 
metallicity and ISM pressure affect the characteristic stellar mass.
High-metallicity, high-pressure environments favor lower 
characteristic masses, providing a possible route to bottom-heavy IMFs in massive 
galaxies \citep{Sharda2022}.  These models support a connection between 
metallicity and IMF variation, but they also imply that metallicity is not the 
only relevant physical parameter.

\citet{Guszejnov2022} used STARFORGE radiation-magnetohydrodynamic simulations 
\citep{Guszejnov2021, Grudic2021} to examine how environmental conditions and 
stellar feedback affect the IMF in Milky-Way-like environments.  Their work shows 
that the characteristic mass of the IMF is primarily unaffected by the initial 
cloud mass, surface density, or turbulence, while metallicity and the interstellar 
radiation field can affect the IMF peak through their influence on gas temperature.

Despite differences in theoretical frameworks and observational methodologies, a 
connection between metallicity and IMF variation has been found repeatedly in the 
literature.  These models and our measurements support metallicity as an important 
empirical tracer of the physical conditions associated with IMF variation. However, 
the large scatter in the local $\alpha_{\rm IMF}$--metallicity relations, and the 
lack of significant correlations within individual radial bins, indicate that 
metallicity alone does not determine the IMF.
Because galaxies commonly show radial metallicity gradients, studying the relation 
between local IMF constraints and local metallicity across different types of 
galaxies offers a promising way to identify the physical drivers of IMF variation.

\subsection{Caveats and Future Directions}

The data quality and modeling framework used in this paper provide strong 
constraints on the IMF from optical SPS modeling of ETGs.  Nevertheless, future high-quality 
spectra extended to redder wavelengths would provide important tests of the IMF and its 
gradients measured here and could reduce remaining degeneracies among IMF 
shape, age, and elemental abundances. NIR spectroscopy covering $1$--$2\,\mu$m includes 
additional gravity and abundance-sensitive features, such as Na~I $\lambda1.14\,\mu$m, 
K~I $\lambda1.17\,\mu$m, Ca~I $\lambda1.98\,\mu$m, and Na~I $\lambda2.21\,\mu$m.  These 
features would improve constraints on both elemental abundances and the IMF, helping to 
separate abundance-driven changes in the spectra from those caused by IMF variation 
\citep{Smith2012, Lagattuta2017}.  They would also provide an independent test of the 
role of the Wing-Ford band as an IMF tracer \citep{Parikh2018, Vaughan2018b}.  

Another area for future development is the treatment of response functions.  In this work, 
we use response functions computed for a Kroupa IMF. Response functions calculated for 
different IMF shapes, or response functions coupled self-consistently to a flexible IMF, 
would provide a useful test of this modeling assumption.  

Another limitation is that our 
long-slit spectra sample the stellar populations along a single position angle for each 
galaxy.  IFU spectroscopy would test whether the IMF, abundance and stellar population 
gradients measured here are representative azimuthally, and would help identify cases 
where major/minor axis differences affect the inferred IMF profiles. IFU observations 
with NIR wavelength coverage would be especially valuable because they would combine 
two-dimensional spatial information with  additional IMF and abundance sensitive features. 

\section{Conclusion}

We have presented spatially resolved measurements of stellar populations and the low-mass 
IMF slopes in 37 massive ETGs selected from the volume-limited MASSIVE sample. The spectra 
are extracted from high-quality Magellan/LDSS$-3$ long-slit observations spanning 
$0.4$--$1.01\,\mu$m, with radial coverage extending to $0.2$--$1.1R_{\rm e}$ depending on 
the target.  For the main sample-averaged statistical analyses, we use the fiducial sample 
of 35 old galaxies defined in \S~3.1.  Our main conclusions are as follows.

\begin{itemize}

    \item Massive ETGs generally show negative IMF gradients.  In our fiducial sample,
    30 galaxies have declining IMF radial profiles, while 5 have positive IMF 
    radial gradients. The sample-averaged IMF mismatch parameter, 
    $\alpha_{\rm IMF}=(M/L)/(M/L)_{\rm Kroupa}$,  decreases from $2.16$ within 
    $R_{\rm e}/8$ to $1.74$ in the $R_{\rm e}/4$--$R_{\rm e}/2$ bin, with 
    galaxy-to-galaxy scatters of $0.50$ and $0.42$, respectively.  Since a Salpeter 
    IMF corresponds to $\alpha_{\rm IMF}\simeq1.6$ for old stellar populations, the 
    average IMF remains more bottom-heavy than Kroupa and approximately 
    Salpeter-like or more bottom-heavy over these radii.  This indicates that IMF 
    variation is already present in the inner regions of these galaxies (\S~3.1).

    \item The diversity in central IMF mismatch decreases toward larger radius. 
    The radial gradient of $\log(\alpha_{\rm IMF})$ 
    anti-correlates with the central value of $\alpha_{\rm IMF}$, indicating that 
    galaxies with more bottom-heavy central IMFs tend to have steeper declining IMF 
    profiles.  This suggests that the central IMF variation in massive ETGs weakens 
    toward larger radii, potentially converging to a less bottom-heavy IMF in the outskirts 
    (\S~3.4).  This convergence is also seen when the sample is divided by central
    $\alpha_{\rm IMF}$: galaxies with different central IMF mismatch values have
    more similar $\alpha_{\rm IMF}$ distributions by $R_{\rm e}/4$--$R_{\rm e}/2$.
    
    \item The stellar-population gradients separate into two broad behaviors. 
    In the fiducial sample, $\log(\sigma)$, [Fe/H], [Na/Fe], and
    $\log(\alpha_{\rm IMF})$ decline with radius, while [Mg/Fe], [Ca/Fe],
    [Ti/Fe], and [Si/Fe] are approximately flat.  Overall, the IMF gradient follows
    the radial behavior of metallicity and Na abundance more closely than most 
    [$\alpha$/Fe].

    \item We find a mild positive correlation between local $\alpha_{\rm IMF}$ and local 
    metallicity indicators, [Fe/H] and [Z/H], but no significant correlation with 
    [Na/Fe] or [Mg/Fe].  The lack of a correlation with [Mg/Fe], together with the 
    approximately flat radial profiles of several $\alpha$-element abundance ratios, 
    suggests that the IMF variation is not primarily tied to the star-formation 
    timescale traced by [$\alpha$/Fe] (\S~3.5). The relation with [Na/Fe] is less 
    straightforward: [Na/Fe] shows a negative radial trend similar to 
    $\alpha_{\rm IMF}$, but does not show a significant local correlation with 
    $\alpha_{\rm IMF}$.  Overall, the data favor a closer connection between IMF 
    variation and metallicity than with either [$\alpha$/Fe] or [Na/Fe].
    
    \item The sample averaged $\log(\alpha_{\rm IMF})$ gradient is much 
    steeper than the Kroupa-IMF $M/L_r$ gradient, indicating that the stellar 
    $M/L_r$ gradient is dominated by radial IMF variation rather than by age, 
    metallicity, and abundance gradients at fixed IMF.

    \item Finally, IMF gradients have a direct impact on stellar mass measurements. 
    Assuming a fixed Kroupa IMF underestimates the stellar masses of massive ETGs 
    by factors of $1.71^{+0.07}_{-0.07}$ and $1.53^{+0.09}_{-0.08}$ within apertures 
    of $R_{\rm e}/2$ and $R_{\rm e}$, respectively, compared with measurements 
    that allow a flexible IMF (\S~4.1). Aperture size and radial IMF variation 
    therefore need to be considered when comparing stellar masses inferred from 
    stellar population, dynamical, and lensing methods.
    
\end{itemize}

\begin{acknowledgments}
The MASSIVE survey is supported in part by NSF AST-1815417 and AST-1817100.  This paper 
includes data gathered with the 6.5 meter Magellan Telescopes located at Las Campanas 
Observatory, Chile.  The authors are pleased to acknowledge that the work reported on in 
this paper was substantially performed using the Princeton Research Computing resources 
at Princeton University which is a consortium of groups led by the Princeton Institute for 
Computational Science and Engineering (PICSciE) and Office of Information Technology's 
Research Computing.

The Siena Galaxy Atlas was made possible by funding support from the U.S. Department of 
Energy, Office of Science, Office of High Energy Physics under Award Number DE-SC0020086 
and from the National Science Foundation under grant AST-1616414.

{\it Software}:
{\tt Python} \citep{VanRossum1995, VanRossum2009}, 
{\tt Astropy} \citep{astropy:2013, astropy:2018}, 
{\tt Numpy} \citep{harris2020array}, 
{\tt Scipy} \citep{2020SciPy-NMeth}, 
{\tt emcee} \citep{ForemanMackey2013}, 
{\tt Matplotlib} \citep{Hunter:2007}, 
{\tt Pandas} \citep{mckinney2010data}, 
{\tt statsmodels} \citep{seabold2010statsmodels},
{\tt Molecfit} \citep{Kausch2015, Smette2015}, 
{\tt PypeIt} \citep{pypeit_Zenodo, pypeit_JOSS}, 
{\tt pPXF} \citep{Cappellari2017}, 
{\tt L.A.Cosmic} \citep{2012ascl.soft07005V}.

\end{acknowledgments}

\appendix

\section{Additional Sample and Gradient Checks}

\begin{figure*}[p]
\centering
\includegraphics[height=0.82\textheight]{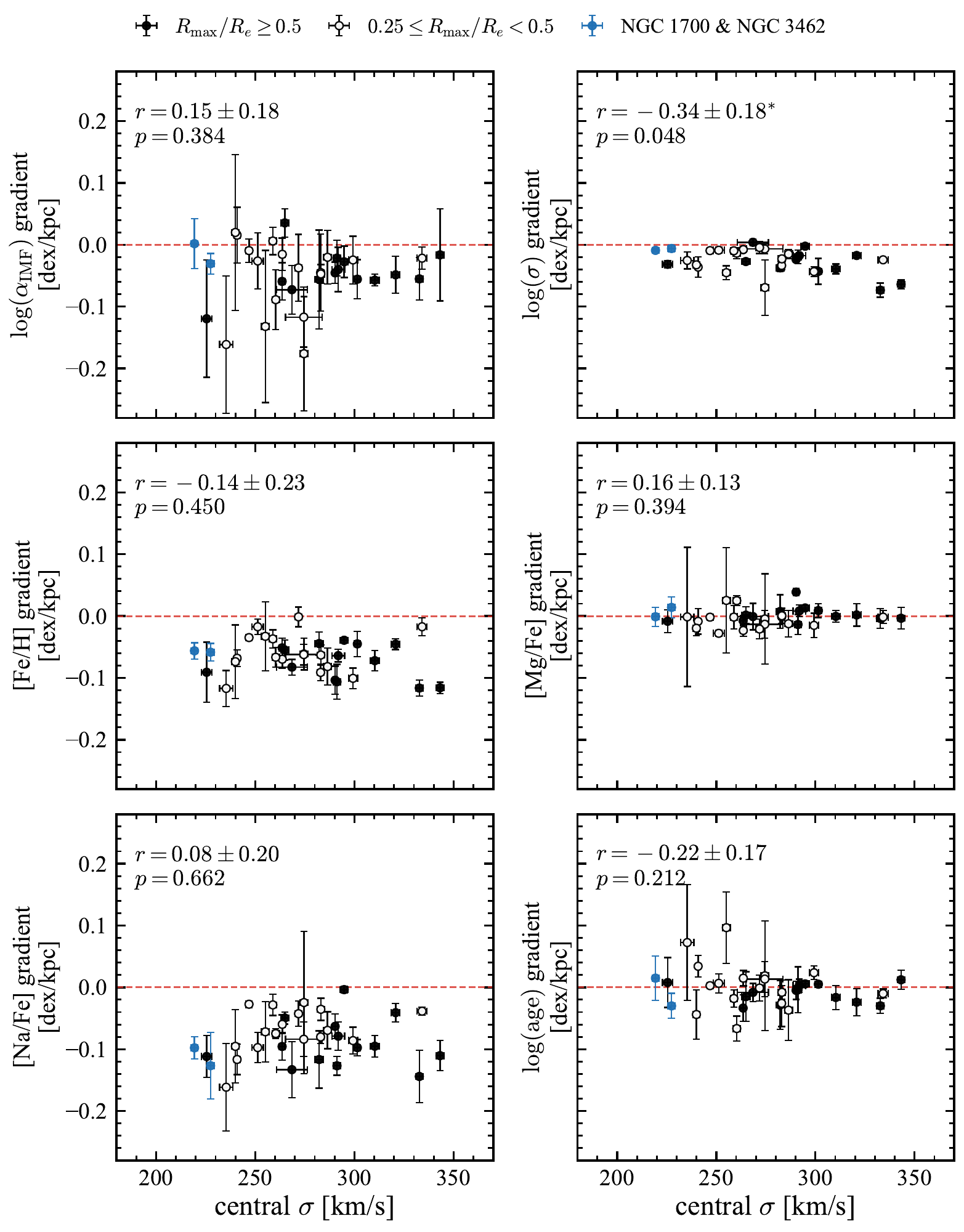}
\caption{
Radial gradients of $\log(\alpha_{\rm IMF})$, $\log(\sigma)$, [Fe/H], [Mg/Fe],
[Na/Fe], and $\log({\rm age})$ as a function of central velocity
dispersion. Filled black symbols show old galaxies with radial coverage beyond
$0.5R_{\rm e}$, while open black symbols show old galaxies with
$0.25\leq R_{\max}/R_{\rm e}<0.5$. Blue points mark the two relatively young
galaxies, NGC~1700 and NGC~3462, shown for reference but excluded from the
Pearson correlation coefficients. Pearson $r$ values and $p$-values are shown
for the fiducial sample.
}
\label{fig_gradient}
\end{figure*}

In this appendix, we present two analyses.  First, we summarize the two relatively
young galaxies that are excluded from the fiducial sample-averaged analyses, 
to show how they compare with the other targets. Second, we examine whether 
the measured gradients depend on central velocity dispersion, a commonly used 
global proxy for galaxy mass.

\subsection{The Two Young Galaxies Excluded from the Fiducial Sample}

\label{app:young}

The main sample-averaged statistical analyses use the fiducial sample defined
in \S~3.1.  We exclude NGC~1700 and
NGC~3462 from those comparisons because their central ages are about 5~Gyr,
making them clear outliers in luminosity-weighted age, and their star-formation 
times are on average later than the other 35 ETGs.  We treat these two galaxies 
separately rather than mixing them into the fiducial sample. 
Here we explore whether their properties are different from those of the other 35 ETGs. 

In their central regions, both young galaxies are metal rich and among the least
$\alpha$-enhanced objects in the sample, with [Fe/H] at the $\geq97$th
percentile and [Mg/Fe] at the $\leq3$rd percentile.  Their [Na/Fe] values are
also relatively low, at the 20th and 14th percentiles.  In contrast, their
$\alpha_{\rm IMF}$ values are not extreme, lying at the 6th and 69th
percentiles.  The central IMF is approximately Salpeter-like in NGC~1700, with
$\alpha_{\rm IMF}\approx1.6$, and more bottom-heavy in NGC~3462, with
$\alpha_{\rm IMF}\approx2.2$.

Their IMF gradients are also not exceptional compared with the rest of the
sample: the $\log(\alpha_{\rm IMF})$ gradient is flat in NGC~3462 and negative
in NGC~1700.  In ETGs, previous studies \citep[e.g.][]{Graves2010} found that younger 
ages are associated with higher metallicity and more compact structure at fixed 
velocity dispersion. Thus, these two young galaxies are chemically distinct from 
the fiducial sample, in line with the scenario proposed by 
\citet{Graves2010} that galaxies with young mean age have more extended 
star-formation histories and are more metal enriched. However, their IMF 
mismatch parameters and IMF gradients are not exceptional.  This suggests that 
younger luminosity-weighted age alone does not produce an obvious offset in 
$\alpha_{\rm IMF}$ for these two objects, although the sample is too small to 
rule out a broader age dependence.

\subsection{Gradients versus Central Velocity Dispersion}
\label{app:gradient_sigma}

Central velocity dispersion is one of the most widely used global predictors of
IMF variation in ETGs.  If the radial gradients were primarily controlled by the
same global quantity that drives the central IMF mismatch, we might expect the
gradient strength to vary systematically with central $\sigma$.  We therefore
examine this relation here.

Figure~\ref{fig_gradient} shows the radial gradients of six fitted quantities: 
$\log(\sigma)$, $\log({\rm age})$, [Fe/H], [Mg/Fe], [Na/Fe], and 
$\log(\alpha_{\rm IMF})$, plotted as a function of central velocity dispersion.  
All gradients are measured with respect to deprojected radius and are reported
in units of dex~kpc$^{-1}$.  
The gradients of $\log(\alpha_{\rm IMF})$, [Fe/H], and [Na/Fe] are predominantly
negative, while the gradients of $\log({\rm age})$ and [Mg/Fe] show both positive
and negative values.  We do not find significant correlations between central
velocity dispersion and the gradients of $\log(\alpha_{\rm IMF})$, [Fe/H],
[Mg/Fe], [Na/Fe], or $\log({\rm age})$.  The $\log(\sigma)$ gradient shows a
marginal anti-correlation with central velocity dispersion.

\bibliographystyle{aasjournalv7.1}
\bibliography{references}

\begin{thebibliography}{}
\expandafter\ifx\csname natexlab\endcsname\relax\def\natexlab#1{#1}\fi
\providecommand{\url}[1]{\href{#1}{#1}}
\providecommand{\dodoi}[1]{doi:~\href{http://doi.org/#1}{\nolinkurl{#1}}}
\providecommand{\doeprint}[1]{\href{http://ascl.net/#1}{\nolinkurl{http://ascl.net/#1}}}
\providecommand{\doarXiv}[1]{\href{https://arxiv.org/abs/#1}{\nolinkurl{https://arxiv.org/abs/#1}}}

\bibitem[{P.~D. {Alton} {et~al.}(2017){Alton}, {Smith}, \& {Lucey}}]{Alton2017}
{Alton}, P.~D., {Smith}, R.~J., \& {Lucey}, J.~R. 2017,
  \bibinfo{title}{{KINETyS: constraining spatial variations of the stellar
  initial mass function in early-type galaxies$^{★}$},} \mnras, 468, 1594,
  \dodoi{10.1093/mnras/stx464}

\bibitem[{P.~D. {Alton} {et~al.}(2018){Alton}, {Smith}, \& {Lucey}}]{Alton2018}
{Alton}, P.~D., {Smith}, R.~J., \& {Lucey}, J.~R. 2018,
  \bibinfo{title}{{KINETyS II: Constraints on spatial variations of the stellar
  initial mass function from K-band spectroscopy},} \mnras, 478, 4464,
  \dodoi{10.1093/mnras/sty1242}

\bibitem[{{}{Astropy Collaboration} {et~al.}(2013){Astropy Collaboration},
  {Robitaille}, {Tollerud}, {Greenfield}, {Droettboom}, {Bray}, {Aldcroft},
  {Davis}, {Ginsburg}, {Price-Whelan}, {Kerzendorf}, {Conley}, {Crighton},
  {Barbary}, {Muna}, {Ferguson}, {Grollier}, {Parikh}, {Nair}, {Unther},
  {Deil}, {Woillez}, {Conseil}, {Kramer}, {Turner}, {Singer}, {Fox}, {Weaver},
  {Zabalza}, {Edwards}, {Azalee Bostroem}, {Burke}, {Casey}, {Crawford},
  {Dencheva}, {Ely}, {Jenness}, {Labrie}, {Lim}, {Pierfederici}, {Pontzen},
  {Ptak}, {Refsdal}, {Servillat}, \& {Streicher}}]{astropy:2013}
{Astropy Collaboration}, {Robitaille}, T.~P., {Tollerud}, E.~J., {et~al.} 2013,
  \bibinfo{title}{{Astropy: A community Python package for astronomy},} \aap,
  558, A33, \dodoi{10.1051/0004-6361/201322068}

\bibitem[{{}{Astropy Collaboration} {et~al.}(2018){Astropy Collaboration},
  {Price-Whelan}, {Sip{\H{o}}cz}, {G{\"u}nther}, {Lim}, {Crawford}, {Conseil},
  {Shupe}, {Craig}, {Dencheva}, {Ginsburg}, {Vand erPlas}, {Bradley},
  {P{\'e}rez-Su{\'a}rez}, {de Val-Borro}, {Aldcroft}, {Cruz}, {Robitaille},
  {Tollerud}, {Ardelean}, {Babej}, {Bach}, {Bachetti}, {Bakanov}, {Bamford},
  {Barentsen}, {Barmby}, {Baumbach}, {Berry}, {Biscani}, {Boquien}, {Bostroem},
  {Bouma}, {Brammer}, {Bray}, {Breytenbach}, {Buddelmeijer}, {Burke},
  {Calderone}, {Cano Rodr{\'\i}guez}, {Cara}, {Cardoso}, {Cheedella}, {Copin},
  {Corrales}, {Crichton}, {D'Avella}, {Deil}, {Depagne}, {Dietrich}, {Donath},
  {Droettboom}, {Earl}, {Erben}, {Fabbro}, {Ferreira}, {Finethy}, {Fox},
  {Garrison}, {Gibbons}, {Goldstein}, {Gommers}, {Greco}, {Greenfield},
  {Groener}, {Grollier}, {Hagen}, {Hirst}, {Homeier}, {Horton}, {Hosseinzadeh},
  {Hu}, {Hunkeler}, {Ivezi{\'c}}, {Jain}, {Jenness}, {Kanarek}, {Kendrew},
  {Kern}, {Kerzendorf}, {Khvalko}, {King}, {Kirkby}, {Kulkarni}, {Kumar},
  {Lee}, {Lenz}, {Littlefair}, {Ma}, {Macleod}, {Mastropietro}, {McCully},
  {Montagnac}, {Morris}, {Mueller}, {Mumford}, {Muna}, {Murphy}, {Nelson},
  {Nguyen}, {Ninan}, {N{\"o}the}, {Ogaz}, {Oh}, {Parejko}, {Parley}, {Pascual},
  {Patil}, {Patil}, {Plunkett}, {Prochaska}, {Rastogi}, {Reddy Janga},
  {Sabater}, {Sakurikar}, {Seifert}, {Sherbert}, {Sherwood-Taylor}, {Shih},
  {Sick}, {Silbiger}, {Singanamalla}, {Singer}, {Sladen}, {Sooley},
  {Sornarajah}, {Streicher}, {Teuben}, {Thomas}, {Tremblay}, {Turner},
  {Terr{\'o}n}, {van Kerkwijk}, {de la Vega}, {Watkins}, {Weaver}, {Whitmore},
  {Woillez}, {Zabalza}, \& {Astropy Contributors}}]{astropy:2018}
{Astropy Collaboration}, {Price-Whelan}, A.~M., {Sip{\H{o}}cz}, B.~M., {et~al.}
  2018, \bibinfo{title}{{The Astropy Project: Building an Open-science Project
  and Status of the v2.0 Core Package},} \aj, 156, 123,
  \dodoi{10.3847/1538-3881/aabc4f}

\bibitem[{C.~E. {Barbosa} {et~al.}(2021){Barbosa}, {Spiniello}, {Arnaboldi},
  {Coccato}, {Hilker}, \& {Richtler}}]{Barbosa2021}
{Barbosa}, C.~E., {Spiniello}, C., {Arnaboldi}, M., {et~al.} 2021,
  \bibinfo{title}{{What does (not) drive the variation of the low-mass end of
  the stellar initial mass function of early-type galaxies},} \aap, 645, L1,
  \dodoi{10.1051/0004-6361/202039810}

\bibitem[{N. {Bastian} {et~al.}(2010){Bastian}, {Covey}, \&
  {Meyer}}]{Bastian2010}
{Bastian}, N., {Covey}, K.~R., \& {Meyer}, M.~R. 2010, \bibinfo{title}{{A
  Universal Stellar Initial Mass Function? A Critical Look at Variations},}
  \araa, 48, 339, \dodoi{10.1146/annurev-astro-082708-101642}

\bibitem[{M. {Bernardi} {et~al.}(2018){Bernardi}, {Sheth}, {Dominguez-Sanchez},
  {Fischer}, {Chae}, {Huertas-Company}, \& {Shankar}}]{Bernardi2018}
{Bernardi}, M., {Sheth}, R.~K., {Dominguez-Sanchez}, H., {et~al.} 2018,
  \bibinfo{title}{{M$_{*}$/L gradients driven by IMF variation: large impact on
  dynamical stellar mass estimates},} \mnras, 477, 2560,
  \dodoi{10.1093/mnras/sty781}

\bibitem[{M. {Cappellari}(2017){Cappellari}}]{Cappellari2017}
{Cappellari}, M. 2017, \bibinfo{title}{{Improving the full spectrum fitting
  method: accurate convolution with Gauss-Hermite functions},} MNRAS, 466, 798,
  \dodoi{10.1093/mnras/stw3020}

\bibitem[{M. {Cappellari} {et~al.}(2012){Cappellari}, {McDermid}, {Alatalo},
  {Blitz}, {Bois}, {Bournaud}, {Bureau}, {Crocker}, {Davies}, {Davis}, {de
  Zeeuw}, {Duc}, {Emsellem}, {Khochfar}, {Krajnovi{\'c}}, {Kuntschner},
  {Lablanche}, {Morganti}, {Naab}, {Oosterloo}, {Sarzi}, {Scott}, {Serra},
  {Weijmans}, \& {Young}}]{Cappellari2012}
{Cappellari}, M., {McDermid}, R.~M., {Alatalo}, K., {et~al.} 2012,
  \bibinfo{title}{{Systematic variation of the stellar initial mass function in
  early-type galaxies},} \nat, 484, 485, \dodoi{10.1038/nature10972}

\bibitem[{M. {Cappellari} {et~al.}(2013){Cappellari}, {Scott}, {Alatalo},
  {Blitz}, {Bois}, {Bournaud}, {Bureau}, {Crocker}, {Davies}, {Davis}, {de
  Zeeuw}, {Duc}, {Emsellem}, {Khochfar}, {Krajnovi{\'c}}, {Kuntschner},
  {McDermid}, {Morganti}, {Naab}, {Oosterloo}, {Sarzi}, {Serra}, {Weijmans}, \&
  {Young}}]{Cappellari2013}
{Cappellari}, M., {Scott}, N., {Alatalo}, K., {et~al.} 2013,
  \bibinfo{title}{{The ATLAS$^{3D}$ project - XV. Benchmark for early-type
  galaxies scaling relations from 260 dynamical models: mass-to-light ratio,
  dark matter, Fundamental Plane and Mass Plane},} \mnras, 432, 1709,
  \dodoi{10.1093/mnras/stt562}

\bibitem[{A.~J. {Cenarro} {et~al.}(2003){Cenarro}, {Gorgas}, {Vazdekis},
  {Cardiel}, \& {Peletier}}]{Cenarro2003}
{Cenarro}, A.~J., {Gorgas}, J., {Vazdekis}, A., {Cardiel}, N., \& {Peletier},
  R.~F. 2003, \bibinfo{title}{{Near-infrared line-strengths in elliptical
  galaxies: evidence for initial mass function variations?},} \mnras, 339, L12,
  \dodoi{10.1046/j.1365-8711.2003.06360.x}

\bibitem[{G. {Chabrier}(2003){Chabrier}}]{Chabrier2003}
{Chabrier}, G. 2003, \bibinfo{title}{{Galactic Stellar and Substellar Initial
  Mass Function},} \pasp, 115, 763, \dodoi{10.1086/376392}

\bibitem[{C.~M. {Cheng} {et~al.}(2023){Cheng}, {Villaume}, {Balogh}, {Brodie},
  {Mart{\'\i}n-Navarro}, {Romanowsky}, \& {van Dokkum}}]{Cheng2023}
{Cheng}, C.~M., {Villaume}, A., {Balogh}, M.~L., {et~al.} 2023,
  \bibinfo{title}{{Initial mass function variability from the integrated light
  of diverse stellar systems},} \mnras, 526, 4004,
  \dodoi{10.1093/mnras/stad2967}

\bibitem[{J. Choi {et~al.}(2016)Choi, Dotter, Conroy, Cantiello, Paxton, \&
  Johnson}]{Choi2016}
Choi, J., Dotter, A., Conroy, C., {et~al.} 2016, \bibinfo{title}{{MESA
  ISOCHRONES AND STELLAR TRACKS (MIST). I. SOLAR-SCALED MODELS},} ApJ, 823, 102

\bibitem[{L. {Ciotti}(1991){Ciotti}}]{Ciotti1991}
{Ciotti}, L. 1991, \bibinfo{title}{{Stellar systems following the {$R^{1/m}$}
  luminosity law},} \aap, 249, 99

\bibitem[{L. {Coccato} {et~al.}(2010){Coccato}, {Gerhard}, \&
  {Arnaboldi}}]{Coccato2010}
{Coccato}, L., {Gerhard}, O., \& {Arnaboldi}, M. 2010,
  \bibinfo{title}{{Distinct core and halo stellar populations and the formation
  history of the bright Coma cluster early-type galaxy NGC 4889},} \mnras, 407,
  L26, \dodoi{10.1111/j.1745-3933.2010.00897.x}

\bibitem[{J.~G. {Cohen}(1978){Cohen}}]{Cohen1978}
{Cohen}, J.~G. 1978, \bibinfo{title}{{Near-infrared luminosity-sensitive
  features in M dwarfs and giants, and in M31 and M32.},} \apj, 221, 788,
  \dodoi{10.1086/156081}

\bibitem[{J.~G. {Cohen}(1979){Cohen}}]{Cohen1979}
{Cohen}, J.~G. 1979, \bibinfo{title}{{Observations and interpretations of
  radial gradients of absorption features in galaxies.},} \apj, 228, 405,
  \dodoi{10.1086/156859}

\bibitem[{C. {Conroy} {et~al.}(2013){Conroy}, {Dutton}, {Graves}, {Mendel}, \&
  {van Dokkum}}]{Conroy2013}
{Conroy}, C., {Dutton}, A.~A., {Graves}, G.~J., {Mendel}, J.~T., \& {van
  Dokkum}, P.~G. 2013, \bibinfo{title}{{Dynamical versus Stellar Masses in
  Compact Early-type Galaxies: Further Evidence for Systematic Variation in the
  Stellar Initial Mass Function},} \apjl, 776, L26,
  \dodoi{10.1088/2041-8205/776/2/L26}

\bibitem[{C. Conroy {et~al.}(2014)Conroy, Graves, \& van Dokkum}]{Conroy2014}
Conroy, C., Graves, G.~J., \& van Dokkum, P.~G. 2014,
  \bibinfo{title}{{Early-type Galaxy Archeology: Ages, Abundance Ratios, and
  Effective Temperatures from Full-spectrum Fitting},} The Astrophysical
  Journal Letters, 780, 33

\bibitem[{C. Conroy {\&} P. van Dokkum(2012)Conroy \& van Dokkum}]{Conroy2012a}
Conroy, C., \& van Dokkum, P. 2012, \bibinfo{title}{{Counting Low-mass Stars in
  Integrated Light},} The Astrophysical Journal Letters, 747, 69

\bibitem[{C. Conroy {\&} P.~G. van Dokkum(2012)Conroy \& van
  Dokkum}]{Conroy2012b}
Conroy, C., \& van Dokkum, P.~G. 2012, \bibinfo{title}{{THE STELLAR INITIAL
  MASS FUNCTION IN EARLY-TYPE GALAXIES FROM ABSORPTION LINE SPECTROSCOPY. II.
  RESULTS},} The Astrophysical Journal Letters, 760, 71

\bibitem[{C. Conroy {et~al.}(2017)Conroy, van Dokkum, \& Villaume}]{Conroy2017}
Conroy, C., van Dokkum, P.~G., \& Villaume, A. 2017, \bibinfo{title}{{The
  Stellar Initial Mass Function in Early-type Galaxies from Absorption Line
  Spectroscopy. IV. A Super-Salpeter IMF in the Center of NGC 1407 from
  Non-parametric Models},} ApJ, 837, 166

\bibitem[{C. Conroy {et~al.}(2018)Conroy, Villaume, van Dokkum, \&
  Lind}]{Conroy2018}
Conroy, C., Villaume, A., van Dokkum, P.~G., \& Lind, K. 2018,
  \bibinfo{title}{{Metal-rich, Metal-poor: Updated Stellar Population Models
  for Old Stellar Systems},} ApJ, 854, 139

\bibitem[{B.~A. Cook {et~al.}(2016)Cook, Conroy, Pillepich, Rodriguez-Gomez, \&
  Hernquist}]{Cook2016}
Cook, B.~A., Conroy, C., Pillepich, A., Rodriguez-Gomez, V., \& Hernquist, L.
  2016, \bibinfo{title}{{THE INFORMATION CONTENT OF STELLAR HALOS: STELLAR
  POPULATION GRADIENTS AND ACCRETION HISTORIES IN EARLY-TYPE ILLUSTRIS
  GALAXIES},} ApJ, 833, 158

\bibitem[{R.~L. {Davies} {et~al.}(1993){Davies}, {Sadler}, \&
  {Peletier}}]{Davies1993}
{Davies}, R.~L., {Sadler}, E.~M., \& {Peletier}, R.~F. 1993,
  \bibinfo{title}{{Line-strength gradients in elliptical galaxies.},} \mnras,
  262, 650, \dodoi{10.1093/mnras/262.3.650}

\bibitem[{T.~A. {Davis} {\&} R.~M. {McDermid}(2017){Davis} \&
  {McDermid}}]{Davis2017}
{Davis}, T.~A., \& {McDermid}, R.~M. 2017, \bibinfo{title}{{Spatially resolved
  variations of the IMF mass normalization in early-type galaxies as probed by
  molecular gas kinematics},} \mnras, 464, 453, \dodoi{10.1093/mnras/stw2366}

\bibitem[{P. Di~Matteo {et~al.}(2009)Di~Matteo, Pipino, Lehnert, Combes, \&
  Semelin}]{DiMatteo2009}
Di~Matteo, P., Pipino, A., Lehnert, M.~D., Combes, F., \& Semelin, B. 2009,
  \bibinfo{title}{{On the survival of metallicity gradients to major
  dry-mergers},} A{\&}A, 499, 427

\bibitem[{P. {Dominiak} {et~al.}(2024){Dominiak}, {Bureau}, {Davis}, {Ma},
  {Greene}, \& {Gu}}]{Dominiak2024}
{Dominiak}, P., {Bureau}, M., {Davis}, T.~A., {et~al.} 2024,
  \bibinfo{title}{{The MASSIVE survey - XIX. Molecular gas measurements of the
  supermassive black hole masses in the elliptical galaxies NGC 1684 and NGC
  0997},} \mnras, 529, 1597, \dodoi{10.1093/mnras/stae314}

\bibitem[{A.~A. {Dutton} {et~al.}(2012){Dutton}, {Mendel}, \&
  {Simard}}]{Dutton2012}
{Dutton}, A.~A., {Mendel}, J.~T., \& {Simard}, L. 2012,
  \bibinfo{title}{{Evidence for a non-universal stellar initial mass function
  in low-redshift high-density early-type galaxies},} \mnras, 422, L33,
  \dodoi{10.1111/j.1745-3933.2012.01230.x}

\bibitem[{I. {Ene} {et~al.}(2019){Ene}, {Ma}, {McConnell}, {Walsh}, {Kempski},
  {Greene}, {Thomas}, \& {Blakeslee}}]{Ene2019}
{Ene}, I., {Ma}, C.-P., {McConnell}, N.~J., {et~al.} 2019, \bibinfo{title}{{The
  MASSIVE Survey XIII. Spatially Resolved Stellar Kinematics in the Central 1
  kpc of 20 Massive Elliptical Galaxies with the GMOS-North Integral Field
  Spectrograph},} \apj, 878, 57, \dodoi{10.3847/1538-4357/ab1f04}

\bibitem[{I. {Ene} {et~al.}(2020){Ene}, {Ma}, {Walsh}, {Greene}, {Thomas}, \&
  {Blakeslee}}]{Ene2020}
{Ene}, I., {Ma}, C.-P., {Walsh}, J.~L., {et~al.} 2020, \bibinfo{title}{{The
  MASSIVE Survey XIV{\textemdash}Stellar Velocity Profiles and Kinematic
  Misalignments from 200 pc to 20 kpc in Massive Early-type Galaxies},} \apj,
  891, 65, \dodoi{10.3847/1538-4357/ab7016}

\bibitem[{S.~M. {Faber} {\&} H.~B. {French}(1980){Faber} \&
  {French}}]{Faber1980}
{Faber}, S.~M., \& {French}, H.~B. 1980, \bibinfo{title}{{Possible M dwarf
  enrichment in the semistellar nucleus of M31},} \apj, 235, 405,
  \dodoi{10.1086/157644}

\bibitem[{I. {Ferreras} {et~al.}(2013){Ferreras}, {La Barbera}, {de La Rosa},
  {Vazdekis}, {de Carvalho}, {Falcon-Barroso}, \&
  {Ricciardelli}}]{Ferreras2013}
{Ferreras}, I., {La Barbera}, F., {de La Rosa}, I.~G., {et~al.} 2013,
  \bibinfo{title}{{Systematic variation of the stellar initial mass function
  with velocity dispersion in early-type galaxies.},} \mnras, 429, L15,
  \dodoi{10.1093/mnrasl/sls014}

\bibitem[{I. {Ferreras} {et~al.}(2019){Ferreras}, {Scott}, {La Barbera},
  {Croom}, {van de Sande}, {Hopkins}, {Colless}, {Barone}, {d'Eugenio},
  {Bland-Hawthorn}, {Brough}, {Bryant}, {Konstantopoulos}, {Lagos}, {Lawrence},
  {L{\'o}pez-S{\'a}nchez}, {Medling}, {Owers}, \& {Richards}}]{Ferreras2019}
{Ferreras}, I., {Scott}, N., {La Barbera}, F., {et~al.} 2019,
  \bibinfo{title}{{The SAMI galaxy survey: stellar population radial gradients
  in early-type galaxies},} \mnras, 489, 608, \dodoi{10.1093/mnras/stz2095}

\bibitem[{C. {Filion} {et~al.}(2024){Filion}, {Wyse}, {Richstein},
  {Kallivayalil}, {van der Marel}, \& {Sacchi}}]{Filion2024}
{Filion}, C., {Wyse}, R. F.~G., {Richstein}, H., {et~al.} 2024,
  \bibinfo{title}{{The Low-mass Stellar Initial Mass Function in Nearby
  Ultrafaint Dwarf Galaxies},} \apj, 967, 165, \dodoi{10.3847/1538-4357/ad4020}

\bibitem[{D. Foreman-Mackey {et~al.}(2013)Foreman-Mackey, Hogg, Lang, \&
  Goodman}]{ForemanMackey2013}
Foreman-Mackey, D., Hogg, D.~W., Lang, D., \& Goodman, J. 2013,
  \bibinfo{title}{{emcee: The MCMC Hammer},} Publications of the Astronomical
  Society of the Pacific, 125, 306

\bibitem[{M. {Geha} {et~al.}(2013){Geha}, {Brown}, {Tumlinson}, {Kalirai},
  {Simon}, {Kirby}, {VandenBerg}, {Mu{\~n}oz}, {Avila}, {Guhathakurta}, \&
  {Ferguson}}]{Geha2013}
{Geha}, M., {Brown}, T.~M., {Tumlinson}, J., {et~al.} 2013,
  \bibinfo{title}{{The Stellar Initial Mass Function of Ultra-faint Dwarf
  Galaxies: Evidence for IMF Variations with Galactic Environment},} \apj, 771,
  29, \dodoi{10.1088/0004-637X/771/1/29}

\bibitem[{M. {Gennaro} {et~al.}(2018){Gennaro}, {Tchernyshyov}, {Brown},
  {Geha}, {Avila}, {Guhathakurta}, {Kalirai}, {Kirby}, {Renzini}, {Simon},
  {Tumlinson}, \& {Vargas}}]{Gennaro2018}
{Gennaro}, M., {Tchernyshyov}, K., {Brown}, T.~M., {et~al.} 2018,
  \bibinfo{title}{{Evidence of a Non-universal Stellar Initial Mass Function.
  Insights from HST Optical Imaging of Six Ultra-faint Dwarf Milky Way
  Satellites},} \apj, 855, 20, \dodoi{10.3847/1538-4357/aaa973}

\bibitem[{D. {Goddard} {et~al.}(2017){Goddard}, {Thomas}, {Maraston},
  {Westfall}, {Etherington}, {Riffel}, {Mallmann}, {Zheng},
  {Argudo-Fern{\'a}ndez}, {Lian}, {Bershady}, {Bundy}, {Drory}, {Law}, {Yan},
  {Wake}, {Weijmans}, {Bizyaev}, {Brownstein}, {Lane}, {Maiolino}, {Masters},
  {Merrifield}, {Nitschelm}, {Pan}, {Roman-Lopes}, {Storchi-Bergmann}, \&
  {Schneider}}]{Goddard2017}
{Goddard}, D., {Thomas}, D., {Maraston}, C., {et~al.} 2017,
  \bibinfo{title}{{SDSS-IV MaNGA: Spatially resolved star formation histories
  in galaxies as a function of galaxy mass and type},} \mnras, 466, 4731,
  \dodoi{10.1093/mnras/stw3371}

\bibitem[{R.~M. {Gonz{\'a}lez Delgado} {et~al.}(2015){Gonz{\'a}lez Delgado},
  {Garc{\'{\i}}a-Benito}, {P{\'e}rez}, {Cid Fernandes}, {de Amorim},
  {Cortijo-Ferrero}, {Lacerda}, {L{\'o}pez Fern{\'a}ndez}, {Vale-Asari},
  {S{\'a}nchez}, {Moll{\'a}}, {Ruiz-Lara}, {S{\'a}nchez-Bl{\'a}zquez},
  {Walcher}, {Alves}, {Aguerri}, {Bekerait{\'e}}, {Bland-Hawthorn}, {Galbany},
  {Gallazzi}, {Husemann}, {Iglesias-P{\'a}ramo}, {Kalinova},
  {L{\'o}pez-S{\'a}nchez}, {Marino}, {M{\'a}rquez}, {Masegosa}, {Mast},
  {M{\'e}ndez-Abreu}, {Mendoza}, {del Olmo}, {P{\'e}rez}, {Quirrenbach}, \&
  {Zibetti}}]{GonzalezDelgado2015}
{Gonz{\'a}lez Delgado}, R.~M., {Garc{\'{\i}}a-Benito}, R., {P{\'e}rez}, E.,
  {et~al.} 2015, \bibinfo{title}{{The CALIFA survey across the Hubble sequence.
  Spatially resolved stellar population properties in galaxies},} \aap, 581,
  A103, \dodoi{10.1051/0004-6361/201525938}

\bibitem[{C.~F. {Goullaud} {et~al.}(2018){Goullaud}, {Jensen}, {Blakeslee},
  {Ma}, {Greene}, \& {Thomas}}]{Goullaud2018}
{Goullaud}, C.~F., {Jensen}, J.~B., {Blakeslee}, J.~P., {et~al.} 2018,
  \bibinfo{title}{{The MASSIVE Survey. IX. Photometric Analysis of 35 High-mass
  Early-type Galaxies with HST WFC3/IR},} \apj, 856, 11,
  \dodoi{10.3847/1538-4357/aab1f3}

\bibitem[{G.~J. Graves {et~al.}(2010)Graves, Faber, \& Schiavon}]{Graves2010}
Graves, G.~J., Faber, S.~M., \& Schiavon, R.~P. 2010,
  \bibinfo{title}{{DISSECTING THE RED SEQUENCE. IV. THE ROLE OF TRUNCATION IN
  THE TWO-DIMENSIONAL FAMILY OF EARLY-TYPE GALAXY STAR FORMATION HISTORIES},}
  \apjl, 721, 278

\bibitem[{J.~E. {Greene} {et~al.}(2015){Greene}, {Janish}, {Ma}, {McConnell},
  {Blakeslee}, {Thomas}, \& {Murphy}}]{Greene2015}
{Greene}, J.~E., {Janish}, R., {Ma}, C.-P., {et~al.} 2015, \bibinfo{title}{{The
  MASSIVE Survey. II. Stellar Population Trends Out to Large Radius in Massive
  Early-type Galaxies},} \apj, 807, 11, \dodoi{10.1088/0004-637X/807/1/11}

\bibitem[{J.~E. {Greene} {et~al.}(2013){Greene}, {Murphy}, {Graves}, {Gunn},
  {Raskutti}, {Comerford}, \& {Gebhardt}}]{Greene2013}
{Greene}, J.~E., {Murphy}, J.~D., {Graves}, G.~J., {et~al.} 2013,
  \bibinfo{title}{{The Stellar Halos of Massive Elliptical Galaxies. II.
  Detailed Abundance Ratios at Large Radius},} \apj, 776, 64,
  \dodoi{10.1088/0004-637X/776/2/64}

\bibitem[{J.~E. {Greene} {et~al.}(2019){Greene}, {Veale}, {Ma}, {Thomas},
  {Quenneville}, {Blakeslee}, {Walsh}, {Goulding}, \& {Ito}}]{Greene2019}
{Greene}, J.~E., {Veale}, M., {Ma}, C.-P., {et~al.} 2019, \bibinfo{title}{{The
  MASSIVE Survey. XII. Connecting Stellar Populations of Early-type Galaxies to
  Kinematics and Environment},} \apj, 874, 66, \dodoi{10.3847/1538-4357/ab01e3}

\bibitem[{M.~Y. {Grudi{\'c}} {et~al.}(2021){Grudi{\'c}}, {Guszejnov},
  {Hopkins}, {Offner}, \& {Faucher-Gigu{\`e}re}}]{Grudic2021}
{Grudi{\'c}}, M.~Y., {Guszejnov}, D., {Hopkins}, P.~F., {Offner}, S. S.~R., \&
  {Faucher-Gigu{\`e}re}, C.-A. 2021, \bibinfo{title}{{STARFORGE: Towards a
  comprehensive numerical model of star cluster formation and feedback},}
  \mnras, 506, 2199, \dodoi{10.1093/mnras/stab1347}

\bibitem[{M. {Gu} {et~al.}(2022){Gu}, {Greene}, {Newman}, {Kreisch},
  {Quenneville}, {Ma}, \& {Blakeslee}}]{Gu2022}
{Gu}, M., {Greene}, J.~E., {Newman}, A.~B., {et~al.} 2022, \bibinfo{title}{{The
  MASSIVE Survey. XVI. The Stellar Initial Mass Function in the Center of
  MASSIVE Early-type Galaxies},} \apj, 932, 103,
  \dodoi{10.3847/1538-4357/ac69ea}

\bibitem[{D. {Guszejnov} {et~al.}(2021){Guszejnov}, {Grudi{\'c}}, {Hopkins},
  {Offner}, \& {Faucher-Gigu{\`e}re}}]{Guszejnov2021}
{Guszejnov}, D., {Grudi{\'c}}, M.~Y., {Hopkins}, P.~F., {Offner}, S. S.~R., \&
  {Faucher-Gigu{\`e}re}, C.-A. 2021, \bibinfo{title}{{STARFORGE: the effects of
  protostellar outflows on the IMF},} \mnras, 502, 3646,
  \dodoi{10.1093/mnras/stab278}

\bibitem[{D. {Guszejnov} {et~al.}(2022){Guszejnov}, {Grudi{\'c}}, {Offner},
  {Faucher-Gigu{\`e}re}, {Hopkins}, \& {Rosen}}]{Guszejnov2022}
{Guszejnov}, D., {Grudi{\'c}}, M.~Y., {Offner}, S. S.~R., {et~al.} 2022,
  \bibinfo{title}{{Effects of the environment and feedback physics on the
  initial mass function of stars in the STARFORGE simulations},} \mnras, 515,
  4929, \dodoi{10.1093/mnras/stac2060}

\bibitem[{C.~R. Harris {et~al.}(2020)Harris, Millman, van~der Walt, Gommers,
  Virtanen, Cournapeau, Wieser, Taylor, Berg, Smith, Kern, Picus, Hoyer, van
  Kerkwijk, Brett, Haldane, del R{\'{i}}o, Wiebe, Peterson,
  G{\'{e}}rard-Marchant, Sheppard, Reddy, Weckesser, Abbasi, Gohlke, \&
  Oliphant}]{harris2020array}
Harris, C.~R., Millman, K.~J., van~der Walt, S.~J., {et~al.} 2020,
  \bibinfo{title}{Array programming with {NumPy},} Nature, 585, 357,
  \dodoi{10.1038/s41586-020-2649-2}

\bibitem[{J.~D. Hunter(2007)Hunter}]{Hunter:2007}
Hunter, J.~D. 2007, \bibinfo{title}{Matplotlib: A 2D graphics environment,}
  Computing in Science \& Engineering, 9, 90, \dodoi{10.1109/MCSE.2007.55}

\bibitem[{T. {Je{\v{r}}{\'a}bkov{\'a}} {et~al.}(2018){Je{\v{r}}{\'a}bkov{\'a}},
  {Hasani Zonoozi}, {Kroupa}, {Beccari}, {Yan}, {Vazdekis}, \&
  {Zhang}}]{Jerabkova2018}
{Je{\v{r}}{\'a}bkov{\'a}}, T., {Hasani Zonoozi}, A., {Kroupa}, P., {et~al.}
  2018, \bibinfo{title}{{Impact of metallicity and star formation rate on the
  time-dependent, galaxy-wide stellar initial mass function},} \aap, 620, A39,
  \dodoi{10.1051/0004-6361/201833055}

\bibitem[{W. {Kausch} {et~al.}(2015){Kausch}, {Noll}, {Smette}, {Kimeswenger},
  {Barden}, {Szyszka}, {Jones}, {Sana}, {Horst}, \& {Kerber}}]{Kausch2015}
{Kausch}, W., {Noll}, S., {Smette}, A., {et~al.} 2015,
  \bibinfo{title}{{Molecfit: A general tool for telluric absorption correction.
  II. Quantitative evaluation on ESO-VLT/X-Shooterspectra},} \aap, 576, A78,
  \dodoi{10.1051/0004-6361/201423909}

\bibitem[{C. {Kobayashi}(2004){Kobayashi}}]{Kobayashi2004}
{Kobayashi}, C. 2004, \bibinfo{title}{{GRAPE-SPH chemodynamical simulation of
  elliptical galaxies - I. Evolution of metallicity gradients},} \mnras, 347,
  740, \dodoi{10.1111/j.1365-2966.2004.07258.x}

\bibitem[{P. Kroupa(2001)Kroupa}]{Kroupa2001}
Kroupa, P. 2001, \bibinfo{title}{{On the variation of the initial mass
  function},} Monthly Notices of the Royal Astronomical Society, 322, 231

\bibitem[{H. {Kuntschner} {et~al.}(2010){Kuntschner}, {Emsellem}, {Bacon},
  {Cappellari}, {Davies}, {de Zeeuw}, {Falc{\'o}n-Barroso}, {Krajnovi{\'c}},
  {McDermid}, {Peletier}, {Sarzi}, {Shapiro}, {van den Bosch}, \& {van de
  Ven}}]{Kuntschner2010}
{Kuntschner}, H., {Emsellem}, E., {Bacon}, R., {et~al.} 2010,
  \bibinfo{title}{{The SAURON project - XVII. Stellar population analysis of
  the absorption line strength maps of 48 early-type galaxies},} \mnras, 408,
  97, \dodoi{10.1111/j.1365-2966.2010.17161.x}

\bibitem[{R. {Kurucz}(1993){Kurucz}}]{Kurucz1993}
{Kurucz}, R. 1993, \bibinfo{title}{{SYNTHE Spectrum Synthesis Programs and Line
  Data.},} SYNTHE Spectrum Synthesis Programs and Line Data. Kurucz CD-ROM No.
  18. Cambridge, 18

\bibitem[{R.~L. Kurucz(1970)Kurucz}]{Kurucz1970}
Kurucz, R.~L. 1970, \bibinfo{title}{{Atlas: a Computer Program for Calculating
  Model Stellar Atmospheres},} SAO Special Report, 309

\bibitem[{F. {La Barbera} {et~al.}(2013){La Barbera}, {Ferreras}, {Vazdekis},
  {de la Rosa}, {de Carvalho}, {Trevisan}, {Falc{\'o}n-Barroso}, \&
  {Ricciardelli}}]{LaBarbera2013}
{La Barbera}, F., {Ferreras}, I., {Vazdekis}, A., {et~al.} 2013,
  \bibinfo{title}{{SPIDER VIII - constraints on the stellar initial mass
  function of early-type galaxies from a variety of spectral features},}
  \mnras, 433, 3017, \dodoi{10.1093/mnras/stt943}

\bibitem[{F. {La Barbera} {et~al.}(2019){La Barbera}, {Vazdekis}, {Ferreras},
  {Pasquali}, {Allende Prieto}, {Mart{\'\i}n-Navarro}, {Aguado}, {de Carvalho},
  {Rembold}, {Falc{\'o}n-Barroso}, \& {van de Ven}}]{LaBarbera2019}
{La Barbera}, F., {Vazdekis}, A., {Ferreras}, I., {et~al.} 2019,
  \bibinfo{title}{{IMF radial gradients in most massive early-type galaxies},}
  \mnras, 489, 4090, \dodoi{10.1093/mnras/stz2192}

\bibitem[{D.~J. {Lagattuta} {et~al.}(2017){Lagattuta}, {Mould}, {Forbes},
  {Monson}, {Pastorello}, \& {Persson}}]{Lagattuta2017}
{Lagattuta}, D.~J., {Mould}, J.~R., {Forbes}, D.~A., {et~al.} 2017,
  \bibinfo{title}{{Evidence of a Bottom-heavy Initial Mass Function in Massive
  Early-type Galaxies from Near-infrared Metal Lines},} \apj, 846, 166,
  \dodoi{10.3847/1538-4357/aa8563}

\bibitem[{R.~B. {Larson}(1974){Larson}}]{Larson1974}
{Larson}, R.~B. 1974, \bibinfo{title}{{Dynamical models for the formation and
  evolution of spherical galaxies},} \mnras, 166, 585,
  \dodoi{10.1093/mnras/166.3.585}

\bibitem[{R. {Lasker} {et~al.}(2013){Lasker}, {van den Bosch}, {van de Ven},
  {Ferreras}, {La Barbera}, {Vazdekis}, \& {Falcon-Barroso}}]{Lasker2013}
{Lasker}, R., {van den Bosch}, R.~C.~E., {van de Ven}, G., {et~al.} 2013,
  \bibinfo{title}{{Bottom-heavy initial mass function in a nearby compact
  l\{star\} galaxy.},} \mnras, 434, L31, \dodoi{10.1093/mnrasl/slt070}

\bibitem[{D. {Leier} {et~al.}(2016){Leier}, {Ferreras}, {Saha}, {Charlot},
  {Bruzual}, \& {La Barbera}}]{Leier2016}
{Leier}, D., {Ferreras}, I., {Saha}, P., {et~al.} 2016, \bibinfo{title}{{Strong
  gravitational lensing and the stellar IMF of early-type galaxies},} \mnras,
  459, 3677, \dodoi{10.1093/mnras/stw885}

\bibitem[{H. {Li} {et~al.}(2017){Li}, {Ge}, {Mao}, {Cappellari}, {Long}, {Li},
  {Emsellem}, {Dutton}, {Li}, {Bundy}, {Thomas}, {Drory}, \& {Lopes}}]{Li2017}
{Li}, H., {Ge}, J., {Mao}, S., {et~al.} 2017, \bibinfo{title}{{SDSS-IV MaNGA:
  Variation of the Stellar Initial Mass Function in Spiral and Early-type
  Galaxies},} \apj, 838, 77, \dodoi{10.3847/1538-4357/aa662a}

\bibitem[{J. {Li} {et~al.}(2023){Li}, {Liu}, {Zhang}, {Tian}, {Fu}, {Li}, \&
  {Yan}}]{Li2023}
{Li}, J., {Liu}, C., {Zhang}, Z.-Y., {et~al.} 2023, \bibinfo{title}{{Stellar
  initial mass function varies with metallicity and time},} \nat, 613, 460,
  \dodoi{10.1038/s41586-022-05488-1}

\bibitem[{C.~M. {Liepold} {et~al.}(2020){Liepold}, {Quenneville}, {Ma},
  {Walsh}, {McConnell}, {Greene}, \& {Blakeslee}}]{Liepold2020}
{Liepold}, C.~M., {Quenneville}, M.~E., {Ma}, C.-P., {et~al.} 2020,
  \bibinfo{title}{{The MASSIVE Survey. XV. A Stellar Dynamical Mass Measurement
  of the Supermassive Black Hole in Massive Elliptical Galaxy NGC 1453},} \apj,
  891, 4, \dodoi{10.3847/1538-4357/ab6f71}

\bibitem[{E.~R. {Liepold} {\&} C.-P. {Ma}(2024){Liepold} \& {Ma}}]{Liepold2024}
{Liepold}, E.~R., \& {Ma}, C.-P. 2024, \bibinfo{title}{{Big Galaxies and Big
  Black Holes: The Massive Ends of the Local Stellar and Black Hole Mass
  Functions and the Implications for Nanohertz Gravitational Waves},} \apjl,
  971, L29, \dodoi{10.3847/2041-8213/ad66b8}

\bibitem[{I. {Lonoce} {et~al.}(2021){Lonoce}, {Feldmeier-Krause}, \&
  {Freedman}}]{Lonoce2021}
{Lonoce}, I., {Feldmeier-Krause}, A., \& {Freedman}, W.~L. 2021,
  \bibinfo{title}{{The Stellar Initial Mass Function and Population Properties
  of M89 from Optical and NIR Spectroscopy: Addressing Biases in Spectral Index
  Analysis},} \apj, 920, 93, \dodoi{10.3847/1538-4357/ac11f9}

\bibitem[{I. {Lonoce} {et~al.}(2023){Lonoce}, {Freedman}, \&
  {Feldmeier-Krause}}]{Lonoce2023}
{Lonoce}, I., {Freedman}, W., \& {Feldmeier-Krause}, A. 2023,
  \bibinfo{title}{{The Initial Mass Function and Other Stellar Properties
  Across the Core of the Hydra I Cluster},} arXiv e-prints, arXiv:2303.00044,
  \dodoi{10.48550/arXiv.2303.00044}

\bibitem[{S.~I. {Loubser} {\&} P. {S{\'a}nchez-Bl{\'a}zquez}(2012){Loubser} \&
  {S{\'a}nchez-Bl{\'a}zquez}}]{Loubser2012}
{Loubser}, S.~I., \& {S{\'a}nchez-Bl{\'a}zquez}, P. 2012,
  \bibinfo{title}{{Stellar population gradients in brightest cluster
  galaxies},} \mnras, 425, 841, \dodoi{10.1111/j.1365-2966.2012.21079.x}

\bibitem[{C.-P. {Ma} {et~al.}(2014){Ma}, {Greene}, {McConnell}, {Janish},
  {Blakeslee}, {Thomas}, \& {Murphy}}]{Ma2014}
{Ma}, C.-P., {Greene}, J.~E., {McConnell}, N., {et~al.} 2014,
  \bibinfo{title}{{The MASSIVE Survey. I. A Volume-limited Integral-field
  Spectroscopic Study of the Most Massive Early-type Galaxies within 108 Mpc},}
  \apj, 795, 158, \dodoi{10.1088/0004-637X/795/2/158}

\bibitem[{I. {Mart{\'\i}n-Navarro}
  {et~al.}(2015{\natexlab{a}}){Mart{\'\i}n-Navarro}, {La Barbera}, {Vazdekis},
  {Falc{\'o}n-Barroso}, \& {Ferreras}}]{Martin-Navarro2015c}
{Mart{\'\i}n-Navarro}, I., {La Barbera}, F., {Vazdekis}, A.,
  {Falc{\'o}n-Barroso}, J., \& {Ferreras}, I. 2015{\natexlab{a}},
  \bibinfo{title}{{Radial variations in the stellar initial mass function of
  early-type galaxies},} \mnras, 447, 1033, \dodoi{10.1093/mnras/stu2480}

\bibitem[{I. {Mart{\'\i}n-Navarro}
  {et~al.}(2015{\natexlab{b}}){Mart{\'\i}n-Navarro}, {La Barbera}, {Vazdekis},
  {Ferr{\'e}-Mateu}, {Trujillo}, \& {Beasley}}]{Martin-Navarro2015a}
{Mart{\'\i}n-Navarro}, I., {La Barbera}, F., {Vazdekis}, A., {et~al.}
  2015{\natexlab{b}}, \bibinfo{title}{{The initial mass function of a massive
  relic galaxy},} \mnras, 451, 1081, \dodoi{10.1093/mnras/stv1022}

\bibitem[{I. {Mart{\'\i}n-Navarro} {et~al.}(2018){Mart{\'\i}n-Navarro},
  {Vazdekis}, {Falc{\'o}n-Barroso}, {La Barbera}, {Y{\i}ld{\i}r{\i}m}, \& {van
  de Ven}}]{Martin-Navarro2018}
{Mart{\'\i}n-Navarro}, I., {Vazdekis}, A., {Falc{\'o}n-Barroso}, J., {et~al.}
  2018, \bibinfo{title}{{Timing the formation and assembly of early-type
  galaxies via spatially resolved stellar populations analysis},} \mnras, 475,
  3700, \dodoi{10.1093/mnras/stx3346}

\bibitem[{I. {Mart{\'\i}n-Navarro} {et~al.}(2019){Mart{\'\i}n-Navarro},
  {Lyubenova}, {van de Ven}, {Falc{\'o}n-Barroso}, {Coccato}, {Corsini},
  {Gadotti}, {Iodice}, {La Barbera}, {McDermid}, {Pinna}, {Sarzi}, {Viaene},
  {de Zeeuw}, \& {Zhu}}]{Martin-Navarro2019}
{Mart{\'\i}n-Navarro}, I., {Lyubenova}, M., {van de Ven}, G., {et~al.} 2019,
  \bibinfo{title}{{Fornax 3D project: a two-dimensional view of the stellar
  initial mass function in the massive lenticular galaxy FCC 167},} \aap, 626,
  A124, \dodoi{10.1051/0004-6361/201935360}

\bibitem[{N.~J. {McConnell} {et~al.}(2013){McConnell}, {Chen}, {Ma}, {Greene},
  {Lauer}, \& {Gebhardt}}]{McConnell2013}
{McConnell}, N.~J., {Chen}, S.-F.~S., {Ma}, C.-P., {et~al.} 2013,
  \bibinfo{title}{{The Effect of Spatial Gradients in Stellar Mass-to-light
  Ratio on Black Hole Mass Measurements},} \apjl, 768, L21,
  \dodoi{10.1088/2041-8205/768/1/L21}

\bibitem[{N.~J. {McConnell} {et~al.}(2012){McConnell}, {Ma}, {Murphy},
  {Gebhardt}, {Lauer}, {Graham}, {Wright}, \& {Richstone}}]{McConnell2012}
{McConnell}, N.~J., {Ma}, C.-P., {Murphy}, J.~D., {et~al.} 2012,
  \bibinfo{title}{{Dynamical Measurements of Black Hole Masses in Four
  Brightest Cluster Galaxies at 100 Mpc},} \apj, 756, 179,
  \dodoi{10.1088/0004-637X/756/2/179}

\bibitem[{W. McKinney {et~al.}(2010)McKinney {et~al.}}]{mckinney2010data}
McKinney, W., {et~al.} 2010, \bibinfo{title}{Data structures for statistical
  computing in python,} in Proceedings of the 9th Python in Science Conference,
  Vol. 445, Austin, TX, 51--56

\bibitem[{D. {Mehlert} {et~al.}(2003){Mehlert}, {Thomas}, {Saglia}, {Bender},
  \& {Wegner}}]{Mehlert2003}
{Mehlert}, D., {Thomas}, D., {Saglia}, R.~P., {Bender}, R., \& {Wegner}, G.
  2003, \bibinfo{title}{{Spatially resolved spectroscopy of Coma cluster
  early-type galaxies. III. The stellar population gradients},} \aap, 407, 423,
  \dodoi{10.1051/0004-6361:20030886}

\bibitem[{K. {Mehrgan} {et~al.}(2024){Mehrgan}, {Thomas}, {Saglia}, {Parikh},
  {Neureiter}, {Erwin}, \& {Bender}}]{Mehrgan2024}
{Mehrgan}, K., {Thomas}, J., {Saglia}, R., {et~al.} 2024,
  \bibinfo{title}{{Dynamical Stellar Mass-to-light Ratio Gradients: Evidence
  for Very Centrally Concentrated IMF Variations in ETGs?},} \apj, 961, 127,
  \dodoi{10.3847/1538-4357/acfe09}

\bibitem[{J.~T. {Mendel} {et~al.}(2020){Mendel}, {Beifiori}, {Saglia},
  {Bender}, {Brammer}, {Chan}, {F{\"o}rster Schreiber}, {Fossati}, {Galametz},
  {Momcheva}, {Nelson}, {Wilman}, \& {Wuyts}}]{Mendel2020}
{Mendel}, J.~T., {Beifiori}, A., {Saglia}, R.~P., {et~al.} 2020,
  \bibinfo{title}{{The Kinematics of Massive Quiescent Galaxies at 1.4 < z <
  2.1: Dark Matter Fractions, IMF Variation, and the Relation to Local
  Early-type Galaxies},} \apj, 899, 87, \dodoi{10.3847/1538-4357/ab9ffc}

\bibitem[{J. {Moustakas} {et~al.}(2023){Moustakas}, {Lang}, {Dey}, {Juneau},
  {Meisner}, {Myers}, {Schlafly}, {Schlegel}, {Valdes}, {Weaver}, \&
  {Zhou}}]{Moustakas2023}
{Moustakas}, J., {Lang}, D., {Dey}, A., {et~al.} 2023, \bibinfo{title}{{Siena
  Galaxy Atlas 2020},} \apjs, 269, 3, \dodoi{10.3847/1538-4365/acfaa2}

\bibitem[{A.~B. {Newman} {et~al.}(2017){Newman}, {Smith}, {Conroy}, {Villaume},
  \& {van Dokkum}}]{Newman2017}
{Newman}, A.~B., {Smith}, R.~J., {Conroy}, C., {Villaume}, A., \& {van Dokkum},
  P. 2017, \bibinfo{title}{{The Initial Mass Function in the Nearest Strong
  Lenses from SNELLS: Assessing the Consistency of Lensing, Dynamical, and
  Spectroscopic Constraints},} \apj, 845, 157, \dodoi{10.3847/1538-4357/aa816d}

\bibitem[{L. {Oldham} {\&} M. {Auger}(2018){Oldham} \& {Auger}}]{Oldham2018}
{Oldham}, L., \& {Auger}, M. 2018, \bibinfo{title}{{Galaxy structure from
  multiple tracers - III. Radial variations in M87's IMF},} \mnras, 474, 4169,
  \dodoi{10.1093/mnras/stx2969}

\bibitem[{L. Oser {et~al.}(2012)Oser, Naab, Ostriker, \& Johansson}]{Oser2012}
Oser, L., Naab, T., Ostriker, J.~P., \& Johansson, P.~H. 2012,
  \bibinfo{title}{{The Cosmological Size and Velocity Dispersion Evolution of
  Massive Early-type Galaxies},} ApJ, 744, 63

\bibitem[{T. {Parikh} {et~al.}(2024){Parikh}, {Saglia}, {Thomas}, {Mehrgan},
  {Bender}, \& {Maraston}}]{Parikh2024}
{Parikh}, T., {Saglia}, R., {Thomas}, J., {et~al.} 2024,
  \bibinfo{title}{{Stellar populations of massive early-type galaxies observed
  by MUSE},} \mnras, 528, 7338, \dodoi{10.1093/mnras/stae448}

\bibitem[{T. {Parikh} {et~al.}(2018){Parikh}, {Thomas}, {Maraston}, {Westfall},
  {Goddard}, {Lian}, {Meneses-Goytia}, {Jones}, {Vaughan}, {Andrews},
  {Bershady}, {Bizyaev}, {Brinkmann}, {Brownstein}, {Bundy}, {Drory},
  {Emsellem}, {Law}, {Newman}, {Roman-Lopes}, {Wake}, {Yan}, \&
  {Zheng}}]{Parikh2018}
{Parikh}, T., {Thomas}, D., {Maraston}, C., {et~al.} 2018,
  \bibinfo{title}{{SDSS-IV MaNGA: the spatially resolved stellar initial mass
  function in {\ensuremath{\sim}}400 early-type galaxies},} \mnras, 477, 3954,
  \dodoi{10.1093/mnras/sty785}

\bibitem[{J. {Pilawa} {et~al.}(2025){Pilawa}, {Liepold}, {Ma}, {Walsh}, \&
  {Greene}}]{Pilawa2025}
{Pilawa}, J., {Liepold}, E.~R., {Ma}, C.-P., {Walsh}, J.~L., \& {Greene}, J.~E.
  2025, \bibinfo{title}{{The MASSIVE Survey. XX. A Triaxial Stellar Dynamical
  Measurement of the Supermassive Black Hole Mass and Intrinsic Galaxy Shape of
  Giant Radio Galaxy NGC 315},} \apj, 989, 98, \dodoi{10.3847/1538-4357/adee1e}

\bibitem[{J.~D. {Pilawa} {et~al.}(2022){Pilawa}, {Liepold}, {Delgado Andrade},
  {Walsh}, {Ma}, {Quenneville}, {Greene}, \& {Blakeslee}}]{Pilawa2022}
{Pilawa}, J.~D., {Liepold}, E.~R., {Delgado Andrade}, S.~C., {et~al.} 2022,
  \bibinfo{title}{{The MASSIVE Survey. XVII. A Triaxial Orbit-based
  Determination of the Black Hole Mass and Intrinsic Shape of Elliptical Galaxy
  NGC 2693},} \apj, 928, 178, \dodoi{10.3847/1538-4357/ac58fd}

\bibitem[{S. {Posacki} {et~al.}(2015){Posacki}, {Cappellari}, {Treu},
  {Pellegrini}, \& {Ciotti}}]{Posacki2015}
{Posacki}, S., {Cappellari}, M., {Treu}, T., {Pellegrini}, S., \& {Ciotti}, L.
  2015, \bibinfo{title}{{The stellar initial mass function of early-type
  galaxies from low to high stellar velocity dispersion: homogeneous analysis
  of ATLAS$^{3D}$ and Sloan Lens ACS galaxies},} \mnras, 446, 493,
  \dodoi{10.1093/mnras/stu2098}

\bibitem[{J. {Prochaska} {et~al.}(2020){Prochaska}, {Hennawi}, {Westfall},
  {Cooke}, {Wang}, {Hsyu}, {Davies}, {Farina}, \& {Pelliccia}}]{pypeit_JOSS}
{Prochaska}, J., {Hennawi}, J., {Westfall}, K., {et~al.} 2020,
  \bibinfo{title}{{PypeIt: The Python Spectroscopic Data Reduction Pipeline},}
  The Journal of Open Source Software, 5, 2308, \dodoi{10.21105/joss.02308}

\bibitem[{J.~X. {Prochaska} {et~al.}(2020){Prochaska}, {Hennawi}, {Cooke},
  {Westfall}, {Wang}, {EmAstro}, {Tiffanyhsyu}, {Wasserman}, {Villaume},
  {Marijana777}, {Schindler}, {Young}, {Simha}, {Wilde}, {Tejos}, {Isbell},
  {Fl{\"o}rs}, {Sandford}, {Vasovi{\'c}}, {Betts}, \& {Holden}}]{pypeit_Zenodo}
{Prochaska}, J.~X., {Hennawi}, J., {Cooke}, R., {et~al.} 2020, {pypeit/PypeIt:
  Release 1.0.0}, v1.0.0 Zenodo, \dodoi{10.5281/zenodo.3743493}

\bibitem[{M.~E. {Quenneville} {et~al.}(2024){Quenneville}, {Blakeslee}, {Ma},
  {Greene}, {Gwyn}, {Ciccone}, \& {Nyiri}}]{Quenneville2024}
{Quenneville}, M.~E., {Blakeslee}, J.~P., {Ma}, C.-P., {et~al.} 2024,
  \bibinfo{title}{{The MASSIVE survey - XVIII. Deep wide-field K-band
  photometry and local scaling relations for massive early-type galaxies},}
  \mnras, 527, 249, \dodoi{10.1093/mnras/stad3137}

\bibitem[{T.~D. {Rawle} {et~al.}(2010){Rawle}, {Smith}, \& {Lucey}}]{Rawle2010}
{Rawle}, T.~D., {Smith}, R.~J., \& {Lucey}, J.~R. 2010,
  \bibinfo{title}{{Stellar population gradients in early-type cluster
  galaxies},} \mnras, 401, 852, \dodoi{10.1111/j.1365-2966.2009.15722.x}

\bibitem[{V. {Rodriguez-Gomez} {et~al.}(2016){Rodriguez-Gomez}, {Pillepich},
  {Sales}, {Genel}, {Vogelsberger}, {Zhu}, {Wellons}, {Nelson}, {Torrey},
  {Springel}, {Ma}, \& {Hernquist}}]{Rodriguez-Gomez2016}
{Rodriguez-Gomez}, V., {Pillepich}, A., {Sales}, L.~V., {et~al.} 2016,
  \bibinfo{title}{{The stellar mass assembly of galaxies in the Illustris
  simulation: growth by mergers and the spatial distribution of accreted
  stars},} \mnras, 458, 2371, \dodoi{10.1093/mnras/stw456}

\bibitem[{G. {Rosani} {et~al.}(2018){Rosani}, {Pasquali}, {La Barbera},
  {Ferreras}, \& {Vazdekis}}]{Rosani2018}
{Rosani}, G., {Pasquali}, A., {La Barbera}, F., {Ferreras}, I., \& {Vazdekis},
  A. 2018, \bibinfo{title}{{The influence of galaxy environment on the stellar
  initial mass function of early-type galaxies},} \mnras, 476, 5233,
  \dodoi{10.1093/mnras/sty528}

\bibitem[{E.~E. {Salpeter}(1955){Salpeter}}]{Salpeter1955}
{Salpeter}, E.~E. 1955, \bibinfo{title}{{The Luminosity Function and Stellar
  Evolution.},} \apj, 121, 161, \dodoi{10.1086/145971}

\bibitem[{G. {Santucci} {et~al.}(2020){Santucci}, {Brough}, {Scott}, {Montes},
  {Owers}, {van Sande}, {Bland-Hawthorn}, {Bryant}, {Croom}, {Ferreras},
  {Lawrence}, {L{\'o}pez-S{\'a}nchez}, \& {Richards}}]{Santucci2020}
{Santucci}, G., {Brough}, S., {Scott}, N., {et~al.} 2020, \bibinfo{title}{{The
  SAMI Galaxy Survey: Stellar Population Gradients of Central Galaxies},} \apj,
  896, 75, \dodoi{10.3847/1538-4357/ab92a9}

\bibitem[{M. {Sarzi} {et~al.}(2018){Sarzi}, {Spiniello}, {La Barbera},
  {Krajnovi{\'c}}, \& {van den Bosch}}]{Sarzi2018}
{Sarzi}, M., {Spiniello}, C., {La Barbera}, F., {Krajnovi{\'c}}, D., \& {van
  den Bosch}, R. 2018, \bibinfo{title}{{MUSE observations of M87: radial
  gradients for the stellar initial-mass function and the abundance of
  sodium},} \mnras, 478, 4084, \dodoi{10.1093/mnras/sty1092}

\bibitem[{J.~M. {Scalo}(1986){Scalo}}]{Scalo1986}
{Scalo}, J.~M. 1986, \bibinfo{title}{{The Stellar Initial Mass Function},}
  \fcp, 11, 1

\bibitem[{M. {Schwarzschild}(1979){Schwarzschild}}]{Schwarzschild1979}
{Schwarzschild}, M. 1979, \bibinfo{title}{{A numerical model for a triaxial
  stellar system in dynamical equilibrium.},} \apj, 232, 236,
  \dodoi{10.1086/157282}

\bibitem[{S. Seabold {\&} J. Perktold(2010)Seabold \&
  Perktold}]{seabold2010statsmodels}
Seabold, S., \& Perktold, J. 2010, \bibinfo{title}{statsmodels: Econometric and
  statistical modeling with python,} in 9th Python in Science Conference

\bibitem[{J.~L. {S{\'e}rsic}(1963){S{\'e}rsic}}]{Sersic1963}
{S{\'e}rsic}, J.~L. 1963, \bibinfo{title}{{Influence of the atmospheric and
  instrumental dispersion on the brightness distribution in a galaxy},} Boletin
  de la Asociacion Argentina de Astronomia La Plata Argentina, 6, 41

\bibitem[{P. {Sharda} {\&} M.~R. {Krumholz}(2022){Sharda} \&
  {Krumholz}}]{Sharda2022}
{Sharda}, P., \& {Krumholz}, M.~R. 2022, \bibinfo{title}{{When did the initial
  mass function become bottom-heavy?},} \mnras, 509, 1959,
  \dodoi{10.1093/mnras/stab2921}

\bibitem[{S. {Shetty} {\&} M. {Cappellari}(2014){Shetty} \&
  {Cappellari}}]{Shetty2014}
{Shetty}, S., \& {Cappellari}, M. 2014, \bibinfo{title}{{Salpeter Normalization
  of the Stellar Initial Mass Function for Massive Galaxies at z
  \raisebox{-0.5ex}\textasciitilde 1},} \apjl, 786, L10,
  \dodoi{10.1088/2041-8205/786/2/L10}

\bibitem[{D.~A. {Simon} {et~al.}(2024){Simon}, {Cappellari}, \&
  {Hartke}}]{Simon2024}
{Simon}, D.~A., {Cappellari}, M., \& {Hartke}, J. 2024,
  \bibinfo{title}{{Supermassive black hole mass in the massive elliptical
  galaxy M87 from integral-field stellar dynamics using OASIS and MUSE with
  adaptive optics: assessing systematic uncertainties},} \mnras, 527, 2341,
  \dodoi{10.1093/mnras/stad3309}

\bibitem[{A. {Smette} {et~al.}(2015){Smette}, {Sana}, {Noll}, {Horst},
  {Kausch}, {Kimeswenger}, {Barden}, {Szyszka}, {Jones}, {Gallenne}, {Vinther},
  {Ballester}, \& {Taylor}}]{Smette2015}
{Smette}, A., {Sana}, H., {Noll}, S., {et~al.} 2015, \bibinfo{title}{{Molecfit:
  A general tool for telluric absorption correction. I. Method and application
  to ESO instruments},} \aap, 576, A77, \dodoi{10.1051/0004-6361/201423932}

\bibitem[{R.~J. {Smith}(2014){Smith}}]{Smith2014}
{Smith}, R.~J. 2014, \bibinfo{title}{{Variations in the initial mass function
  in early-type galaxies: a critical comparison between dynamical and
  spectroscopic results.},} \mnras, 443, L69, \dodoi{10.1093/mnrasl/slu082}

\bibitem[{R.~J. {Smith}(2020){Smith}}]{Smith2020}
{Smith}, R.~J. 2020, \bibinfo{title}{{Evidence for Initial Mass Function
  Variation in Massive Early-Type Galaxies},} \araa, 58, 577,
  \dodoi{10.1146/annurev-astro-032620-020217}

\bibitem[{R.~J. Smith {et~al.}(2012)Smith, Lucey, \& Carter}]{Smith2012}
Smith, R.~J., Lucey, J.~R., \& Carter, D. 2012, \bibinfo{title}{The stellar
  initial mass function in red-sequence galaxies: 1-μm spectroscopy of {Coma}
  cluster galaxies with {Subaru}/{FMOS}: {The} stellar {IMF} in red-sequence
  galaxies,} Monthly Notices of the Royal Astronomical Society, 426, 2994,
  \dodoi{10.1111/j.1365-2966.2012.21922.x}

\bibitem[{A. {Sonnenfeld} {et~al.}(2019){Sonnenfeld}, {Jaelani}, {Chan},
  {More}, {Suyu}, {Wong}, {Oguri}, \& {Lee}}]{Sonnenfeld2019}
{Sonnenfeld}, A., {Jaelani}, A.~T., {Chan}, J., {et~al.} 2019,
  \bibinfo{title}{{Survey of gravitationally-lensed objects in HSC imaging
  (SuGOHI). III. Statistical strong lensing constraints on the stellar IMF of
  CMASS galaxies},} \aap, 630, A71, \dodoi{10.1051/0004-6361/201935743}

\bibitem[{A. {Sonnenfeld} {et~al.}(2015){Sonnenfeld}, {Treu}, {Marshall},
  {Suyu}, {Gavazzi}, {Auger}, \& {Nipoti}}]{Sonnenfeld2015}
{Sonnenfeld}, A., {Treu}, T., {Marshall}, P.~J., {et~al.} 2015,
  \bibinfo{title}{{The SL2S Galaxy-scale Lens Sample. V. Dark Matter Halos and
  Stellar IMF of Massive Early-type Galaxies Out to Redshift 0.8},} \apj, 800,
  94, \dodoi{10.1088/0004-637X/800/2/94}

\bibitem[{C. {Spiniello} {et~al.}(2011){Spiniello}, {Koopmans}, {Trager},
  {Czoske}, \& {Treu}}]{Spiniello2011}
{Spiniello}, C., {Koopmans}, L.~V.~E., {Trager}, S.~C., {Czoske}, O., \&
  {Treu}, T. 2011, \bibinfo{title}{{The X-Shooter Lens Survey - I. Dark matter
  domination and a Salpeter-type initial mass function in a massive early-type
  galaxy},} \mnras, 417, 3000, \dodoi{10.1111/j.1365-2966.2011.19458.x}

\bibitem[{C. {Spiniello} {et~al.}(2014){Spiniello}, {Trager}, {Koopmans}, \&
  {Conroy}}]{Spiniello2014}
{Spiniello}, C., {Trager}, S., {Koopmans}, L. V.~E., \& {Conroy}, C. 2014,
  \bibinfo{title}{{The stellar IMF in early-type galaxies from a non-degenerate
  set of optical line indices},} \mnras, 438, 1483,
  \dodoi{10.1093/mnras/stt2282}

\bibitem[{M. {Spolaor} {et~al.}(2010){Spolaor}, {Kobayashi}, {Forbes}, {Couch},
  \& {Hau}}]{Spolaor2010}
{Spolaor}, M., {Kobayashi}, C., {Forbes}, D.~A., {Couch}, W.~J., \& {Hau},
  G.~K.~T. 2010, \bibinfo{title}{{Early-type galaxies at large galactocentric
  radii - II. Metallicity gradients and the [Z/H]-mass, [{\$\alpha\$}/Fe]-mass
  relations},} \mnras, 408, 272, \dodoi{10.1111/j.1365-2966.2010.17080.x}

\bibitem[{M. {Spolaor} {et~al.}(2009){Spolaor}, {Proctor}, {Forbes}, \&
  {Couch}}]{Spolaor2009}
{Spolaor}, M., {Proctor}, R.~N., {Forbes}, D.~A., \& {Couch}, W.~J. 2009,
  \bibinfo{title}{{The Mass-Metallicity Gradient Relation of Early-Type
  Galaxies},} \apjl, 691, L138, \dodoi{10.1088/0004-637X/691/2/L138}

\bibitem[{B. {Tang} {\&} G. {Worthey}(2017){Tang} \& {Worthey}}]{Tang2017}
{Tang}, B., \& {Worthey}, G. 2017, \bibinfo{title}{{Optical spectroscopy and
  initial mass function of z = 0.4 red galaxies},} \mnras, 467, 674,
  \dodoi{10.1093/mnras/stx099}

\bibitem[{S. {Thater} {et~al.}(2023){Thater}, {Lyubenova}, {Fahrion},
  {Mart{\'\i}n-Navarro}, {Jethwa}, {Nguyen}, \& {van de Ven}}]{Thater2023}
{Thater}, S., {Lyubenova}, M., {Fahrion}, K., {et~al.} 2023,
  \bibinfo{title}{{Effect of the initial mass function on the dynamical SMBH
  mass estimate in the nucleated early-type galaxy FCC 47},} \aap, 675, A18,
  \dodoi{10.1051/0004-6361/202245362}

\bibitem[{J. {Thomas} {et~al.}(2011){Thomas}, {Saglia}, {Bender}, {Thomas},
  {Gebhardt}, {Magorrian}, {Corsini}, {Wegner}, \& {Seitz}}]{Thomas2011}
{Thomas}, J., {Saglia}, R.~P., {Bender}, R., {et~al.} 2011,
  \bibinfo{title}{{Dynamical masses of early-type galaxies: a comparison to
  lensing results and implications for the stellar initial mass function and
  the distribution of dark matter},} \mnras, 415, 545,
  \dodoi{10.1111/j.1365-2966.2011.18725.x}

\bibitem[{Y.-S. {Ting} {et~al.}(2018{\natexlab{a}}){Ting}, {Conroy}, {Rix}, \&
  {Asplund}}]{Ting2018}
{Ting}, Y.-S., {Conroy}, C., {Rix}, H.-W., \& {Asplund}, M. 2018{\natexlab{a}},
  \bibinfo{title}{{Measuring Oxygen Abundances from Stellar Spectra without
  Oxygen Lines},} \apj, 860, 159, \dodoi{10.3847/1538-4357/aac6c9}

\bibitem[{B.~M. Tinsley(1979)Tinsley}]{Tinsley1979}
Tinsley, B.~M. 1979, \bibinfo{title}{{Stellar lifetimes and abundance ratios in
  chemical evolution},} ApJ, 229, 1046

\bibitem[{C. {Tortora} {et~al.}(2014){Tortora}, {Napolitano}, {Saglia},
  {Romanowsky}, {Covone}, \& {Capaccioli}}]{Tortora2014}
{Tortora}, C., {Napolitano}, N.~R., {Saglia}, R.~P., {et~al.} 2014,
  \bibinfo{title}{{Evolution of central dark matter of early-type galaxies up
  to z {\ensuremath{\sim}} 0.8},} \mnras, 445, 162,
  \dodoi{10.1093/mnras/stu1712}

\bibitem[{C. {Tortora} {et~al.}(2013){Tortora}, {Romanowsky}, \&
  {Napolitano}}]{Tortora2013}
{Tortora}, C., {Romanowsky}, A.~J., \& {Napolitano}, N.~R. 2013,
  \bibinfo{title}{{An Inventory of the Stellar Initial Mass Function in
  Early-type Galaxies},} \apj, 765, 8, \dodoi{10.1088/0004-637X/765/1/8}

\bibitem[{T. {Treu}(2010){Treu}}]{Treu2010}
{Treu}, T. 2010, \bibinfo{title}{{Strong Lensing by Galaxies},} \araa, 48, 87,
  \dodoi{10.1146/annurev-astro-081309-130924}

\bibitem[{A. van~der Wel {et~al.}(2014)van~der Wel, Franx, van Dokkum, Skelton,
  Momcheva, Whitaker, Brammer, Bell, Rix, Wuyts, Ferguson, Holden, Barro,
  Koekemoer, Chang, McGrath, H{\"a}ussler, Dekel, Behroozi, Fumagalli, Leja,
  Lundgren, Maseda, Nelson, Wake, Patel, Labb{\'e}, Faber, Grogin, \&
  Kocevski}]{vanderWel2014}
van~der Wel, A., Franx, M., van Dokkum, P.~G., {et~al.} 2014,
  \bibinfo{title}{{3D-HST+CANDELS: The Evolution of the Galaxy Size-Mass
  Distribution since z = 3},} \apjl, 788, 28

\bibitem[{P. van Dokkum {et~al.}(2017)van Dokkum, Conroy, {Villaume}, {Brodie},
  \& {Romanowsky}}]{vanDokkum2017}
van Dokkum, P., Conroy, C., {Villaume}, A., {Brodie}, J., \& {Romanowsky},
  A.~J. 2017, \bibinfo{title}{{The Stellar Initial Mass Function in Early-type
  Galaxies from Absorption Line Spectroscopy. III. Radial Gradients},} ApJ,
  841, 68, \dodoi{10.3847/1538-4357/aa7135}

\bibitem[{P.~G. {van Dokkum} {et~al.}(2012{\natexlab{a}}){van Dokkum}, {Bloom},
  \& {Tewes}}]{vanDokkum2012}
{van Dokkum}, P.~G., {Bloom}, J., \& {Tewes}, M. 2012{\natexlab{a}},
  {L.A.Cosmic: Laplacian Cosmic Ray Identification}, \doeprint{1207.005}

\bibitem[{P.~G. {van Dokkum} {et~al.}(2012{\natexlab{b}}){van Dokkum}, {Bloom},
  \& {Tewes}}]{2012ascl.soft07005V}
{van Dokkum}, P.~G., {Bloom}, J., \& {Tewes}, M. 2012{\natexlab{b}},
  {L.A.Cosmic: Laplacian Cosmic Ray Identification}, \doeprint{1207.005}

\bibitem[{P.~G. van Dokkum {et~al.}(2010)van Dokkum, Whitaker, Brammer, Franx,
  Kriek, Labb{\'e}, Marchesini, Quadri, Bezanson, Illingworth, Muzzin, Rudnick,
  Tal, \& Wake}]{vanDokkum2010}
van Dokkum, P.~G., Whitaker, K.~E., Brammer, G., {et~al.} 2010,
  \bibinfo{title}{{The Growth of Massive Galaxies Since z = 2},} ApJ, 709, 1018

\bibitem[{G. Van~Rossum {\&} F.~L. Drake(2009)Van~Rossum \&
  Drake}]{VanRossum2009}
Van~Rossum, G., \& Drake, F.~L. 2009, Python 3 Reference Manual (Scotts Valley,
  CA: CreateSpace)

\bibitem[{G. Van~Rossum {\&} F.~L. Drake~Jr(1995)Van~Rossum \&
  Drake~Jr}]{VanRossum1995}
Van~Rossum, G., \& Drake~Jr, F.~L. 1995, Python reference manual (Centrum voor
  Wiskunde en Informatica Amsterdam)

\bibitem[{S.~P. {Vaughan} {et~al.}(2018{\natexlab{a}}){Vaughan}, {Davies},
  {Zieleniewski}, \& {Houghton}}]{Vaughan2018a}
{Vaughan}, S.~P., {Davies}, R.~L., {Zieleniewski}, S., \& {Houghton}, R. C.~W.
  2018{\natexlab{a}}, \bibinfo{title}{{The stellar population and initial mass
  function of NGC 1399 with MUSE},} \mnras, 479, 2443,
  \dodoi{10.1093/mnras/sty1434}

\bibitem[{S.~P. {Vaughan} {et~al.}(2018{\natexlab{b}}){Vaughan}, {Davies},
  {Zieleniewski}, \& {Houghton}}]{Vaughan2018b}
{Vaughan}, S.~P., {Davies}, R.~L., {Zieleniewski}, S., \& {Houghton}, R. C.~W.
  2018{\natexlab{b}}, \bibinfo{title}{{Radial measurements of IMF-sensitive
  absorption features in two massive ETGs},} \mnras, 475, 1073,
  \dodoi{10.1093/mnras/stx3199}

\bibitem[{M. {Veale} {et~al.}(2017{\natexlab{a}}){Veale}, {Ma}, {Greene},
  {Thomas}, {Blakeslee}, {McConnell}, {Walsh}, \& {Ito}}]{Veale2017b}
{Veale}, M., {Ma}, C.-P., {Greene}, J.~E., {et~al.} 2017{\natexlab{a}},
  \bibinfo{title}{{The MASSIVE Survey - VII. The relationship of angular
  momentum, stellar mass and environment of early-type galaxies},} \mnras, 471,
  1428, \dodoi{10.1093/mnras/stx1639}

\bibitem[{M. Veale {et~al.}(2018)Veale, Ma, Greene, Thomas, Blakeslee, Walsh,
  \& Ito}]{Veale2018}
Veale, M., Ma, C.-P., Greene, J.~E., {et~al.} 2018, \bibinfo{title}{{The
  MASSIVE survey {\textendash} VIII. Stellar velocity dispersion profiles and
  environmental dependence of early-type galaxies},} Monthly Notices of the
  Royal Astronomical Society, 473, 5446

\bibitem[{M. {Veale} {et~al.}(2017{\natexlab{b}}){Veale}, {Ma}, {Thomas},
  {Greene}, {McConnell}, {Walsh}, {Ito}, {Blakeslee}, \& {Janish}}]{Veale2017a}
{Veale}, M., {Ma}, C.-P., {Thomas}, J., {et~al.} 2017{\natexlab{b}},
  \bibinfo{title}{{The MASSIVE Survey - V. Spatially resolved stellar angular
  momentum, velocity dispersion, and higher moments of the 41 most massive
  local early-type galaxies},} \mnras, 464, 356, \dodoi{10.1093/mnras/stw2330}

\bibitem[{A. Villaume {et~al.}(2017)Villaume, Conroy, Johnson, Rayner, Mann, \&
  van Dokkum}]{Villaume2017}
Villaume, A., Conroy, C., Johnson, B., {et~al.} 2017, \bibinfo{title}{{The
  Extended IRTF Spectral Library: Expanded Coverage in Metallicity,
  Temperature, and Surface Gravity},} ASTROPHYS J SUPPL S, 230, 23

\bibitem[{P. Virtanen {et~al.}(2020)Virtanen, Gommers, Oliphant, Haberland,
  Reddy, Cournapeau, Burovski, Peterson, Weckesser, Bright, {van der Walt},
  Brett, Wilson, Millman, Mayorov, Nelson, Jones, Kern, Larson, Carey, Polat,
  Feng, Moore, {VanderPlas}, Laxalde, Perktold, Cimrman, Henriksen, Quintero,
  Harris, Archibald, Ribeiro, Pedregosa, {van Mulbregt}, \& {SciPy 1.0
  Contributors}}]{2020SciPy-NMeth}
Virtanen, P., Gommers, R., Oliphant, T.~E., {et~al.} 2020,
  \bibinfo{title}{{{SciPy} 1.0: Fundamental Algorithms for Scientific Computing
  in Python},} Nature Methods, 17, 261, \dodoi{10.1038/s41592-019-0686-2}

\bibitem[{G.~A. {Wegner} {et~al.}(2012){Wegner}, {Corsini}, {Thomas}, {Saglia},
  {Bender}, \& {Pu}}]{Wegner2012}
{Wegner}, G.~A., {Corsini}, E.~M., {Thomas}, J., {et~al.} 2012,
  \bibinfo{title}{{Further Evidence for Large Central Mass-to-light Ratios in
  Early-type Galaxies: The Case of Ellipticals and Lenticulars in the A262
  Cluster},} \aj, 144, 78, \dodoi{10.1088/0004-6256/144/3/78}

\bibitem[{R.~F. {Wing} {\&} J. {Ford}(1969){Wing} \& {Ford}}]{Wing1969}
{Wing}, R.~F., \& {Ford}, W.~Kent, J. 1969, \bibinfo{title}{{The Infrared
  Spectrum of the Cool Dwarf Wolf 359},} \pasp, 81, 527, \dodoi{10.1086/128814}

\bibitem[{S. {Zibetti} {et~al.}(2020){Zibetti}, {Gallazzi}, {Hirschmann},
  {Consolandi}, {Falc{\'o}n-Barroso}, {van de Ven}, \&
  {Lyubenova}}]{Zibetti2020}
{Zibetti}, S., {Gallazzi}, A.~R., {Hirschmann}, M., {et~al.} 2020,
  \bibinfo{title}{{Insights into formation scenarios of massive early-type
  galaxies from spatially resolved stellar population analysis in CALIFA},}
  \mnras, 491, 3562, \dodoi{10.1093/mnras/stz3205}

\end{thebibliography}

\end{document}